# NASA Innovative Advanced Concepts PHASE I FINAL REPORT

# A Lunar Long-Baseline UV/Optical Imaging Interferometer:

# Artemis-enabled Stellar Imager (AeSI)

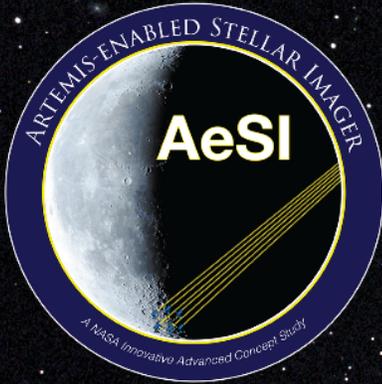
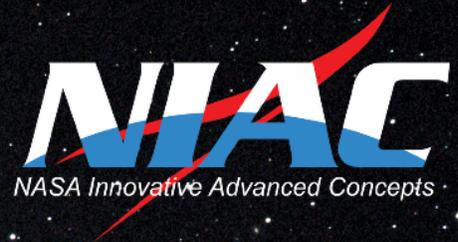
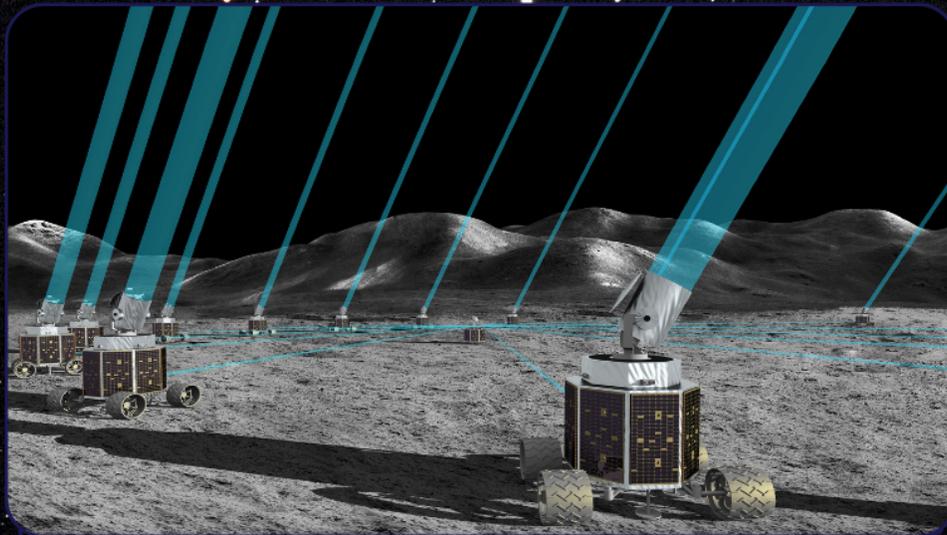
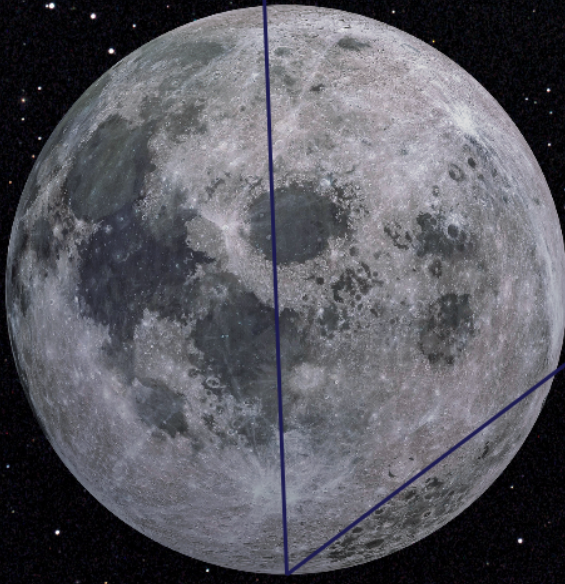

February 2025

Dr. Kenneth G. Carpenter

and the AeSI Team

# *Dedications*

For

Carolus J. Schrijver (LMATC)

Co-creator of the Stellar Imager (SI)
Vision Mission Concept

and

Richard G. Lyon (NASA's GSFC) and Ron Allen (STScI)

Two additional key players in the development of SI,
also gone too soon.

# The AeSI Team (*in alphabetical order*)

| Name | Affiliation | Role |
|---|---|---|
| *Ken Carpenter* | NASA/GSFC | NIAC Fellow, Mission Implementation Lead |
| *Tabetha Boyajian* | Louisiana State University | Ground Interferometry Expert |
| *Derek Buzasi* | University of Chicago | Asteroseismology expert |
| *Jim Clark* | Naval Research Lab | Mechanical Engineer |
| *Michelle Creech-Eakman* | New Mexico Tech | Ground Interferometry Expert |
| *Bruce Dean* | NASA/GSFC | Optical Engineer/WS&C, AI/ML |
| *Ashley Elliott* | Louisiana State University | Graduate Student |
| *Julianne Foster* | BAE Systems | System Engineer |
| *Qian Gong* | NASA/GSFC | Optical Engineer |
| *Margarita Karovska* | CfA \| Harvard & Smithsonian | Science Definition Co-Lead |
| *David Kim* | NASA/GSFC | Power Systems Engineer |
| *Jon Hulberg* | CUA | Science Definition |
| *David Leisawitz* | NASA/GSFC | Space Interferometry Expert |
| *Mike Maher* | BAE Systems | System Engineer |
| *Jon Morse* | California Institute of Technology | Senior Advisor, Lunar Science & Infrastructure |
| *Dave Mozurkewich* | Seabrook Engineering | Lead System Engineer, Time Evolution of Observatory |
| *Sarah Peacock* | UMBC/NASA/GSFC | Science Definition, Study Co-Manager, Outreach Co-Lead |
| *Noah Petro* | NASA/GSFC | Artemis Expert |
| *Gioia Rau* | NSF, NASA/GSFC | Science Definition Co-Lead, Study Co-Manager, Outreach Co-Lead |
| *Paul Scowen* | NASA/GSFC | Science Definition |
| *Len Seals* | NASA/GSFC | Scattered Light/Optical Engineer |
| *Walter Smith* | NASA/GSFC | Mechanical Engineer |
| *Max Smuda* | Florida Gulf Coast University | Graduate Student |
| *Breann Sitarski* | NASA/GSFC | Optical Engineer |
| *Buddy Taylor* | NASA/GSFC | Mechanical Engineer |
| *Gerard van Belle* | Lowell Observatory | Interferometry Expert, Mission Design Lead |
| *Erik Wilkinson* | BAE Systems | System Engineer |

# TABLE OF CONTENTS

Image credits for Cover Art:
*Background image*: Open Source Image via Canva
*AeSI logo*: Gladys Kober
*AeSI artist rendering*: Britt Griswold
*Full Design:* Ashley Elliott

# LIST OF ABBREVIATIONS

A – Angstrom (1x10$^{-10}$ meters)
*AeSI – Artemis enabled Stellar Imager*
AGB – Asymptotic Giant Branch
AGN – Active Galactic Nuclei
APG – Annealed Pyrolytic Graphite
bps – bits per second
BELR – Broad Emission-Line Region
BSnum – Beam Splitter in beam train
C – Centrigrade temperature
CAD – Computer Aided Design
CH – Camera Head
Chandra – NASA X-Ray observatory
CHARA – Center for High Angular Resolution Astronomy
CLPS – Commercial Lunar Payload Services
CMOS – Complementary Metal-Oxide Semiconductor
COTS – Commercial off-the-shelf
CUA – Catholic University of America
CV – Cataclysmic Variable
DPU – Data Processing Unit
e- – electron
EMCCD - Electron Multiplying Charge Coupled Device
EUV – Extreme Ultraviolet wavelengths
FPA – Focal Plane Array
FSS – Farside Seismic Suite
FUSE – Far Ultraviolet Spectroscopic Explorer – NASA observatory
GB – Gigabytes
GRC – Glenn Research Center
GSFC – Goddard Space Flight Center
H$_{num}$ – Numbered mirror in the hub
HITL – Human-in-the-loop
HLS – Human Landing System
HST – Hubble Space Telescope – NASA Observatory
HWO – Habitable Words Observatory
IDC – Integrated Design Center
IDL – Instrument Design Lab
ILSA – Instrument for Lunar Seismic Activity
IMU – Inertial measurement unit
IOC-C – Initial Operating Capability - C

IRAD – internal research and development
K – Kelvin temperature
kbps – kilobits per second
kg -- kilogram
km -- kilometer
kpc – kiloparsec
kW – kilowatt
l – liter
L2 – Lagrange Point 2
LCRNS – Lunar Communications Relay and Navigation Systems
LEMS – Lunar Environment Monitoring Station
LMC – Large Magellanic Cloud
LUT – Chinese Lunar based telescope
Lya – Lyman Alpha
λ – lambda -- wavelength
m – meter
$m_v$ – visual magnitude
$M_o$ – 1 solar mass
MOC – Mission Operations Center
$M_{num}$ – Numbered mirror in the beam train
MAPP – Mobile Autonomous Prospecting Platform
mas – milliarcsecond
Mbps – Megabits per second
MDL – Mission Design Lab
MHD – magnetohydrodynamic
MIRC-X – Michigan Infrared Combiner-eXeter -- infrared beam combiner at CHARA
mm – millimeter
Mpc – megaparsec
MYSTIC – Michigan Young Star Imager -- optical beam combiner at CHARA
N – some number, n
NASA – National Aeronautics and Space Administration
NIAC – NASA Innovative Advanced Concepts
nm – nanometer
NPOI – Navy Precision Optical Interferometer
NSF – National Science Foundation
NUV – Near Ultraviolet wavelengths
Ω – solid angle
pc – parsec
pix – pixels
PNT – pointing, navigation and timing

QE – Quantum Efficiency
RAC-1 – Regolith Adherence Characterization
Resel – Resolution element
RF – Radio Frequency
RFI – Radio frequency interference
RLOF – Roche Lobe Overflow
RSG – Red supergiant
SI – Stellar Imager
SLS – Space Launch System
SN – Supernova
SNR – Signal-to-Noise Ratio
STIS – Space Telescope Imaging Spectrograph on Hubble Space Telescope
STMD – Space Technology Mission Directorate
TLS – transit light source effect
TRL – Technology Readiness Level
UMBC – University of Maryland, Baltimore County
UV – Ultraviolet wavelengths
V – Visibility amplitude
VISION – Visible Imaging System for Interferometric Observations -- optical beam combiner at
VLBI – Very Long Baseline Interferometry
VLTI – Very Large Telescope Interferometer
VM – Vision Mission
W – Watt
WD – White dwarf
XMM – Newton – X-ray Multi-Mirror Mission – Newton observatory of the European Space Agency
YSO – Young Stellar Object

# EXECUTIVE SUMMARY

This report presents the findings of a 9-month NIAC Phase I feasibility study for the *Artemis-enabled Stellar Imager (AeSI)*, a proposed high-resolution ultraviolet (UV)/optical interferometer designed for deployment on the lunar surface. Its primary science goal is to image the surfaces and interiors of stars with unprecedented detail, revealing new details about their magnetic processes and dynamic evolution. This capability will transform our understanding of stellar physics and has broad applicability across astrophysics, from resolving the cores of Active Galactic Nuclei (AGN) to studying supernovae, planetary nebulae, and the late stages of stellar evolution. By leveraging the stable vacuum environment of the Moon and the infrastructure being established for the Artemis Program, *AeSI* presents a compelling case for a lunar-based interferometer.

In this study, the *AeSI* Team, working with NASA's Goddard Space Flight Center's (GSFC) Integrated Design Center (IDC), has firmly established the feasibility of building and operating a reconfigurable, dispersed aperture telescope (i.e., an interferometer) on the lunar surface. The collaboration produced a credible Baseline design featuring 15 primary mirrors arranged in a 1 km diameter array, with the potential to expand to 30 mirrors and larger array sizes through staged deployments. Additionally, this study identified numerous opportunities for optimization and the necessary trade studies to refine the design further. These will be pursued in follow-up investigations, such as a NIAC Phase II study, to advance the concept toward implementation.

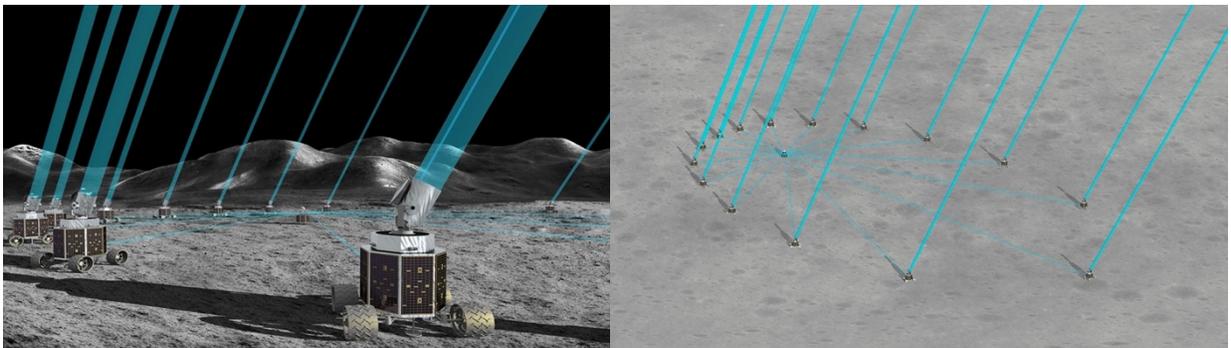

**Figure 0-1:** The Baseline design for the *Artemis-enabled Stellar Imager* (*AeSI*). Image Credit: Britt Griswold

# 1 Study Background and Assumptions

## 1.1 Introduction and Motivation

The desire to expand our understanding of the Universe, both near and far, has always driven advancements in astronomical instrumentation. Higher sensitivity and higher angular resolution are essential to answering fundamental astrophysical questions. NASA's return to the Moon[1] offers a transformative opportunity to take practical steps toward realizing high-impact scientific capabilities, particularly in the realm of interferometric imaging.

Traditional imaging methods fall into two categories: direct imaging and interferometric imaging. Direct imaging, which relies on a lens to form an image directly on a detector, has inherent resolution limits constrained by the aperture size. To push beyond these limits, interferometric imaging becomes essential, as it measures correlations of light from widely separated telescopes to reconstruct an image via the Fourier Transform. The scientific need for ultra-high angular resolution requires apertures ranging from hundreds of meters to kilometers in diameter, making dispersed aperture interferometers an inevitable part of the future of astronomy.

The science enabled by interferometry is compelling across all wavelengths, and deep-space observations will achieve resolutions far beyond what is possible with single-aperture telescopes. In the X-ray, it can resolve black hole event horizons. In the UV/optical, it can resolve stellar magnetic activity occurring on the stellar surface and probe its impact on planetary habitability. Far-infrared studies can reveal the processes governing the formation of habitable planets.

A lunar interferometer presents unique advantages over both ground-based and free-flying space-based facilities. The Moon's vacuum environment simplifies infrastructure needs, as optical delay lines do not require enclosures. Its longer instrumental coherence times are orders of magnitude higher than those on the Earth, (10–100+ seconds vs. the atmospheric limit of ~1 millisecond) which drastically enhance sensitivity. Additionally, a lunar interferometer provides access to UV wavelengths that are blocked by Earth's atmosphere, while avoiding the engineering challenges of millimeter-accurate free-flying station keeping required for space-based systems.

Previous studies examined the feasibility of space-based versus lunar-based interferometers. In particular, Bely et al. 1996 concluded that, in the absence of a supporting lunar infrastructure, free-flying interferometers were the better option. As a result, past efforts focused on free-flyer designs. However, with the Artemis program establishing a sustainable human presence on the Moon, it is now both compelling and timely to revisit the idea of a lunar-based interferometer.

---

[1] https://www.nasa.gov/directorates/spacetech/Lunar_Surface_Innovation_Initiative

Our NIAC Phase I study is dedicated to investigating the feasibility of a long-baseline, high-resolution UV/Optical interferometer on the lunar surface in cooperation with the Artemis program. Our goal, by the end of Phase II, is to conduct a study as detailed as those performed for large baseline, free-flying interferometers during the 2003–2005 Vision Mission Studies (Allen 2008). To guide our lunar-based facility design, we leverage insights from the 2003–2005 study of *Stellar Imager (SI)* (Carpenter et al. 2008), a proposed 0.5 km diameter UV/optical free-flying space interferometer.

In Phase I, we explored the feasibility of constructing the *Artemis-enabled Stellar Imager (AeSI)* on the lunar surface. This mission would enable revolutionary science (Figure 1.1), including: imaging the surfaces of nearby (~4 pc) solar-type stars and more distant (>2 kpc) supergiants to study magnetically driven activity (plages, starspots, convection), imaging accretion disks around nascent stars, and resolving the central engines of Active Galactic Nuclei (AGN).

Constructing an interferometer beyond Earth's surface is technically challenging, but the Moon offers an ideal testing ground. Smaller precursor missions or full-scale deployments on the lunar surface would benefit from a mix of human and robotic support for assembly, debugging, and maintenance. With Artemis paving the way for a sustained lunar presence, the time is right to evaluate and develop the technologies needed to bring ultra-high-resolution interferometry to the next frontier.

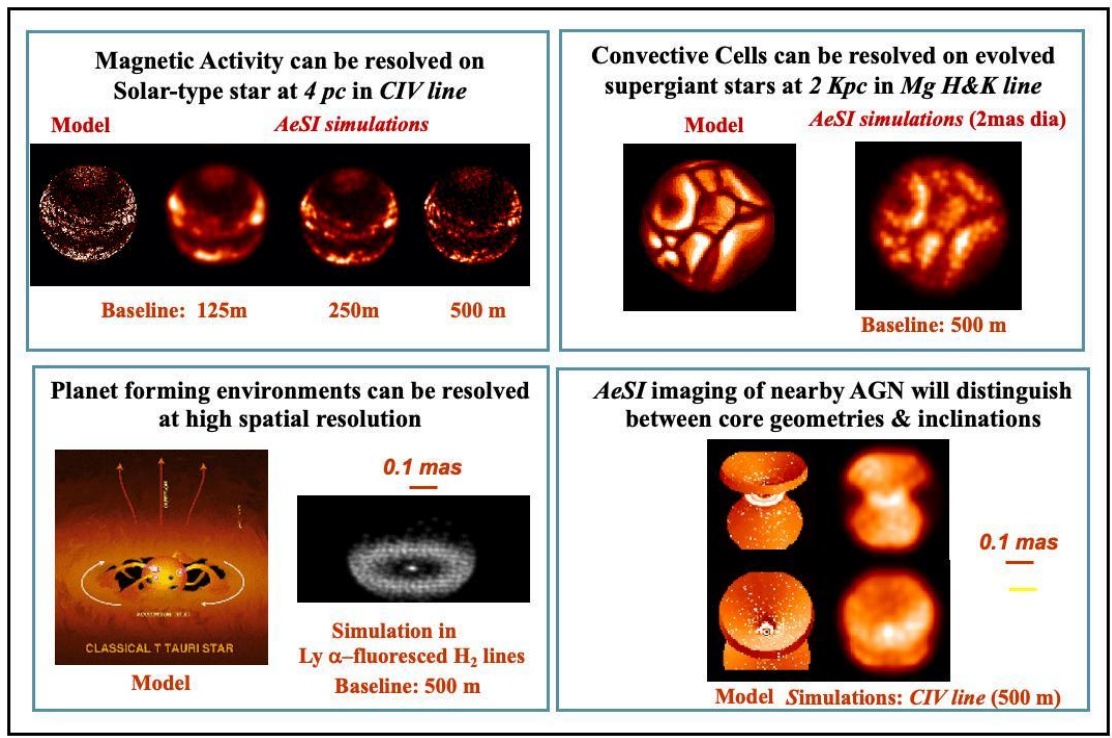

**Figure 1.1** Simulations of *AeSI* observations.

### 1.1.1 The Legacy of *Stellar Imager* (*SI*)

From 2003-2005, NASA funded the study of a suite of advanced mission concepts (Allen 2008) in order to extend the analysis of scientific objectives, system design, and operations of potential future visionary space science missions and to identify technology development needs for these missions. One of these mission concepts was a free-flying UV/Optical interferometer, with 30 primary mirrors, each on a "mirrorsat," spread across a sparse virtual aperture with an outer diameter typically ~0.5 km, though adjustable over the 100 to 1,000 m range. These mirrorsats point to the same target at the same time and direct the target light to a beam combiner 5 km distant when the array is set to a 0.5 km diameter. All 31 elements fly together using precision formation-flying technology and the observatory is located near the Sun-Earth L2 point. An artist's concept of the design is shown in Figure 1.2. The Vision Mission (VM) Study determined that Stellar Imager (SI) was feasible but required challenging technological development to enable the necessary precision flying of many spacecraft over relatively large separations. A lunar variant was not considered to be competitive given the absence of any supporting lunar infrastructure. The full results of the VM study are available at: https://hires.gsfc.nasa.gov/si.

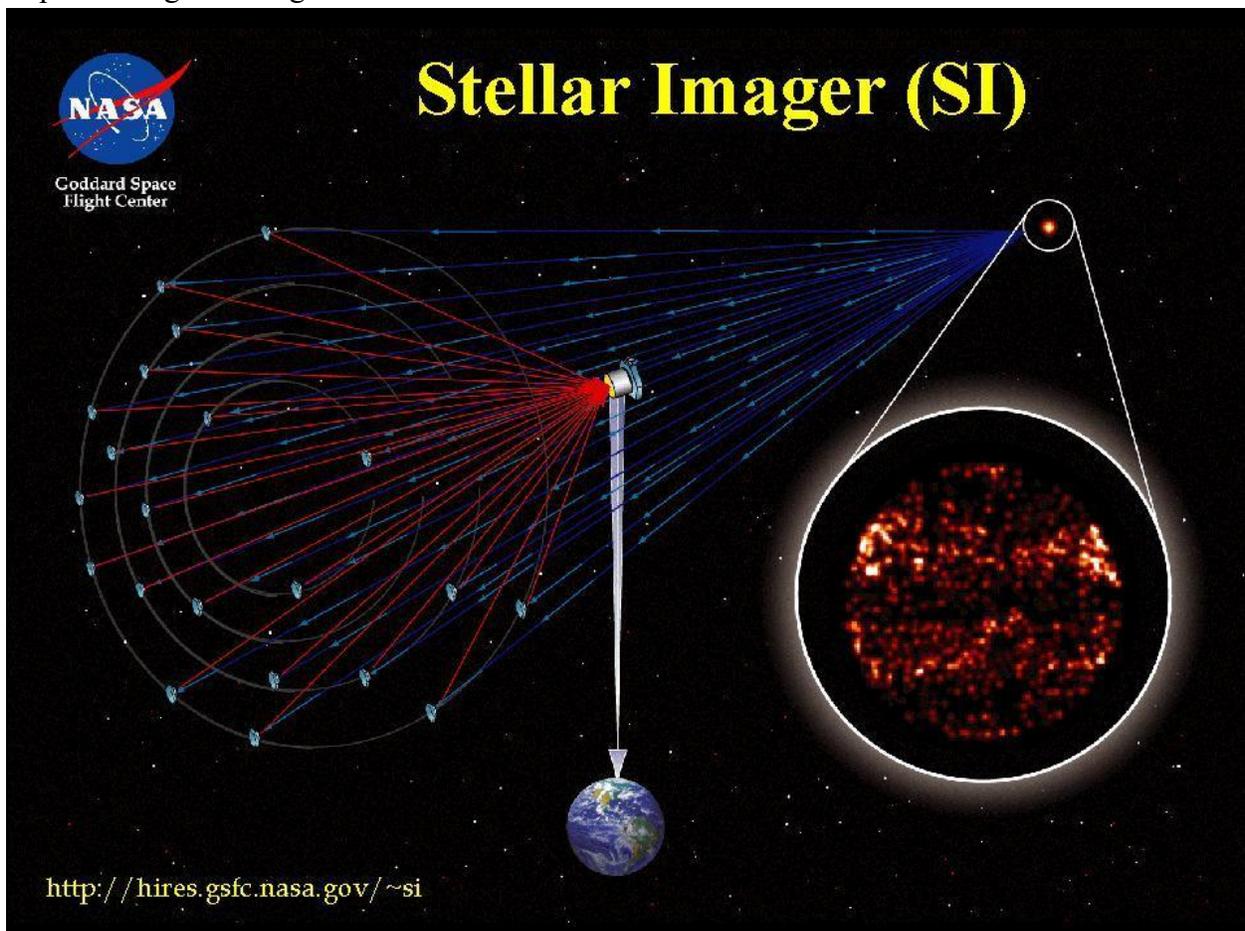

**Figure 1.2** Artist's concept of the free-flying Stellar Imager (SI) Vision Mission, developed as part of the NASA Space Science Vision Mission Studies in 2004-2005.

## 1.2 Mission Overview

In this NIAC Phase I study, we have designed a variant of the free-flying version of *SI* to obtain a resolution 100x that of the Hubble Space Telescope (HST) at UV and Optical wavelengths (1200 – 6600 A). There are a few significant differences between what is needed in a free-flying design versus what is needed on the lunar surface. The most significant is that, because the array cannot be tilted to be normal to the incoming light, optical delay lines and/or novel array reconfigurations are needed on the Moon, as on Earth-based facilities. This requires mobile platforms that can move over the lunar surface and enable variable optical path lengths for each primary mirror element, to keep the target-mirror-hub path lengths for all elements the same. Other concerns unique to the Moon are: dust and molecular contamination, seismic activity, and temperature variations. We address these issues in Sections 3.2 and 3.4.10, respectively. An advantage is that the lunar variant does not require the difficult task of precision formation flying 30+ spacecraft to the cm-level and there is support and resources for the lunar variant available from the Artemis Program.

Figure 1.3 (top) illustrates our initial (pre-Phase I) 6-element array concept for the lunar-based *AeSI* design, which included a second set of rovers for each primary mirror to accommodate long delay lines. However, through the course of this study, we refined the initial concept to a 15-element array (Figure 1.3, bottom) and eliminated the need for the secondary set of rovers. This was achieved by adopting innovative, non-symmetric configurations—specifically, an ellipsoidal layout with the elongated axis oriented toward the target—to minimize large differences in optical path length from target to primary mirror element to hub. Additionally, we chose to begin with 15 elements (instead of 6) to enhance science productivity from the outset, with the capability to expand to 30+ elements in one or more subsequent deployments. These refinements ensure that the design meets our science requirements while simplifying the system architecture.

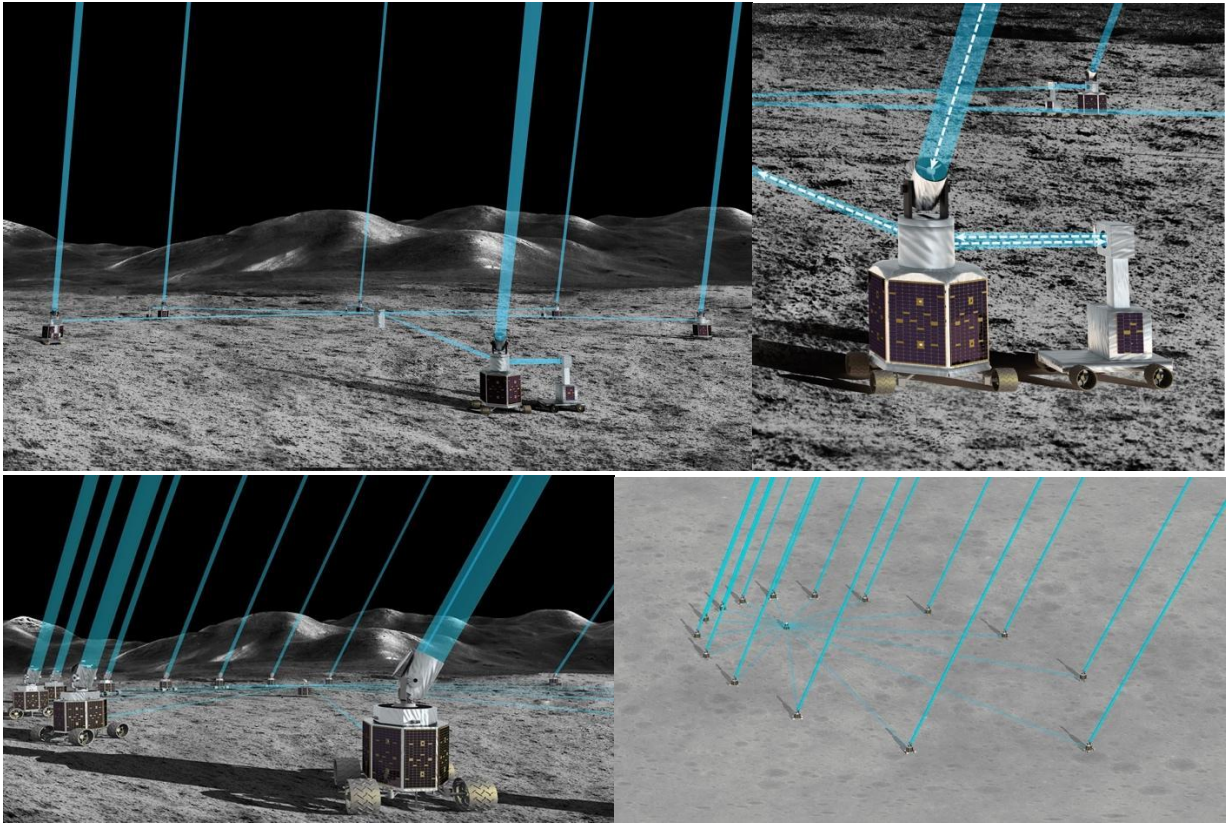

**Figure 1.3** The pre-Phase I concept (*top line*) vs. the Phase I Baseline design (*bottom line*) for the *Artemis-enabled Stellar Imager* (*AeSI*). Artist illustrations: Britt Griswold

## 1.3 Operational and Scientific Benefits
### 1.3.1 Innovation and Importance

*AeSI* will not only provide fundamentally unique science results, but its development will help pave the way to a fleet of space-based interferometers, both on the Moon and free-flying in open space. These future interferometers, which together will provide observations over a broad range of wavelengths from the X-ray to the far-infrared, will be required to obtain the high angular resolution that is not obtainable with monolithic or segmented mirror designs.

The innovations provided by *AeSI* include:
- Primary mirror elements that move on rovers (a.k.a. carts) to positions anywhere inside the virtual diameter of the interferometer's sparse aperture (500 m to 1 km), while providing the required stability. This is necessary to enable array reconfiguration, path length equalization, and adjustment of baselines.
- A beam-combining hub that accommodates beams from any configuration of the carts and accommodates increasing numbers of primary mirror elements.
- Improvements to existing dust, molecular contamination, and seismic mitigation concepts (Calle et al. 2008, Calle et al 2011, Farr 2020) and an exploration of the trade-space between designing to prevent contamination or scheduling routine maintenance to keep the optics clean.
- Identification of the kinds of support needed from lunar infrastructure and the development of an example plan for deploying, maintaining, and evolving such a facility using a mix of humans and robots.

### 1.3.2 Potential Impacts

By leveraging the unique advantages of the lunar environment and integrating with NASA's Artemis program, *AeSI* will drive advancements in technology, inspire collaborations across disciplines, and set the stage for a new era of space-based astronomy and exploration. The development of *AeSI* provides advancements in the following areas:

**On new missions**: *AeSI* presents a transformative approach for future missions, capitalizing on the unique advantages of the lunar surface to advance astronomy in unprecedented ways. By utilizing extended coherence times, access to UV wavelengths, and simplified infrastructure requirements, *AeSI* addresses challenges that hinder terrestrial and free-flying space-based interferometers. The Moon's stable surface enables precise alignment and eliminates the need for millimeter-accuracy station-keeping, paving the way for reconfigurable, dispersed aperture interferometers. This approach marks a significant step toward developing larger arrays on the lunar surface and in space, unlocking new scientific possibilities across diverse wavelengths — from X-ray imaging of black hole event horizons to high-resolution imaging of terrestrial-type exoplanets, and exploring signs of life in distant environments. (see e.g., https://www.blackholeexplorer.org/ or https://life-space-mission.com)

**On science**: With unprecedented imaging capabilities, *AeSI* will unlock new dimensions in our understanding of dynamic astrophysical phenomena. With an unprecedented sub-milliarcsecond resolution, *AeSI* will reveal intricate details of main sequence stars, achieving, e.g., >80 resolution elements (resels) across α Cen A (G2); >50 resels across α Cen B (K1), Procyon A (F5) and Sirius A (A0); and ~20 resels across ε Eridani (K2), τ Ceti (G8) and β Hydri (G2). These time-resolved surface views coupled with the spatially resolved asteroseismology of the interiors of sun-like and other stars (e.g., red supergiants at 2 kpc and beyond) will enable a major increase in our understanding of solar/stellar magnetic activity. **At its core, *AeSI*'s mission is to decipher the role of magnetism across the universe, with a particular focus on magnetic activity in stars like our Sun.** By advancing long-term forecasting of solar activity and the resulting space weather, *AeSI* directly addresses the Astro2020 Science Theme "Worlds and Suns in Context" (NASEM 2021). Beyond Sun-like stars, *AeSI* will revolutionize our understanding of the high energy environments, climates and potential habitability of exoplanets, and a vast range of magneto-hydrodynamic processes. Through in-depth studies of stellar magnetic activity, *AeSI* will illuminate how space weather from host stars influences exoplanets, providing critical insights into the conditions that support habitability in distant worlds. *AeSI* will image the surfaces of cool evolved stars with unprecedented high-angular resolution, revealing key features of their surface, transforming our understanding of mass loss and stellar evolution. It will also enable major advances in our understanding of AGN cores, supernovae, planetary nebulae, interacting binaries, stellar winds and pulsations, forming stars and disks, and evolved, dying stars.

**On NASA and the professional community:** *AeSI* represents a critical step in establishing the Moon as a platform for transformative scientific capabilities and ensures NASA is poised to capitalize on the Artemis program's investments to enable groundbreaking science. It also will drive innovation in interferometry, adaptive optics, and UV/optical instrumentation, technologies that can be applied to future space missions and terrestrial observatories.

**On the public:** The opportunity to develop and operate a major facility on the Moon, in tandem with the human spaceflight program, promises to capture the imagination of the public and scientific community alike. Much like the collaboration between the Space Shuttle Program and HST, this initiative will foster enthusiasm and engagement through its groundbreaking science and ambitious goals.

This study will deepen the astronomical community's understanding of the challenges and solutions involved in developing interferometers, whether on the Moon or free-flying in space, and will provide critical insights for the next generation of space-based observatories.

# 2 Science Goals and Objectives

*AeSI*'s primary scientific goal is to image Sun-like stars with unparalleled precision, offering new insights into their magnetic activity and the interior dynamos that generate it. With its extraordinary sub-milliarcsecond resolution, *AeSI* will capture intricate details, transforming distant point sources into complex structures and evolving visual narratives. Beyond studying Sun-like stars, *AeSI* will revolutionize our understanding of high-energy environments around exoplanets—including their climates, potential habitability, and magneto-hydrodynamic processes—and unveil key features of surface processes in cool evolved stars, reshaping our understanding of mass loss and stellar evolution. With its broad applicability, *AeSI* has the potential to not just transform our understanding of stellar surfaces and planetary systems, but also offer groundbreaking insights into AGN cores, supernovae, planetary nebulae, interacting binaries, stellar winds, pulsations, star formation, and the evolution of dying stars.

## 2.1 Magnetic Processes in Stars

The high spatial- and temporal-resolution imaging capabilities of *AeSI* will enable imaging of the surfaces of nearby solar-type stars, resolving magnetically driven activity such as starspots, plages, and convection. Active regions are very bright and dominate the solar spectral irradiance at wavelengths most important for space weather. *AeSI*'s observations will improve our understanding of both the solar and stellar dynamo and our ability to forecast future solar and stellar activity and mitigate effects of space weather (including its impact on Earth and the habitability of extrasolar planets; see Section 2.3).

### 2.1.1 Long-Term Activity

Current models struggle to predict solar activity due to the non-linear, non-local nature of stellar dynamos, and the sheer scale of the Sun's dynamo region (the outer 200,000 km of the solar interior) (e.g., Figure 2.1). The scientific community is uncertain where the main dynamo action occurs or what key processes are involved. To develop a dynamo model with predictive value, we must establish which processes are involved, and which approximations are allowed. Given the Sun's irregular 11-year cycle and long-term modulations, observing only this one star at high resolution isn't sufficient for developing predictive models. Studying similar stars' magnetic activity offers a more efficient way to validate these models. For example, surface magnetic activity records of stars on or near the lower main sequence (e.g. from the Mount Wilson Observatory Ca II H and K survey, Soon & Yaskell 2003) show variability similar to the solar variability, including Maunder minimum-like phases, on time scales of many decades. Figure 2.2 shows a K2 star (HD 166620) with an activity cycle like that of the Sun, showing a rapid decline in its chromospheric activity as it enters into a Maunder-minimum state. The key to successfully navigating the route to a workable, predictive dynamo model is the realization that in order to understand the solar dynamo, we need a population study: we need to study the dynamo-driven activity in a sample of stars like the Sun, and compare it to observations of young stars, old stars, binary stars, and any other obvious use cases. The potential for a breakthrough in our understanding and our ability to predict future behavior lies in spatially-resolved imaging of the dynamo-driven activity patterns on a variety of stars. These patterns, and how they depend on stellar properties (including convection, differential rotation and meridional circulation, evolutionary stage/age), are crucial for dynamo theorists to explore the sensitive dependencies on many poorly known parameters, to investigate bifurcations in a nonlinear 3- dimensional dynamo theory, and to validate the ultimate model.

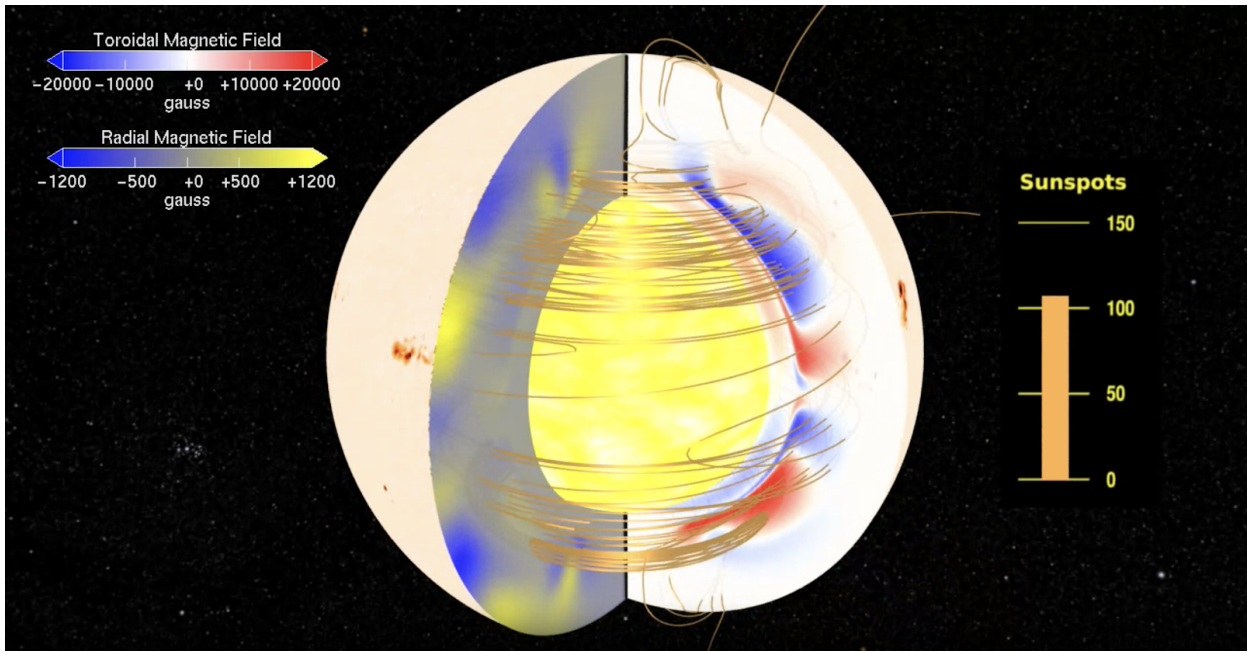

**Figure 2.1** This solar dynamo model simulates how the Sun's magnetic field grows and sustains itself by channeling convective energy. This simplified model reproduces key solar magnetic behaviors, including cyclic oscillations, magnetic migration toward the equator (seen in the "Butterfly Diagram"), and polarity reversals, though it cannot represent non-axisymmetric features like active longitudes. (image credit: NASA/SVS/Tom Bridgman 2008)

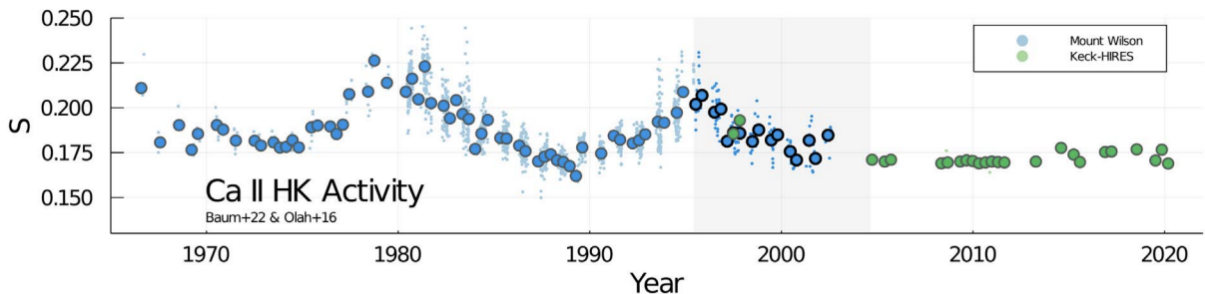

**Figure 2.2** $S_{HK}$ time series for HD 166620 showing transition into Maunder minimum (image credit: Luhn et al. 2022).

Direct interferometric imaging with *AeSI* is essential for capturing detailed information on dynamo patterns in stars with Sun-like activity. Other observational techniques, though useful for more magnetically active stars, fall short when it comes to stars like the Sun because:

- Rotationally-induced Doppler shifts are too subtle relative to the line width to enable effective Zeeman-Doppler imaging
- Magnetic activity levels are insufficient to cause significant spectral changes from magnetic line splitting
- Rotational modulation measurements suffer from deconvolution issues, creating ambiguities in spot locations, sizes, and latitude distributions, and are ineffective for studying facular patterns in quiet regions affected by differential rotation and field dispersal

*AeSI*'s direct imaging approach bypasses these limitations. Additionally, planned asteroseismic observations will reveal the internal structure and rotation of stars (Section 2.2), offering invaluable insights into the physical mechanisms of stellar dynamos. Imaging magnetically active stars and their surroundings will also provide a retrospective glimpse of the Sun's evolution—beginning from its formation in a molecular cloud, through phases of waning activity, and culminating in its final stages beyond the red-giant phase when it will expand to Earth's orbit.

### 2.1.2 Short-Term Activity

*AeSI* will provide simultaneous measurements across UV and Optical wavelengths, capturing stellar flares in both bands while also offering unprecedented spatial resolution (Figure 2.3). This unique combination will allow researchers not only to correlate UV and Optical flare emissions but also to do so with a detailed view of where these emissions originate on a star's surface. By capturing spatially resolved images of flaring regions, *AeSI* will reveal how energy is distributed across the stellar surface and how different areas contribute to flare evolution in real-time. This will provide critical insights into the spatial structure of flares, the distribution and intensity of magnetic fields, and how energy release mechanisms vary across a star's surface. Such detailed observations are key to advancing our understanding of flare dynamics, improving models of magnetic reconnection, plasma heating, and particle acceleration. Ultimately, *AeSI*'s ability to map flare activity spatially and across wavelengths will enhance predictions about flare impacts on surrounding exoplanets, helping to assess the conditions that might influence exoplanet atmospheres and their potential habitability.

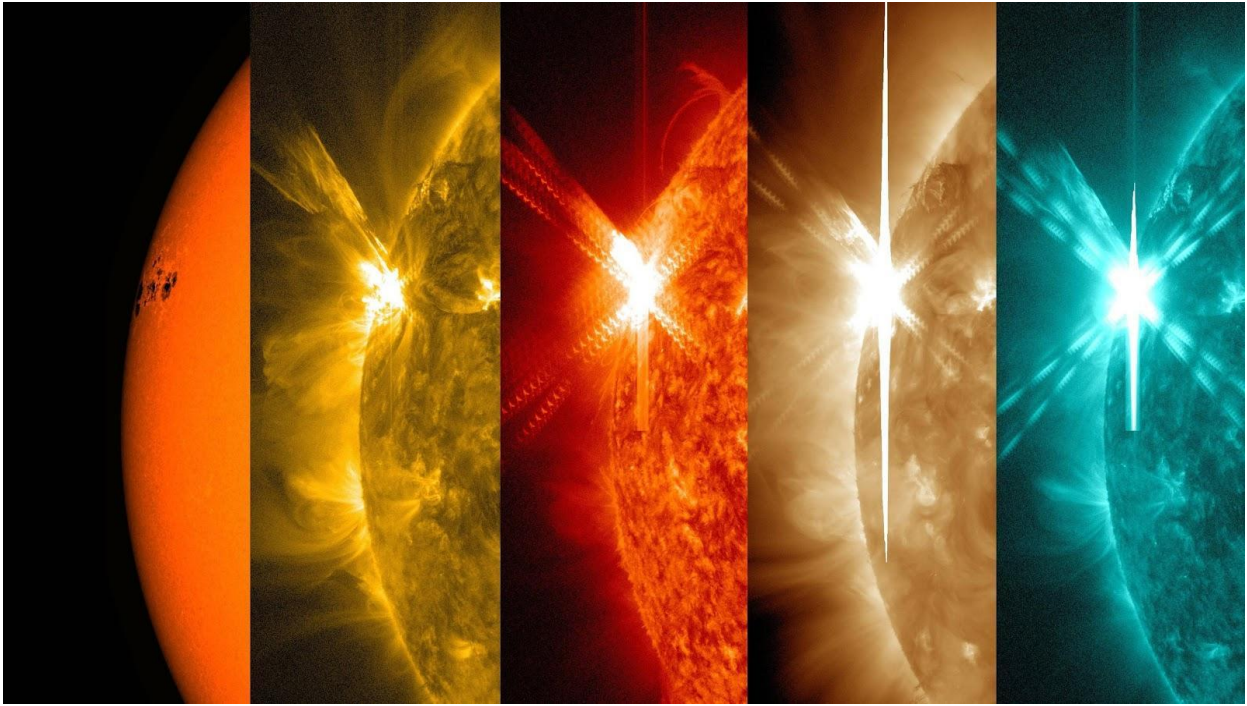

**Figure 2.3** Solar flare images in the Optical and UV (left to right: 6173 A, 171 A, 304 A, 193 A, 131 A) . *AeSI* will be capable of producing resolved images on several nearby stars. (image credit: NASA/SDO)

## 2.2  Stellar Interiors

Asteroseismology has extended and revitalized the study of stellar structure in recent years, because the detection and measurement of stellar oscillation frequencies provides a nearly model-independent window into stellar interior (Garcia & Ballot 2019, Aerts 2021, Bowman & Bugnet 2024). In solar-like stars, oscillations are driven stochastically, as acoustic noise arising from convection propagates through the interior, exciting resonant frequencies which lead to signals at the photosphere, which in turn can be detected using variations in either radial velocity or intensity. These resonant oscillations can occur with either gravity or pressure as the restoring force and are characterized as g-modes and p-modes, respectively. In sun-like stars, p-modes propagate in surface regions while g-modes are found near the core.

Because stars are spherical (to first order), oscillations are generally parameterized using spherical harmonics: radial order *n*, degree *l* and azimuthal order *m* (Aerts 2010; Figure 2.4). Numerous modes are excited simultaneously, so the velocity and intensity maps reflect the sum of all of these.

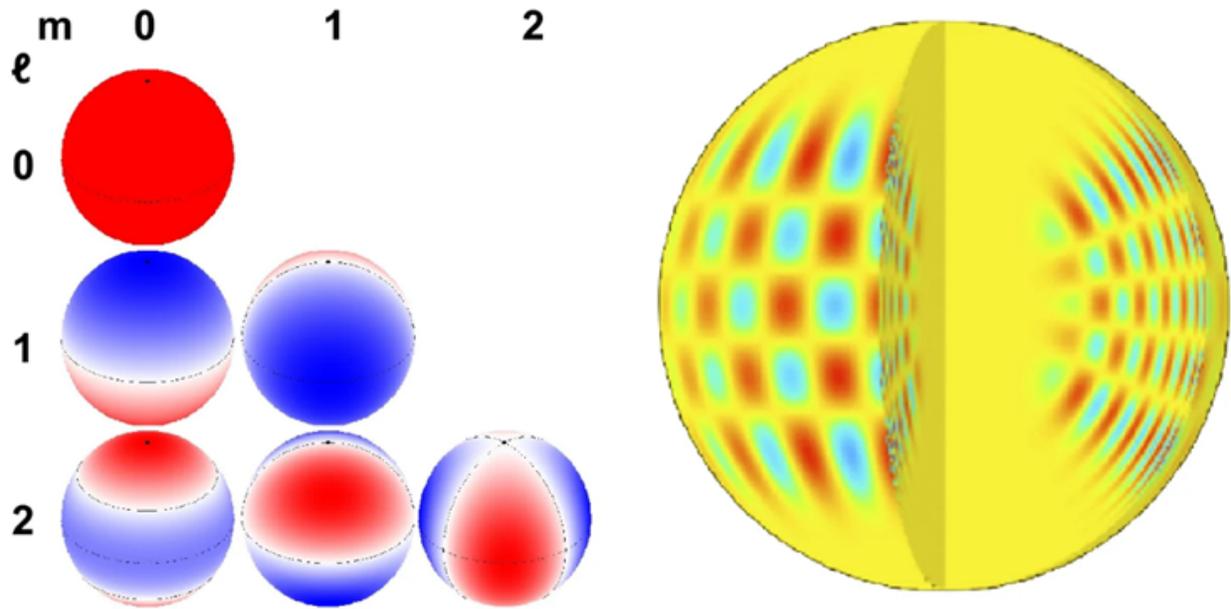

Figure 2.4 Represented on the left by spherical harmonics are modes of degree ℓ=0, 1, 2 and azimuthal order m=0, 1, 2. Blue regions represent regions instantaneously moving towards the observer, while red those moving away. The figure on the right shows a mode with ℓ=20, m=16 and n=14. (image credit: Garcia & Ballot 2019).

The information derived from individual modes can be intuitively understood through a ray diagram (Figure 2.5), which illustrates the path of modes of different angular degree *l* through the interior of the star. Any given mode is confined in a resonant cavity between upper and lower turning points, bounded at radii where the mode is no longer oscillatory but evanescent. The observed frequency of a particular mode is then determined by a convolution of the sound speed along its path, and the sound speed is in turn set by the physical conditions along that path. Accordingly, while detection of even one or a few modes gives insight into the stellar internal structure, characterization of a larger number of modes of different angular degree allows a much more fine-grained picture of the interior and more accurate values for fundamental physical parameters such as mass, radius, and age.

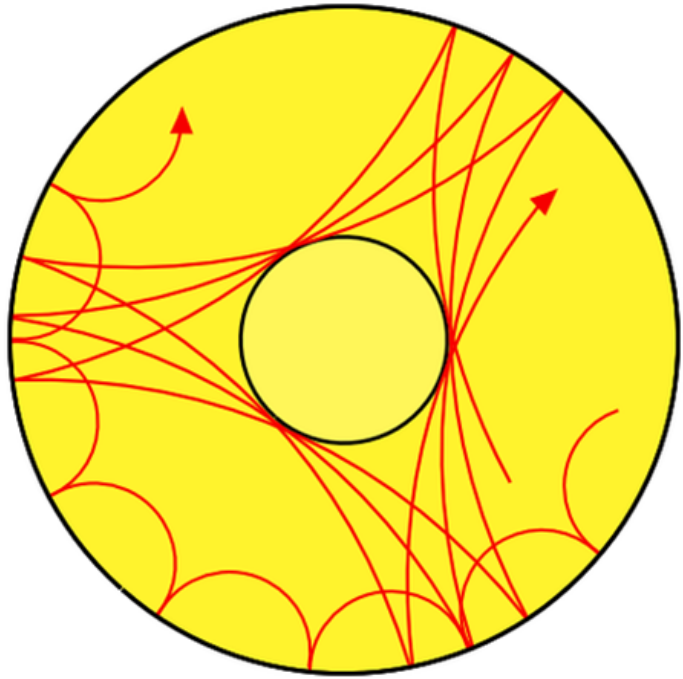

**Figure 2.5** Ray paths for two different oscillation p-modes in a solar-like star, showing the dependence of the lower turning point on angular degree. Higher degree modes sample only the outer layers, while lower degree modes primarily sample the deeper interior.

To date, it has been impossible to detect large numbers of modes of *l*>3 in stars other than the Sun. This is because when any star is observed as an unresolved source, the contributions from high-degree modes essentially cancel out, both in radial velocity and intensity. *AeSI*'s broad wavelength coverage, from 1200-6600 A, offers a significant advantage in probing both the surface and outer layers of a star with varying sensitivities. UV wavelengths are particularly effective at probing the outermost layers, while optical wavelengths provide deeper insight into the more dynamic regions of the photosphere. This wavelength coverage enables the detection of a wide range of oscillation modes at different depths within the star, facilitating the identification of higher-degree modes (l>3) and enhancing our understanding of the star's internal structure.

A key component of *AeSI's* observing strategy is conducting month-long observations of stars for asteroseismology. Long-duration observations are essential for accurately detecting stellar oscillations, particularly those associated with higher-degree modes, which are typically low-frequency signals. To achieve a sufficient signal-to-noise ratio and resolve these modes, extended observation periods are necessary. This approach allows us to accumulate the required data to detect even subtle oscillations, thereby enabling a more precise characterization of the star's internal structure. *AeSI's* ability to resolve the stellar surface and detect modes of significantly higher angular degree, in turn will enable construction of far more detailed internal density and rotation profiles, as well as the study of internal magnetic fields (Fuller et al., 2015;

Stello et al., 2016, Lecoanet et al. 2022). Red giant stars are also commonly observed to display stochastically excited pulsations, and as they ascend the red giant branch they display mixed pressure-gravity modes. Detection of a wide range of mixed modes will allow direct access to the physics of the deep stellar interior, including the effects of angular momentum transport and core magnetic fields and their impact on stellar structure and evolution (Chaplin & Miglio 2013, Hekker & Christensen-Dalsgaard 2017, Gehan et al. 2018, Garcıa & Ballot 2019, Kurtz 2022).

## 2.3 Exoplanet Host Stars

*AeSI* offers unique advantages for studying exoplanet host stars and the interactions between those stars and their planets' atmospheres. The ultraviolet imaging capabilities are particularly advantageous for addressing phenomena including the "transit light source effect", escaping atmospheres from gas giants, and H II fluorescence in hot Jupiter atmospheres.

### 2.3.1 Host Star Spectra

Stellar ultraviolet radiation is critical in driving processes like photochemistry, heating, and atmospheric escape on exoplanets. Magnetic activity leads to oscillating levels of UV flux during stellar cycles and short, intense bursts of UV radiation during flares, all of which plays a crucial role in shaping the environments of orbiting planets. Each emission line in the UV probes a different temperature plasma or depth in the stellar atmosphere, allowing us to constrain the temperature-pressure profile (e.g., Figure 2.6). The wavelength coverage of *AeSI* (1200 - 6600 A) includes emission lines that form at temperatures ranging from the photosphere to the upper transition region:
- Lya (1216 A) - 2,000 - 80,000 K
- N V (1240 A) - 200,000 K
- C II (1334 A) - 6,000 - 20,000 K
- Si IV (1400 A) - 60,000 K
- C IV (1550 A) - 100,000 K
- He II (1643 A) - 100,000 K
- Al II (1670 A) - 3,500 K
- Mg II h and k (2800 A) - 3,500-6000 K in wings, ~16,000 K in core

The resolved UV spectral line data that *AeSI* will measure can be used to inform semi-empirical models that predict the full stellar UV spectrum from extreme ultraviolet (EUV) to near-ultraviolet (NUV) wavelengths (e.g., Peacock et al. 2019,2020, Hintz et al. 2023). These semi-empirical stellar models serve as essential inputs for studying photochemistry and escape in exoplanet atmospheres. Understanding these effects is vital for assessing the habitability of planets, particularly those around low-mass stars, which tend to exhibit increased UV activity and frequent flares.

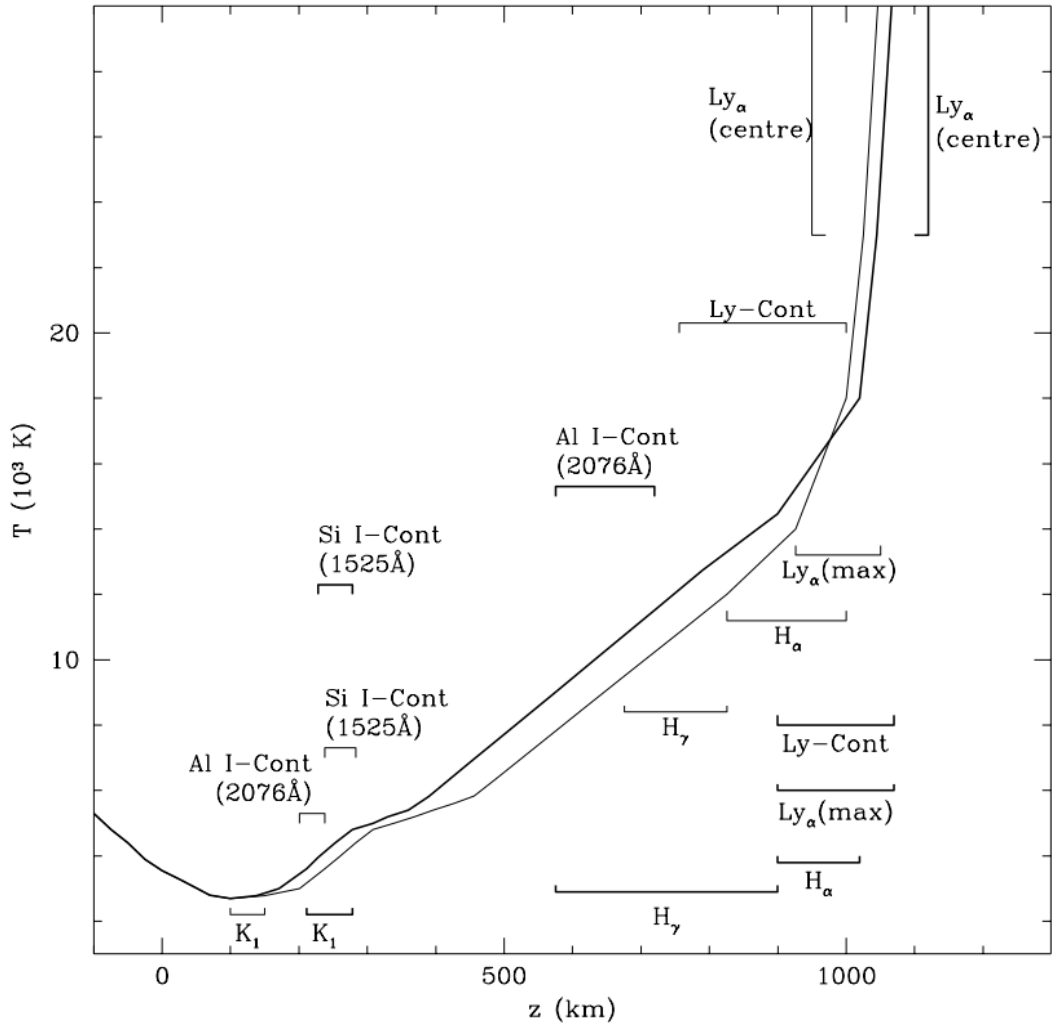

**Figure 2.6** Temperature distribution for a stellar chromosphere with approximate depths where various continua and lines originate are indicated (image credit: Falchi & Mauas 1998).

### 2.3.2 Transit Light Source Effect

The "transit light source effect" (TLS; Pont et al. 2013, Oshagh et al. 2014, Llama et al. 2015, McCullough et al. 2014, Rackham et al. 2018) refers to variations in the stellar surface during a planetary transit. Host stars are neither static nor flat, and their variability and features can influence the extracted planet spectrum when disentangling the stellar signal from the planet. In particular, starspots and faculae can mimic or obscure atmospheric signals plausible in the planet's atmosphere that are instead coming from the star. With *AeSI*'s high spatial resolution, the inhomogeneous features will be mappable on the stellar disk at the time of transit, allowing for a more precise correction for the TLS and a more definitive conclusion of whether certain detected molecules are inherent to the planet's atmosphere.

### 2.3.3 Escaping Atmospheres

Close-in planets undergoing hydrodynamic escape have a comet-like tail of ionized gas trailing behind them as they transit their host star. This tail is typically detected via absorption in Lyman-alpha (1215.67 A), as hydrogen is the most commonly escaping species, but heavier atoms and ions can be dragged into the exosphere and their escape has been observed in NUV Fe II lines and the Mg II h and k (~2800 A) doublet (e.g., Koskinen et al. 2013, Sing et al. 2019, Huang et al. 2023, Sreejith et al. 2024). In the Baseline design described further in this report, the capabilities of *AeSI* will allow for the tracing of extended magnesium clouds surrounding and tailing behind close-in ultra-hot Jupiters, providing insights into the structure and mass-loss rates of these exospheres. With future enhancements to the sensitivity of *AeSI*, more atmospheric escape targets will become available as measuring the extended hydrogen cloud in Lyman-alpha becomes accessible.

### 2.3.4 H II Fluorescence

H II fluorescence in hot Jupiter atmospheres is often triggered by the host star's high-energy radiation. Detecting, measuring and monitoring this phenomenon can provide clues about the temperature, ionization state, and chemical composition of these atmospheres. Free from the limitations of Earth's atmosphere, *Ae*could monitor this fluorescence with exceptional clarity, helping elucidate the complex interactions of hot Jupiters with their host stars (including energy transfer mechanisms and how stellar UV radiation impacts the planet's dynamics).

## 2.4 Cool, Evolved Giants and Super Giants

Asymptotic Giant Branch (AGB) stars and red supergiant (RSG) stars are among the most important contributors of enriched materials to the interstellar medium through their strong winds. In AGB stars, stellar winds are primarily driven by pulsations that generate shock waves in the upper atmosphere. These shocks propel material into the surrounding circumstellar environment. As this material travels away from the star, it cools sufficiently to condense into dust. The dust is radiatively accelerated and drags gas with it though friction, resulting in a strong stellar wind. Thanks to its high-angular resolution and imaging capabilities, *AeSI* will be able to image the effects of pulsations and shocks in the photospheres and chromospheres of AGB stars, providing an unprecedented view of the mass-loss process (Rau et al. 2024).

However, for higher mass stars such as RSG, their mass loss mechanism is less well understood, and likely different from AGB stars. For example, Arroyo-Torres et al. 2015 shows that pulsation-based models do not predict the extension of the atmospheres of RSG stars. In 2019, the great dimming of Betelgeuse provided an opportunity to better understand the RSG mass loss process. Montarges et al. (2021) showed the dimming may have been caused by the formation of a clump of dust in the line of sight between Betelgeuse and the Earth, indicating that RSG mass loss can be episodic and inhomogeneous, and proposed that the star ejected a bubble of gas which later condensed into the dust cloud from its surface prior to the great dimming, and suggested this bubble came from a large convection cell. The connection between episodic mass loss and the photospheric features of RSG stars must be explored to understand RSG mass loss. Hydrodynamic models of RSG atmospheres indicate their surfaces are dominated by a few massive convection cells (e.g., Freytag et al. 2002, Chiavassa et al. 2011). Similar convection cells occur on AGB stars (e.g., Rosales-Guzmán et al., 2024). Figure 2.7 shows a simulated image of an RSG with convection cells and a dark spot (Freytag et al. 2024). *AeSI* will be capable of imaging the photospheres of RSG and AGB stars with an angular resolution high enough to resolve their convection cells, clarifying the effect of photospheric features on mass loss.

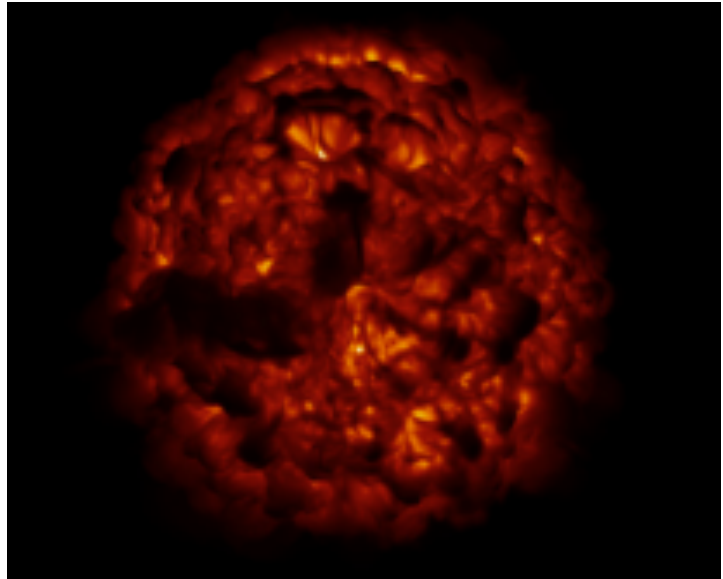

**Figure 2.7** Simulated surface of a red supergiant star from Figure 1 of Freytag et al. 2024. Model number st35gm04n045[2] (M = 5 M☉, Teff = 3368 K. t= 14.514 yr).

Magnetic activity in cool, evolved stars is another possible driver of mass loss (Falceta-Gonçalves & Jatenco-Pereira 2002). Although the pulsation-driven wind models of AGB stars generally explain their mass loss and many interferometric observations of their outer atmosphere have been done (e.g., Rau et al. 2015, 2017, 2019; Hulberg et al. 2025), Rau et al. (2019) showed that additional contributions from magneto-hydrodynamic (MHD) effects such as Alfvén waves may explain discrepancies in gas velocities between observations and pulsation only models. Alfvén waves may originate in the chromosphere and depend on properties such as its width (Rau et al. 2018).

Red giant branch (RGB) stars, such as K-giants, are not typically strong pulsators compared to their asymptotic giant branch or red supergiant counterparts. Instead, their mass loss is thought to be driven primarily by magnetohydrodynamic processes, including the excitation of Alfvén waves in their outer envelopes (e.g., Matt et al. 2012). These Alfvén waves are believed to propagate outward from the star's surface, carrying away energy and momentum, which in turn accelerates the stellar wind and drives mass loss (e.g., Airapetian et al. 2000, Airapetian et al 2010).

---

[2] https://www.astro.uu.se/~bf/

Chromospheric emission in the ultraviolet (UV) is an important diagnostic tool for probing the properties of the outer layers of cool evolved stars. Studies by e.g., Carpenter et al. (1994) and Carpenter et al. (2018) have shown that UV chromospheric emission lines can provide direct constraints on the physical conditions in the chromosphere and the surrounding circumstellar material. These observations, which target both the UV emission spectra and variability of these stars, are crucial for understanding how mass loss is driven in these cool, less evolved stars.

*AeSI* will provide a transformative view of the chromospheric emission from AGB, RGB, and RSG stars at unprecedented angular resolutions, potentially on the order of sub-milliarcseconds[3]. This will allow for detailed imaging and UV/optical spectral analysis of the chromospheres and surrounding winds of these stars. In particular, the high spatial resolution offered by *AeSI* will offer new insights into how MHD-driven effects, including the role of Alfvén waves, contribute to the observed mass loss rates in these stars. Such observations will help distinguish the contributions of magnetized winds from those driven purely by radiation pressure and stellar pulsations.

In addition to its imaging capabilities, *AeSI* will focus on spectroscopic observations of chromospheric emission lines, which are key for probing the structure and dynamics of the winds of M- and K-giants. Photon scattering in the stellar wind can alter the observed profiles of UV and optical emission lines (e.g., Rau et al. 2018), providing a way to study the properties of the cool winds around giant stars. The scattering process involves the interaction of ultraviolet photons emitted from the chromosphere with atoms and dust in the stellar wind, and the resulting changes in the line profiles can reveal information about wind acceleration, density, and velocity structure. By combining these spectroscopic data with high-resolution imaging, *AeSI* will significantly improve our understanding of how these winds are structured and how they contribute to the overall mass loss.

These observations will be instrumental in refining theoretical models of mass loss in cool evolved stars, and they may offer new constraints on the mechanisms driving such mass loss in stars that are not strong pulsators but still undergo significant mass loss, such as K-giants on the red giant branch.

---

[3] 1,000 m baseline at 2,500 Å would be 0.0516 mas

Stars with masses greater than 1.5 M₀ do not have a convective envelope during their lifetime on the main sequence and as such do not display signs of magnetic activity. On the red giant branch, these stars develop a dynamo due to their convective envelopes. The onset of magnetic activity may lead to magnetic braking and different rotation rates between the interior and the convective envelope (Schrijver & Pols, 1993). Photospheric and chromospheric imaging as well as asteroseismic observations with *AeSI* will be able to provide detail on the dynamo in these stars (Carpenter et al. 2019).

## 2.5 Supernovae

The ejecta of young supernovae undergo unconstrained expansion for a few hundred years after the supernova itself. The initial expansion itself may have ejecta velocities of 5,000-10,000 km s$^{-1}$. With *AeSI*, this expansion may be observed from the very initial stages, i.e., from when the shell is only 0.005 pc in diameter (assuming a distance of 0.6 Mpc) in galaxies in the Local Group. Using velocity information on the ejecta from spectroscopy, the distance to the supernova and therefore to the host galaxy is easily inferred. This technique has already been applied with success to infer the distance to the LMC (with SN 1987A) using high resolution HST data, and to M81 (SN 1993J) using radio VLBI. Radio VLBI, although successfully applied in the case of M81, is not a guaranteed technique because several supernovae do not become powerful radio emitters until much later in their dynamical expansion stage. Shell diameter measurements to the Virgo Cluster at 16 Mpc might be possible.

*Known Targets*: A project of this nature will require the fortuitous occurrence of a supernova in a nearby galaxy, e.g., SN 1987A could be directly imaged upon with *AeSI* (peak visual magnitude of ~3.26, Bouchet & Danziger 1993). *AeSI* could be utilized to make a shell diameter measurement of a much dimmer supernova. At a distance of 0.6 Mpc (roughly M31), the diameter of a shell 0.005 pc in diameter could be measured. Ten years after the 1987A explosion, the fireball was 0.05 pc in diameter (Pun & Kirshner 1996). From that, a size of 0.005 pc after one year may be inferred; at that time 1987A was $m_v \sim 9.0$, indicating $m_v \sim 14.5$ at 0.6 Mpc. Based upon expected supernovae rates (van den Bergh & Tammann 1991, which admittedly are quite uncertain), there should be an observable supernova within the Local Group once every 5 to 20 years.

## 2.6 Accretion in Symbiotics and other Interacting Binaries

Almost all high-energy sources in the universe are powered through the potential energy released via accretion. Understanding accretion driven flows in interacting binaries will directly affect our understanding of similar flows around Young Stellar Objects (YSOs), including the formation of planets in the circumstellar disk as well as the much larger scale accretion flows in Active Galactic Nuclei (AGNs).

Compact, mass transferring binaries provide us with laboratories for testing energetic processes such as magnetically driven accretion and accretion geometries, various evolutionary scenarios, and conditions for induced stellar activity.

In close binary stars the flow of material from one component into the potential well of the other is a key in determining the future evolutionary histories of each component and the system itself, and particularly the production of degenerate companions and supernovae. Our cosmological standard candles, the Type Ia supernovae, for example, may be a consequence of accretion onto a white dwarf (WD) in a close binary, whereas the WD mass is approaching the Chandrasekhar limit.

Currently, most of our accretion paradigms are based on time-resolved spectroscopic observations. For example, in Cataclysmic Variables (CVs) the picture of accretion onto compact objects via an extended accretion disc is solidly based on spectral and timing information. However, several objects challenge our standard picture and there are significant gaps in our understanding of their formation and evolution.

Large uncertainties exist in our quantitative understanding of accreting processes in many interacting systems. The interaction between the components in close binaries is believed to occur via Roche lobe overflow and/or wind accretion Roche Lobe Overflow (RLOF).

3-D hydrodynamic simulations show that the accretion processes in tidally interacting systems are very complex. Wind accretion is even more complicated. The amount of the accreted material depends on the characteristics of both components including stellar activity and wind properties (e.g. density and velocity), the binary parameters (e.g. orbital period and separation), and the dynamics of the flow. In the case of RLOF, the accretion may form an extended accretion disk whose turbulent magnetic dynamo drives the flow through it. Stellar activity of the rapidly rotating donors and their impact on the binary remains poorly constrained despite being crucial in regulating the mass transfer rate and setting the long-term evolution. The key to further advance in accretion studies is resolving a wide range of interacting binaries and studying their components and mass flows. The *AeSI* sub-milliarcsecond resolution in the UV will lead to unprecedented opportunities for detailed studies of accretion phenomena in many interacting systems including symbiotics, Algol type binaries, Cataclysmic Variables (CVs) and their progenitors. *AeSI* will be able to resolve the components of a variety of interacting systems and will therefore provide a unique laboratory for studying accretion processes. The binary components can be studied individually at many wavelengths including Ly α, C IV, and Mg h and k lines, and the geometry of accretion can be imaged directly, giving us the first direct constraints on the accretion geometries for a range of systems. This in turn will allow us to benchmark crucial accretion paradigms that affect any stellar population and even the structure evolution of galaxies whose central black-holes are steadily accreting, shaping the long term evolution.

Symbiotic binary systems are some of the most fascinating interacting systems because of their dramatic transformations, and an extremely complex circumbinary environment. Their spectra show signatures of a late-type giant and a high-temperature component, often a compact object in the form of a white dwarf or even a neutron star.

Symbiotics are very important systems because they are potential progenitors of asymmetric Planetary Nebulae, and they have been invoked as progenitors of at least a fraction of Supernovae type Ia - a key cosmological distance indicator. *AeSI* will be able to separate the components of nearby (a few kpc) currently unresolved symbiotic systems.

So far, the individual components have been resolved in only a couple nearby symbiotic systems: Mira AB (Karovska et al. 2005) and R Aqr (Bujarrabal et al. 2018). *AeSI* will have the capability of resolving both a nearby and more distant symbiotic systems (Figure 2.8) and will be able to image directly the individual binary components and the dynamical accretion flows.

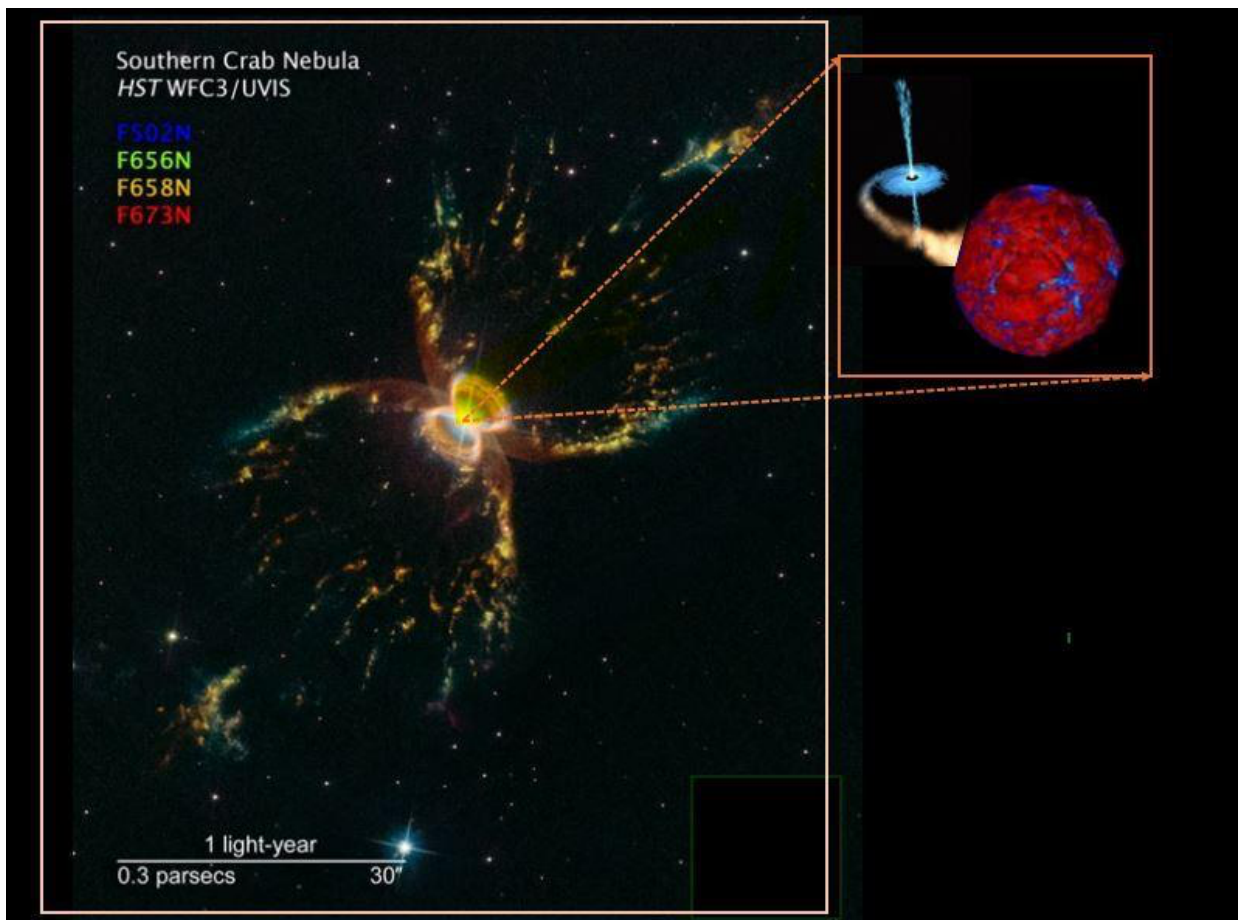

**Figure 2.8** HST composite image of the Symbiotic system Southern Crab (Hen2-104) (distance ~3000 pc) obtained in several spectral lines showing the complex morphology of the outflow,

outbursts, and jet extending over 30". The binary itself and the close circumbinary environment is not resolved with the HST (0.05" resolution). *AeSI* sub-milliarcsecond resolution imaging in the UV will separate the components of this interacting binary and allow imaging of the individual components (an evolved mass losing red giant and accreting white dwarf), and the details of the accretion flow, as well as the origin and the inner morphology of the jet itself (see the insert in the display).

## 2.7 Active Galactic Nuclei

*AeSI* images of nearby AGN will resolve the transition zone between the broad and narrow emission line regions and provide key information on the origin and orientation of powerful, relativistic jets. *AeSI* sub-milliarcsec resolution images of the transition zone between broad-line and narrow-line regions would answer the question: "is material being stripped from the broad-line clouds, which are close to the nucleus, and driven out to the narrow-line region?" This is best studied in the UV/optical emission lines within a parsec scale from the nucleus. The problems of the origin and orientation of the jets could be addressed with detailed broad-band images of the inner regions. It would also enable a search for electron scattering by outflowing plasma associated with jets or with massive winds driven off accretion disks. Such images could also provide an answer to the question: "Do type 1 Seyferts have molecular tori?". Broad-band imaging at sub-parsec scales could tell us if tori are obscuring starlight.

One major area of AGN research is the AGN winds. Evidence for powerful winds comes from high-resolution spectra, e.g., from Chandra and XMM-Newton, and from their UV counterparts, STIS and FUSE. These AGN winds carry a substantial mass loss rate compared with the mass loss needed to power the AGN continuum itself. They are therefore important to understanding the dynamics and structure of AGN. Because these winds enrich the surrounding intergalactic medium they have larger implications for cosmology. Though least ambiguously seen in absorption, AGN winds also produce emission lines. The location of the gas producing these emission lines is much debated. Suggestions range from the size of the optical/UV broad emission line region (BELR) of a few 10s of light-days, to the size of the "obscuring torus" at a few parsecs.

*AeSI* UV/Optical observations of high ionization emission lines (e.g. Fe lines and Fe emission line "bump" in near UV) suggest both a small and a large scale region, differentiated by their velocity structure. *AeSI* sub-milliarcsecond resolution in the UV will allow imaging at the 'obscuring torus' scale and should yield telling images of the torus itself and of the AGN BELR and wind. If the CIV remains point-like at this level, the more radical BELR-scale hypothesis will be greatly strengthened.

Resolving the angular size of the BELR will provide a key input to cosmological models; It will allow determining direct geometrical distances to quasars that can measure the cosmological constant, Λ, with minimal assumptions; This method (Elvis-Karovska method) is equivalent to geometric parallax (as described in Elvis and Karovska, 2002), The ``standard length'' would be (BELR) as determined from the light-travel time measurements of reverberation mapping. Elvis and Karovska (2002) suggested that the effect of nonzero Λ on angular diameter is large, 40% at z=2, so mapping angular diameter distances versus redshift will give Λ with (relative) ease. These measurements would be made in UV-optical, with a resolution of less than 0.01 arcsec to measure e.g., the size of the BELR in z=2 quasars.

Figure 2.9 shows a simulation of *AeSI* capabilities for differentiating between different AGN broad emission line region morphologies and inclinations. The hourglass shape model (Elvis 2000) can be discerned in this simulated observation and its tilt is clearly visible as well.

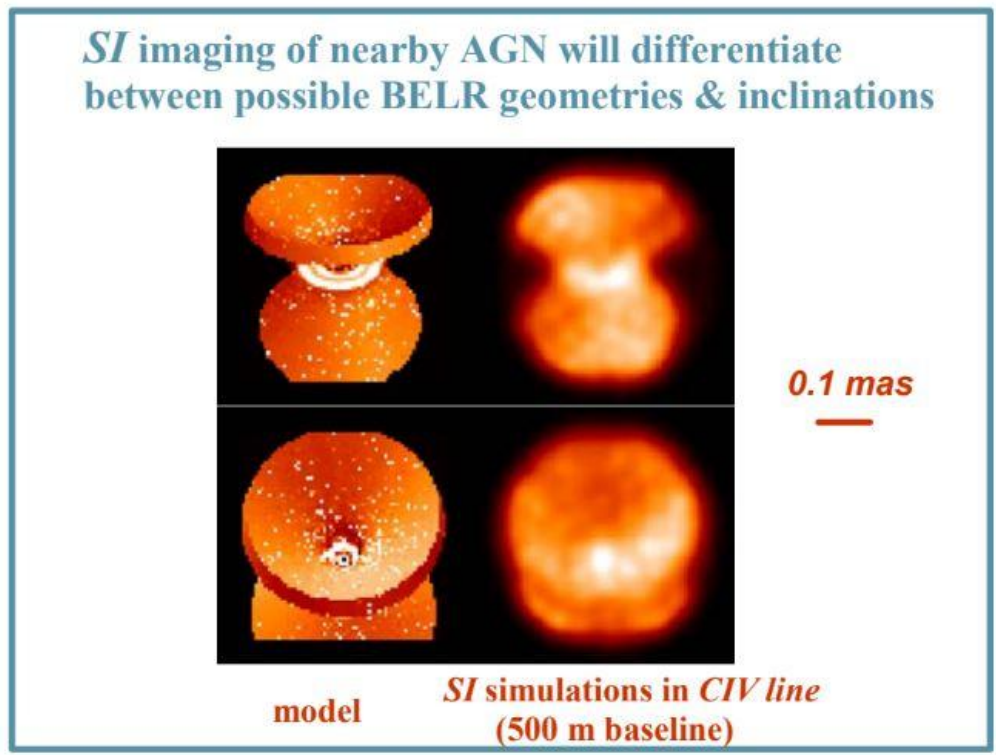

**Figure 2.9** *AeSI* imaging of AGN morphologies of broad emission line regions at various inclinations. These simulations were produced using the SISIM code (Rajagopal et al. 2003) assuming 30 mirror elements distributed in a non-redundant pattern, within a 500-meter circle and using a 10 Å bandpass around the C IV doublet near 1550 Å.

# 3 Detailed Mission Analysis

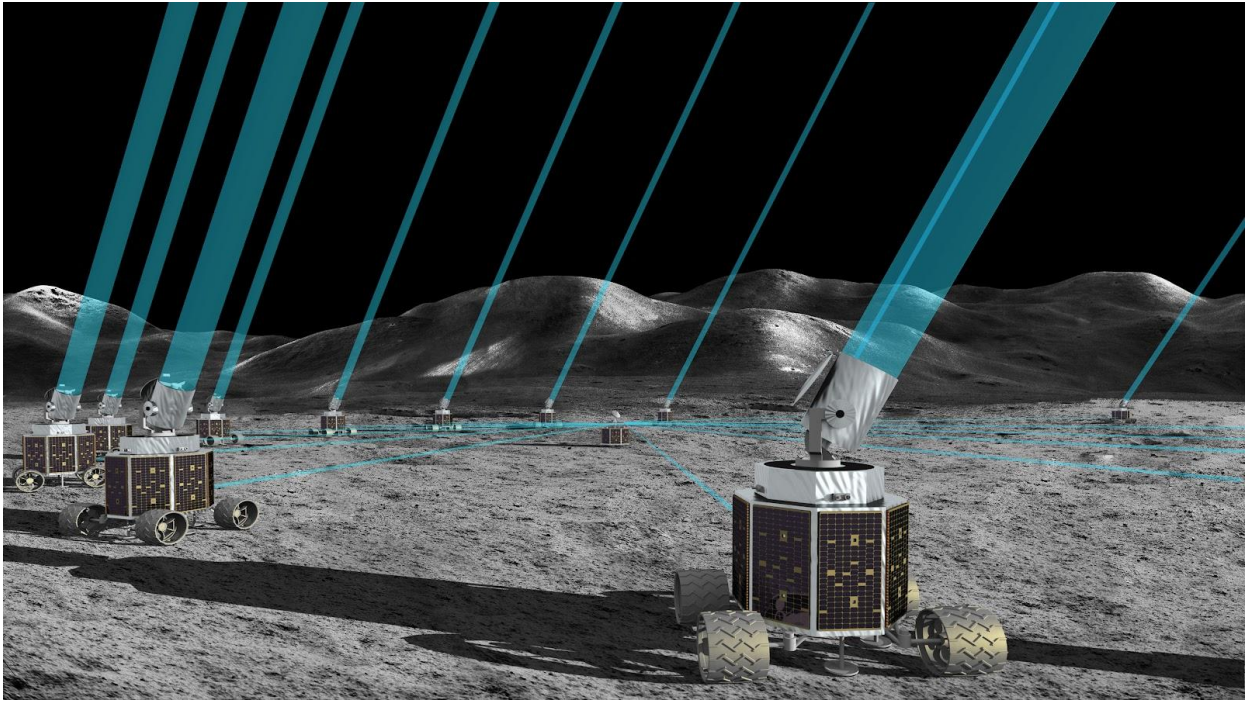

**Figure 3.1** Artist rendition of the Baseline design for *AeSI*. (image credit: Britt Griswold)

## 3.1 Mission Design

*AeSI* is a separated aperture, long-baseline UV/Optical interferometer (Figure 3.1). The facility collects light at each of the individual primary mirror elements, relays it to a central hub, and recombines the light in a multi-beam Michelson combiner (Figure 3.2). The data from the interference signals of the combined beams provides information sufficient for reconstruction of stellar surface images. Many of the basic technical principles for separated aperture imaging involved in *AeSI* have already been well-established with Earth-based facilities such as CHARA (Gies et al. 2024), the VLTI (Haubois et al. 2022), and NPOI (van Belle et al. 2022).

*AeSI*'s location on the lunar surface has some critical advantages. The lack of an atmosphere greatly simplifies the facility: no adaptive optics are needed for the individual systems, and the elaborate vacuum systems necessary for beam relay are eliminated. The lack of an atmosphere fundamentally enables the science operations: *AeSI* can operate at much shorter wavelengths than any Earth-based facility, and the coherence times are limited by the instrument and not the atmosphere.

The individual apertures will be one meter-class telescopes; at this size, sufficient UV photons can be collected in support of the facility's science drivers. These telescopes will be individually positioned relative to the central hub, such that for a given observation, the gross path length from target to telescope to hub is roughly equivalent to that of the other individual apertures. Generally speaking, for a given pointing of *AeSI,* a plane wave of light can be expected to come from the target which, projected onto the sky, will be a circle the size of desired synthetic aperture. Intersecting the ground plane on which *AeSI* resides, that circle will inscribe an ellipse for non-zenith pointings; the apertures of *AeSI* will each move to reside on the perimeter of that ellipse, and the hub will be at one of the ellipse foci (e.g., Figure 3.3). These telescopes will compress the collected starlight from their aperture size down to a relay beam that is a few centimeters across. Slight changes in the target sky position - e.g. as it tracks across the sky due to lunar rotation - will be accommodated by meter-class delay lines for pathlength compensation that are part of the beam train from sky to hub.

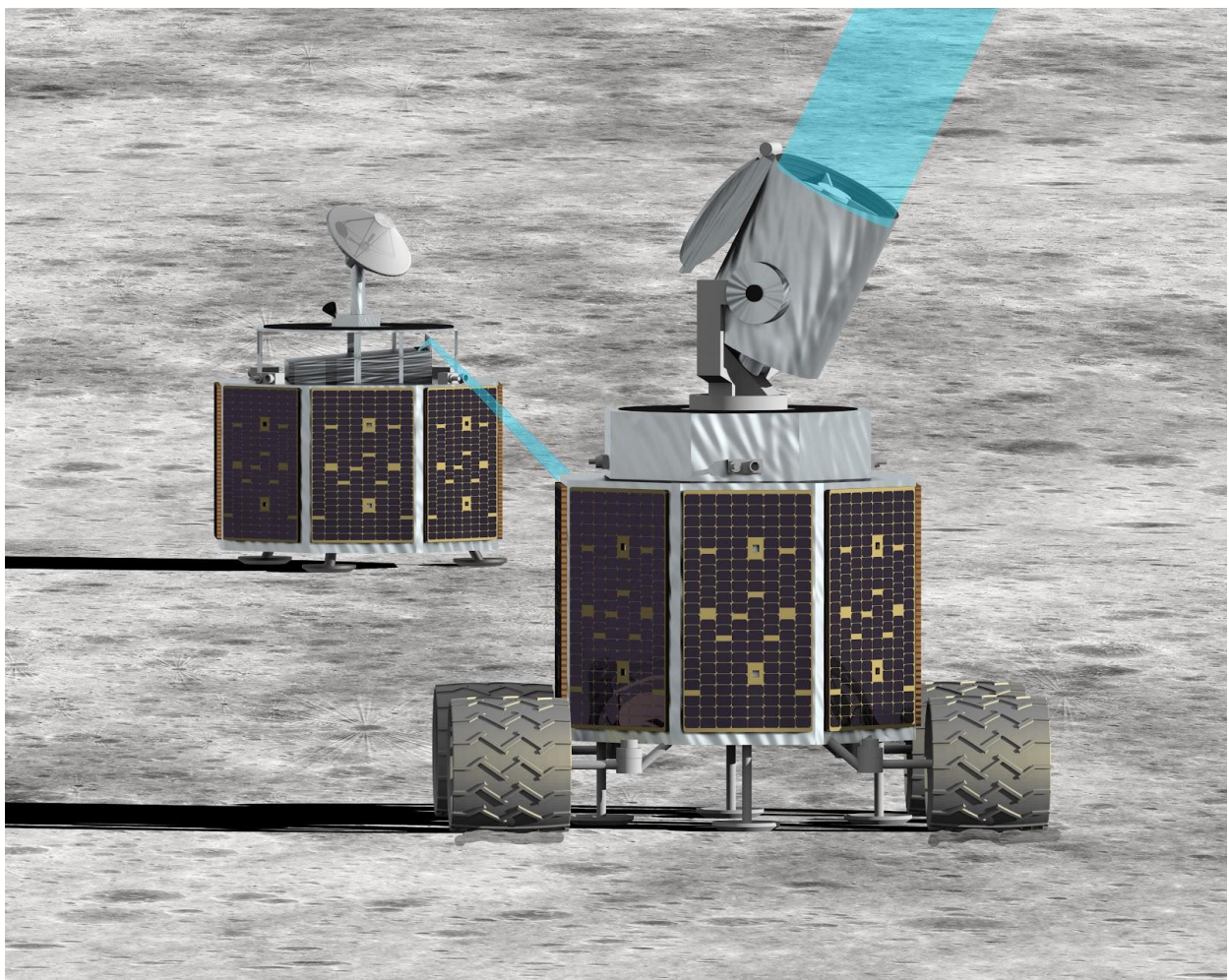

**Figure 3.2** Artist rendition of one of the primary mirror elements directing a beam to the central hub. (image credit: Britt Griswold)

The beams that arrive at the hub will be fully pathlength-compensated, ready for recombination. The individual aperture's beams will be arranged in a linear non-redundant array, with those beam inputs then mapped onto an image plane. In the axis along that linear array, interference fringes will form for each possible pair of telescopes from the array; the non-redundant nature of the spacings means the fringe frequencies will be unique to each pair and retrievable via a Fourier transform. In the orthogonal axis, a dispersive element can be used to enable the wavelength-dependent nature of the on-sky scene to be revealed. This recombination scheme has been used successfully at the CHARA MIRC-X (Anugu et al. 2020) and MYSTIC (Setterholm et al. 2022) instruments and the NPOI VISION (Garcia et al. 2016), and is expected to scale well for *AeSI*.

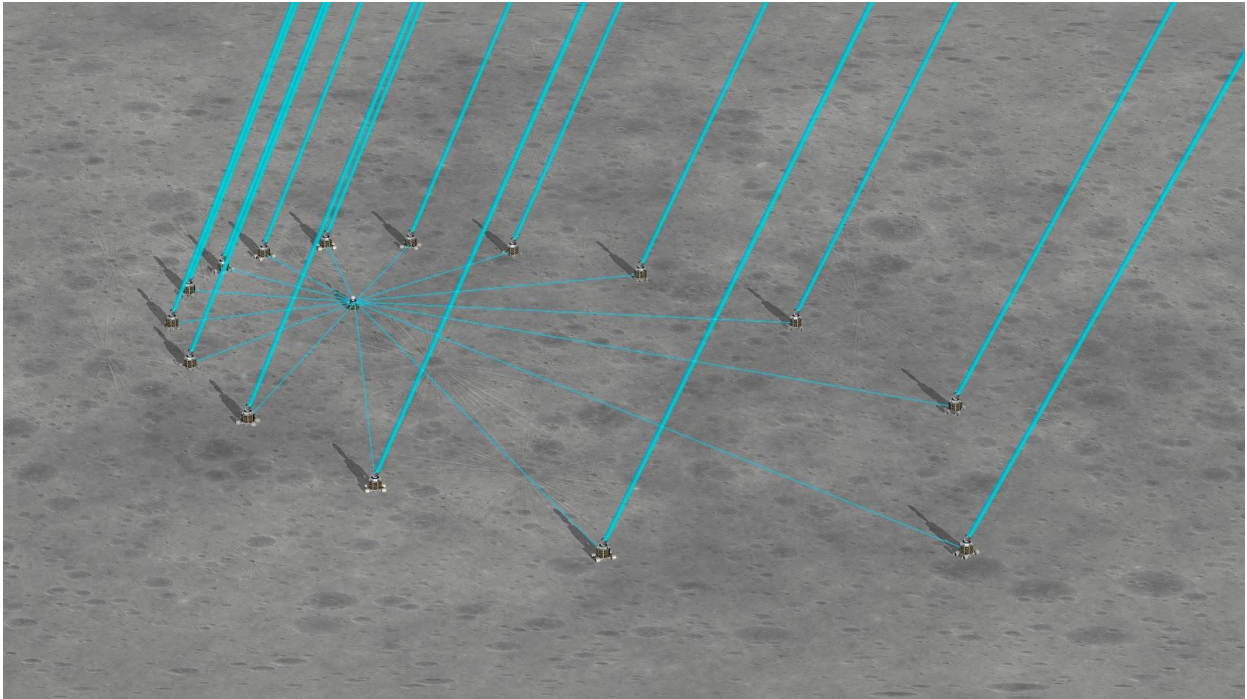

**Figure 3.3** Artist rendition of the *AeSI* array in an elliptical configuration. Each of the primary mirror elements collects light and relays it to the central hub. (image credit: Britt Griswold)

## 3.2  Approach to Evaluating the Mission Concept

There are many significant technical challenges to putting an interferometer on the Moon, but all of them have solutions that could be implemented as soon as the Artemis infrastructure is available. Optical interferometers have been operational on Earth since the 1970's (Labeyrie 1975). Making an operational one on the Moon requires optics, actuators, sensors, a control system, and power, all of which have space heritage. They can easily be deposited on the Moon with a lunar lander, such as the SpaceX Starship Human Landing System (HLS, see details in Section 3.3.3), to easily deliver far more than the 16 pieces (15 primaries +1 combiner hub) of our Baseline designed interferometer.

The lunar surface is harsher on surface hardware than Earth's due to larger temperature swings and the jagged nature of lunar dust. Heaters and sun shields can mitigate temperature swings. Dust and perhaps molecular contamination (for the UV) may be an issue, but accommodations can be made to mitigate those concerns. We plan to address the dust by strategically placing all optics above the lunar dust levitation layer (~30 cm; Rennilson & Criswell 1974; Katzan et al., 1991), with optics positioned more than a meter above the surface (Figure 3.4). To further mitigate dust accumulation, we are exploring the use of dust-repellent coatings and active dust mitigation strategies, such as electrostatic or mechanical dust removal systems. The Chinese Lunar-based Ultraviolet telescope (LUT) has been successfully operating on the Moon since 2013 (Wang et al. 2015), and with the RAC-1 (Regolith Adherence Characterization) instrument flying aboard the 2025 Commercial Lunar Payload Services (CLPS) launch, we will soon learn more about the scope of the dust problem and how to deal with it.

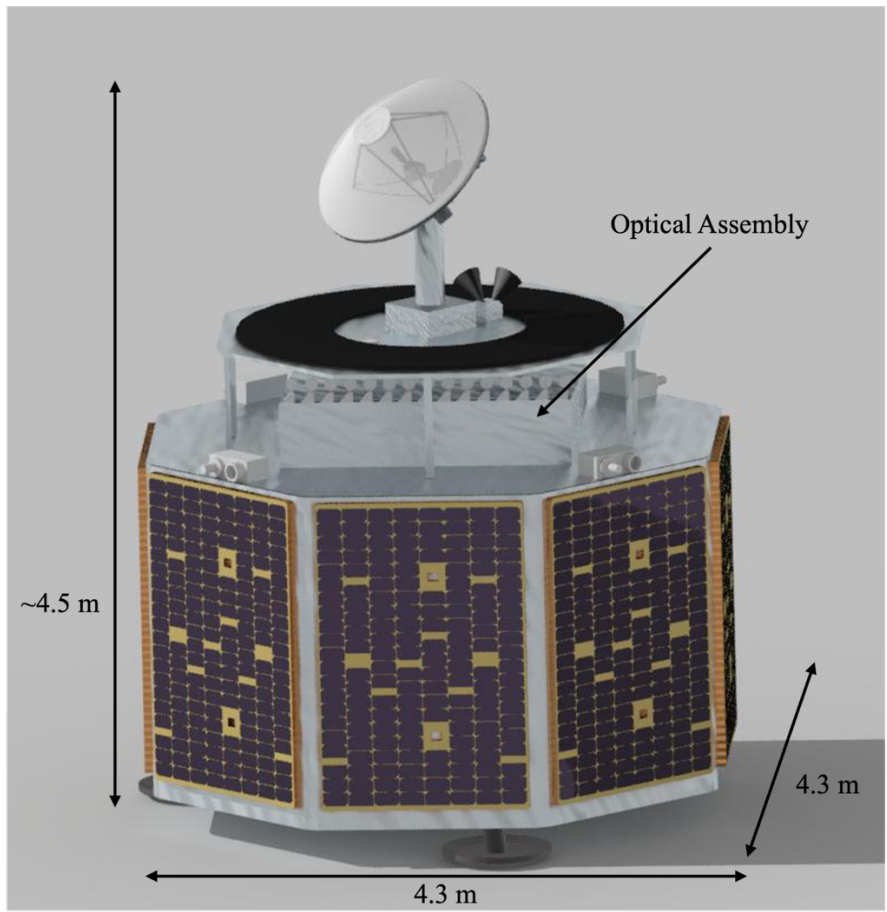

**Figure 3.4** Artist rendering of the hub, where the optical assembly is located well above the lunar dust levitation layer (~30 cm).

Seismic activity from internal sources (e.g., thermal contraction) are of low concern (Nunn et al., 2021); however, seismic noise due to micrometeorite impacts and vibrations from Artemis infrastructure are being investigated.  We will learn further details regarding the specific surface vibration environment for Artemis infrastructure from the Instrument for Lunar Seismic Activity (ILSA) payload on the Chandrayaan-3 mission (John et al., 2024) that landed near the Lunar South Pole in 2023, the Farside Seismic Suite (FSS) mission seismometers (Panning et al., 2022; a payload on the 2026 CLPS launch), and the Lunar Environment Monitoring Station[4] (LEMS) experiment on Artemis III. The findings from these missions/experiments will enable us to optimize our seismic activity mitigation procedures and designs. Our current plan is to place a seismometer on the hub to alert us when observations need to be discarded due to seismic disturbances and to provide lunar geologists with valuable data on moonquakes.

With this background, our initial approach for most technical challenges is to adapt techniques already demonstrated on the Earth.  However, several critical innovations are required to create a working interferometer on the Moon, as discussed above. We have built on the experience of our team in both the study of visionary space science mission concepts and in the design, construction, and use of ground-based interferometers, to execute a logical, strategic course of study with judiciously chosen study topics. This study identified and addressed, in a methodical fashion, significant unknowns and challenges, and the needed technology development. This was done in collaboration with the GSFC IDC[5] to ensure a realistic and credible design and allow an independent assessment of its feasibility. This collaboration executed the following tasks:

1. Assess the science case for a UV/optical interferometer located at the Lunar South Pole
2. Flow down the science requirements into mission design parameters
3. Perform engineering and architecture studies implied by the required innovations and challenges discussed above
4. Identify the support needed from lunar infrastructure
5. Develop a strategic plan for the long-term maintenance and evolution of the facility, leveraging an optimized balance of human and robotic operations. Assess the costs and challenges of constructing a full-scale array with up to 30 primary elements in a single deployment versus a phased approach that begins with a smaller array and expands over time. For the latter, evaluate whether the initial beam-combining hub—housing the detector and core observatory systems—should be designed for full-array compatibility from the outset or upgraded incrementally as the array grows.

During the first four months of the Phase I study, our team conducted initial assessments of all key mission elements, consulting community experts as needed. In the fifth month, we employed a concurrent engineering approach at the GSFC IDC, conducting a three-day hybrid architectural study with personnel from both the Mission Design Lab (MDL) and the Instrument Design Lab (IDL).

---

[4] https://science.nasa.gov/lems/
[5] https://etd.gsfc.nasa.gov/capabilities/integrated-design-center/

Leading up to this study, we held preparatory meetings with IDC engineers to communicate mission requirements and our preliminary architectural design. Regular discussions ensured that the study was structured to effectively assess and refine the mission concept, identifying optimal engineering solutions and defining a clear path for technology development and future mission phases.

The IDC and *AeSI* team addressed all key design considerations, refining estimates and concepts for:
1. Suitable launch vehicles
2. Mass, volume, and power requirements
3. Siting of the telescopes and hub/beam combiner
4. Rover designs, including both GSFC-developed and commercial options (e.g., Lunar Outpost's Mobile Autonomous Prospecting Platform[6] (MAPP)), to navigate uneven terrain and transport primary and delay-line optics
5. *AeSI* deployment strategies
6. Optical system alignment and fringe acquisition (autonomous vs. human-controlled)
7. Telescope mobility and array reconfiguration
8. Communication architecture for handling high data rates
9. Dust and molecular contamination mitigation
10. Observatory evolution over time
11. Overall operations concept

In the final four months of the Phase I study, the *AeSI* team incorporated IDC recommendations to refine our mission architecture and operations concept, defining the best path forward for a potential Phase II study.

*A note on key assumptions made in this study:*

> In the IDC collaborative study, several key assumptions were made when refining the Baseline design to demonstrate that a lunar-based interferometer can be largely self-sufficient, minimizing reliance on underdeveloped lunar infrastructure and supporting its feasibility as a near-term concept.
>
> For communications, we initially assumed a 50 Mbps capability would be available in Lunar Communications Relay and Navigation Systems Initial Operating Capability C (LCRNS IOC-

---

[6] https://www.tothemoon.mit.edu/lunar-outpost-mapp-rover

C, the third operational capability phase) but revised this estimate to 5 Mbps as a more conservative approach. For power, we accounted for a maximum darkness duration of 15 days at the interferometer site and required the use of technology readiness level (TRL) 9 solar and battery cells.

Regarding safety and hazards, we assumed that astronaut access to *AeSI* on the lunar surface would be restricted due to potential risks (e.g., battery-related concerns). To mitigate dust-related issues, we assumed that aperture covers would provide sufficient protection. Additionally, while baffles were deemed necessary, no specific lengths were assumed. For stray light management, we assumed that reusable covers would safeguard dust-sensitive surfaces during potentially contaminating events.

## 3.3 Mission Implementation

The *AeSI* mission concept is based on a collaboration with the human Artemis program to enable a cost-effective and productive scientific facility that is competitive and/or better than the free-flying version (SI). Artemis infrastructure along with astronauts and robots will provide or support activities including launch and landing capabilities, observatory deployment, some aspects of operations, and the servicing and upgrading of the interferometer. We provide details below on the key phases of mission implementation.

### 3.3.1 Site Selection

As an Artemis-enabled mission, the site selection for *AeSI* must prioritize proximity to Artemis stations to facilitate the deployment, servicing, and enhancement of the observatory by humans and/or robots. However, it must also be sufficiently distant to avoid interference and contamination from Artemis activities, such as landings, launches, and routine surface operations.

Key factors driving the site selection for *AeSI* include:

- **Proximity to Artemis**: The site must be close enough to an Artemis base to enable easy access for human and robotic assistance, but far enough to avoid disruptions and contamination from Artemis operations. The ideal distance appears to be in the ~2-10 km range, but this is still under study.
- **Terrain**: The location must be a relatively flat region with slopes less than 15 degrees and large enough to accommodate a 1-2 km diameter array.
- **Power Requirements**:
    - If the observatory is powered by solar arrays and batteries, the ideal site would be relatively flat, with the interferometric array in shadow for a majority of the time, while solar arrays benefit from maximum exposure to sunlight (potentially located on a nearby hill).

> o  If lunar surface fission reactors are used, a flat site with maximized time in darkness is preferred. The reactor could be located on a nearby hill to minimize radiation exposure, though this may not be a strict requirement.

A key consideration for site selection at the Lunar South Pole is solar illumination, which directly impacts power availability, thermal stability, and dust dynamics. Unlike equatorial regions with regular 14-day day-night cycles, the south pole exhibits significant variability in illumination depending on location. Some sites experience no midnight sun, with seasonal variations leading to night durations between 9 and 14 days (Figure 3.5, left). Others have periods of both continuous sunlight and daytime shadowing, with seasonal night variations ranging from 7 to 13 days and additional shorter-duration shadowing events lasting from 0.1 to 3 days (Figure 3.5, right). These variations influence power system design, as prolonged darkness necessitates greater energy storage, while frequent transitions between light and shadow drive more extreme thermal fluctuations and increase instances of dust levitation, all of which must be carefully considered for sustained lunar operations.

Examples of potential locations near early Artemis base candidates are shown in Figure 3.6 For this study, we assumed the site would be located near Connecting Ridge. Similar locations near more recently identified Artemis base candidate locations[7] are available, as well. At present, the quality of lunar data supports identification of multiple candidate regions to support *AeSI*; we have no major concerns of how *AeSI* can fit into long-term Artemis plans.

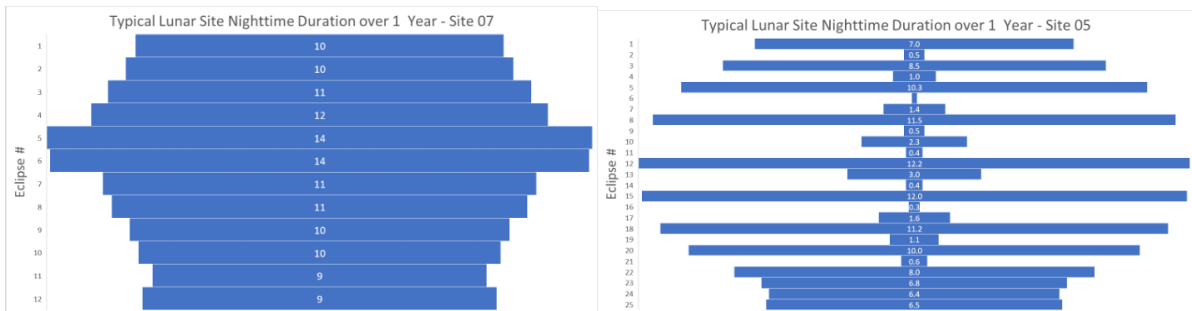

**Figure 3.5** Solar illumination is highly variable and site-specific near the Lunar South Pole, influencing power, thermal conditions, and dust dynamics. (Heritage analysis from Erwan Mazarico; see also, Mazarico et al. 2011)

---

[7] https://www.nasa.gov/news-release/nasa-provides-update-on-artemis-iii-moon-landing-regions/

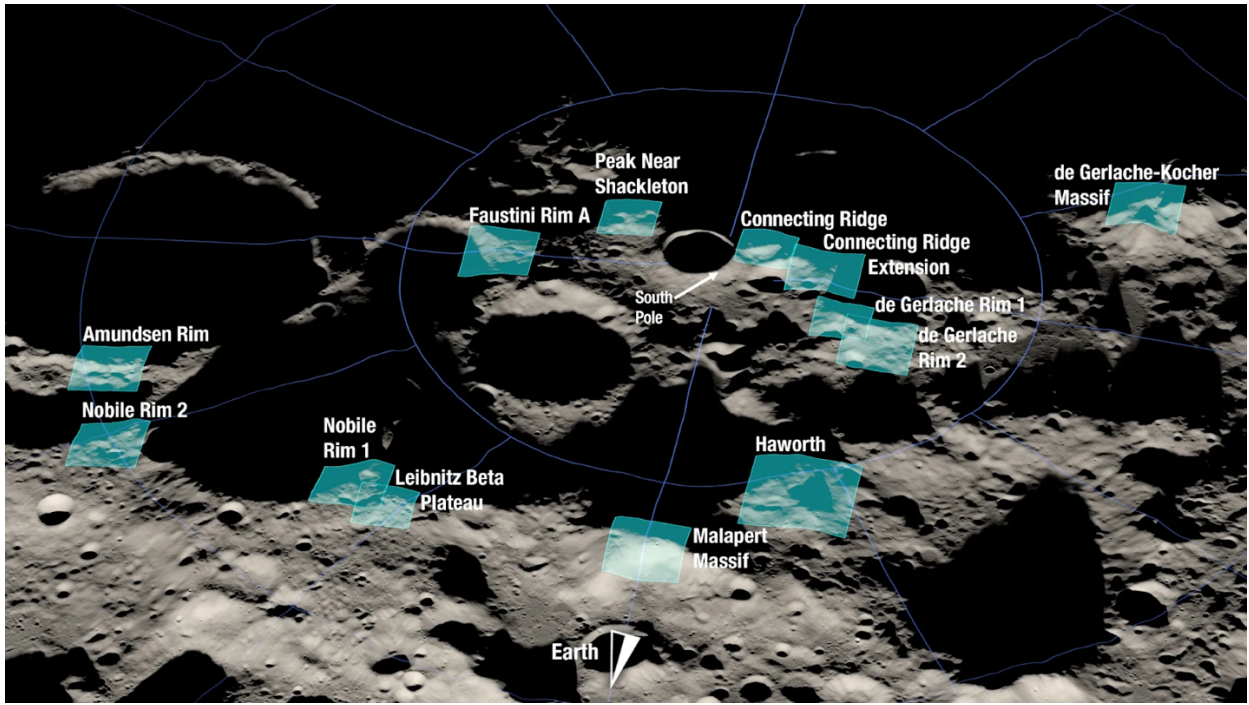

**Figure 3.6** Illustrative candidate locations for *AeSI*, near some of the original Artemis candidate base locations. (image credit: NASA)

### 3.3.2 Pre-Launch Support/Landing Site Preparations

Prior to arrival, pre-launch site preparations may be required, including having the surface pre-paved to accommodate hub set-up and configuring primary mirror elements (carts) into their final positions. A staging area will also need to be considered in the pre-launch support activities and a landing pad, depending on the landing site decision.

### 3.3.3 Launch and Lunar Landing

**Launch and Landing:** The *AeSI* hub and carts will be delivered to the lunar surface via rocket, requiring safe landing capabilities. This assumes the need for a landing pad, beacon infrastructure, position, navigation, and timing (PN&T), and communication systems, which may require infrastructure development at the remote site if not utilizing the Artemis base.

In the Baseline design, the transportation of *AeSI* to the lunar surface near an Artemis base camp is one of the key contributions of the Artemis program. The SpaceX Starship HLS is capable of launching the entire *AeSI* system into orbit and then onward to the lunar surface. Given the Starship's fairing size (9m x 18m) and its payload capacity (100-150 tons to low Earth orbit, 27 tons to geostationary transfer orbit), we anticipate sharing space in the fairing with other payloads. The rough mass estimate for the *AeSI* hub is 2,700 kg, with dimensions of 4.5m tall x 4.3m wide x 4.3m deep—well within the capabilities of the Starship rocket and lander.

**Unpacking:** After landing, *AeSI* will need to be unloaded from the rocket. Given the expected size and mass of the hardware, it is assumed that lifting, transport, and storage resources will be provided as part of the Artemis project.

### 3.3.4 Deployment

Since Starship will land near an Artemis base but not directly on an *AeSI*-suitable site, the observatory components will need to be transported from the landing site to the interferometer site and configured for initial commissioning and observation. The array elements are mounted on carts that can move across the lunar surface, allowing for reconfigurations during operations. These carts may be able to transport the array elements from the landing site to the observatory site, although it might be safer and more efficient to use a larger transport rover, if available through the Artemis Program. The hub, which is not designed to move at the observatory site, will require transportation as well. Once the interferometer components are roughly positioned, the commissioning program will take over to precisely place them for operational use.

### 3.3.5 Array Configuration

The baseline architecture for configuring the *AeSI* array was chosen to use independent lunar rovers for positioning the primary mirror elements relative to the hub and each other. However, this choice represents an initial approach rather than the outcome of a comprehensive, objective trade study. A configuration trade study to be explored in Phase II, including the option of placing the primary mirror elements on rails, is outlined in Section 4.4.

The configuration requirements for the *AeSI* array include:
- Maximum observation time (determined by the time scales of the astrophysical phenomena being studied)
- U-V Plane Coverage (the number of discrete positions within the observation time)
- Range (the maximum relative distance from the hub)
- Number of spokes
- Positional accuracy
- Reconfiguration time (calculated as [observation time/discrete positions] + repositioning time + reconnect time to the hub)

### 3.3.6 Servicing (Maintenance and Upgrades)

One of the great advantages of locating *AeSI* on the Moon is that servicing will be much easier than would be the case for the free-flying *SI* at L2. Our plan is to use the resources of Artemis, including transportation of new hardware from Earth to the Lunar surface and then to the observatory site, and a mixture of human and/or robotic services to perform both maintenance and upgrades to the facility.

**Maintenance**

The regular maintenance requirements should be minimal, with occasional dust removal from station surfaces if needed, and on-demand servicing to repair or replace any failed components. The interferometer is highly modular, and most servicing would likely involve replacing a faulty cart (i.e., a primary mirror station or "array element") with a spare. The malfunctioning cart could be returned to an Artemis site for repair and, if possible, kept as a spare for future use. The observatory is designed to tolerate the temporary loss of one or more array elements, allowing replacement scheduling to align with Artemis' operational requirements.

The hub is a more complex, stationary component without its own mobility system, but it could be designed with modular elements to allow in-situ servicing by robots or astronauts. If a failure occurs that cannot be addressed this way, the hub will need to be transported back to an Artemis site using the same method used for its initial deployment, where it could either be repaired or replaced with a new unit. Having a spare hub and one or more backup carts would be highly beneficial, though this and other contingency strategies need to be explored further.

**Upgrades**

The primary upgrade planned for *AeSI* is expanding the number of array elements from the initial design of 15 elements, to a target of 30, implemented in one or multiple stages. The addition of more primary mirror elements would dramatically increase the efficiency and scientific productivity of the observatory and is the most desired enhancement. The process is relatively straightforward, primarily involving the deployment of additional mobile carts to carry the new array elements. However, a key consideration is whether to design the central hub to support 30 beams from the outset, incorporate modular components that allow for on-site expansion, or replace the hub when scaling up. The current baseline approach is to deploy a hub capable of handling up to 30 elements from the start, though this trade-off will continue to be evaluated.

Other upgrades that may be of interest would be to install new, more efficient detectors and/or mirrors with higher reflectivity if dramatic improvements are made over the years in either or both. These would likely be done by replacement of carts and the hub, but the option of replacing just those components within the units on-site should be considered in future trades.

We believe both maintenance and upgrades could be accomplished by a combination of humans and/or robots and that the ideal mix will depend on the evolution of Artemis plans and the availability of astronauts and robots on-site.

### 3.3.7 Concept of Operations: Design Reference Mission

The Baseline design for *AeSI* assumes lunar daytime operation and power supplied by photovoltaics. This is the current design assumption because the plans for solar deployment are part of the current Artemis base. This means that observations are taken in sunlight, with the potential for scattered light entering the beam trains, which will need to be carefully mitigated for success. The following activities will be undertaken to deploy and configure the telescopes and subsequently undertake observations with *AeSI*.

First, the *AeSI* launch packages will be delivered to the lunar surface at the Artemis base. The packages will be taken to the identified location for the *AeSI* array. A group of nodes for the Hub and the Collection equipment are the two main categories to consider. The Hub node provides power to the array as well as communications to Earth, and it houses the beam combination hardware. It is centrally positioned at the array location, powered up using the existing Artemis power infrastructure, and initiates and consolidates communications to Earth for operations. Next, the Collection nodes (one per telescope) are now powered up and communication with the Hub node is initiated and established. The Collection nodes are finally deployed (via rovers) to their starting locations and sent to their home positions for starting observations.

Each telescope at a Collection node must be stabilized and oriented, physically, in the appropriate location for observing. This can be determined both mechanically and optically on the sky. Each telescope location will allow both optical beam pointing back to the Hub node (for beam combination) as well as communications with the Hub for coordination of the array, so a clear line of sight is required without background confusion. Orientation, homing of the beams, and stabilizing communication must have scripted activities to determine metrics for success, or restarting the process if communications or a beam drops out.

To initiate Science Operations of the facility, we assume the Collection nodes are in locations appropriately selected to achieve the spatial resolutions required for the particular scientific observation. The Hub sends the command to initiate observations, at which point the telescopes point to and track on the science target. The delay lines are appropriately placed for the overall *AeSI* array layout, and then tracking is initiated to mimic lunar aperture baseline synthesis. A search algorithm is initiated until sufficient signal to noise ratio (SNR) is achieved to lock on fringes, at which point the *AeSI* array is "locked" on the target. When sufficient SNR thresholds are achieved, data recording is started. Based on the coherence time of the site, after an appropriate amount of data is recorded for the observation, the data will be recorded locally at the Hub and then transmitted back to Earth for archiving. (Coherence time in this case refers not to atmospheric factors (that we experience on Earth) but likely vibrational and systemic factors of the opto-mechanical system that limit the ability to integrate "forever". If no such factors exist, then the scientific observation data collection and SNR will become the limiting factors for the length of an observation.) Typically, science and calibration object observations are interleaved as many-layered sandwiches of data for Earth-based interferometers. The amount of time for each layer and how many layers are required to properly calibrate the data will depend uniquely on the lunar conditions as well as the anticipated cadence of evolution of scientific observables for the science target itself. (For instance, if a feature on the science target is expected to evolve in a few minutes, then observations of that target will need to be at a cadence typically twice as high as the evolution of the feature.) When this science-calibrator sequencing is completed, the Collection nodes can be commanded to a new location suitable for a new set of scientific observations, and then the Science Operations sequence is repeated.

At the anticipated end of life of the *AeSI* facility, all the Collection nodes should stow the telescopes and return them close to the Hub node so that all the equipment can be easily retrieved for removal. It is anticipated that lunar night will be too difficult to maintain adequate power through and ultimately survive under the conditions where photovoltaics are used as the only power source. Since surface nuclear power stations are being considered for Artemis, we will pursue lunar nighttime operations modes for *AeSI* in NIAC Phase II or other future studies. Advantages here may include ease in locating beams from the Collection nodes to the Hub due to lower confusion, or performance advantages associated with lower noise electronics or detectors at the much colder temperatures.

A functional block diagram for *AeSI* is shown in Figure 3.7.

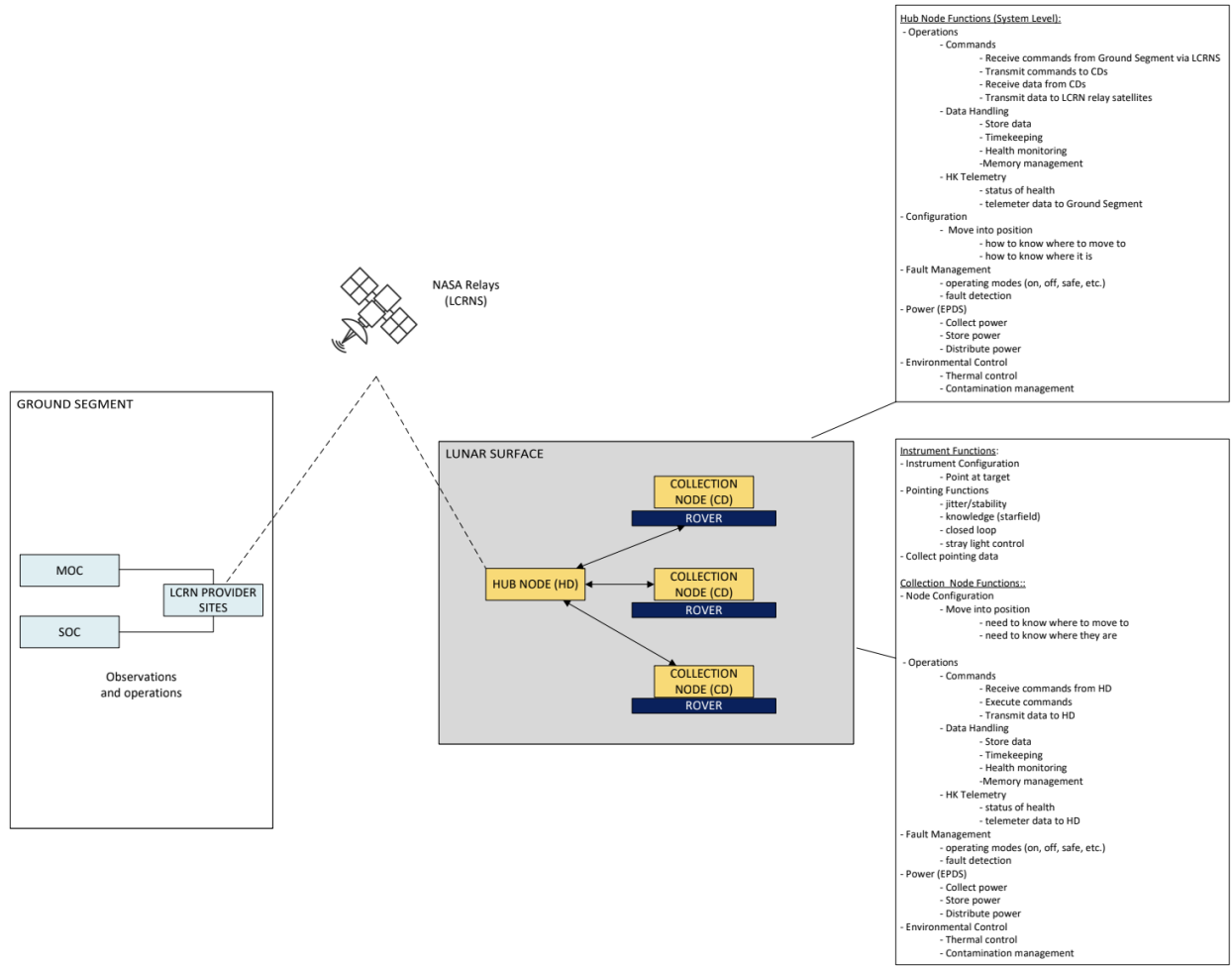

**Figure 3.7** Functional block diagram for *AeSI*.

## 3.4 Instrument Engineering and Architecture

### 3.4.1 System Overview

*AeSI* is designed to achieve the primary science goal of producing 30x30 resolution element images of solar-type stars. To achieve the necessary resolution would require a single-dish telescope with a half-kilometer aperture. Neither current technology nor any foreseeable advancements enable us to build this. Instead, *AeSI* will use interferometric imaging reconstruction techniques to reach this goal. Electric field correlations across the wavefront measure the Fourier transform of the image. By measuring these correlations using multiple small, widely separated apertures, *AeSI* trades collecting area for resolution thereby significantly reducing the system mass and cost while maintaining the required high angular resolution.

To see how this works, consider Figure 3.8. An on-axis source contributes an electric field that has constant amplitude across the aperture. If the source is off axis, the wavefront is tilted producing a sinusoidal variation of amplitude. The larger the angle, the higher the frequency of the sinusoid. It can be shown that integrating over the target gives the Fourier transform of the image. We can measure one Fourier component of the image by interfering light from two small apertures. The Fourier component, called the visibility, is the amplitude and phase of the fringe while the spatial frequency is the baseline (separation of the apertures measured perpendicular to the target) divided by the wavelength. The two-dimensional spatial frequency is expressed in inverse radians and is usually referred to as u and v. To reconstruct an image from Fourier components, one needs to make as many independent measurements as there are unknowns. An N-by-N image contains $N^2$ unknown, independent intensities. This requires measurements on about $N^2/2$ independent baselines since each baseline provides two measurements (amplitude and phase). Therefore, N telescopes provide N(N-1)/2 baselines or $(N-1)^2$ measurements (some phases are not measurable as they are indistinguishable from system phase errors). This problem becomes less important as more telescopes are added to the instrument. System designs can also be modified to accommodate observations of more specialized targets.

Measuring enough visibilities to form a 30 by 30 image with 15 telescopes requires moving the telescopes a couple of times while observing a single target. This is acceptable as long as the target does not vary significantly during the time it takes to collect the data. However, the observing procedure needs to be optimized since the quality of the image depends not only on source structure but also on balancing the number of times the array can be reconfigured, the number of telescopes and how the beams are combined.

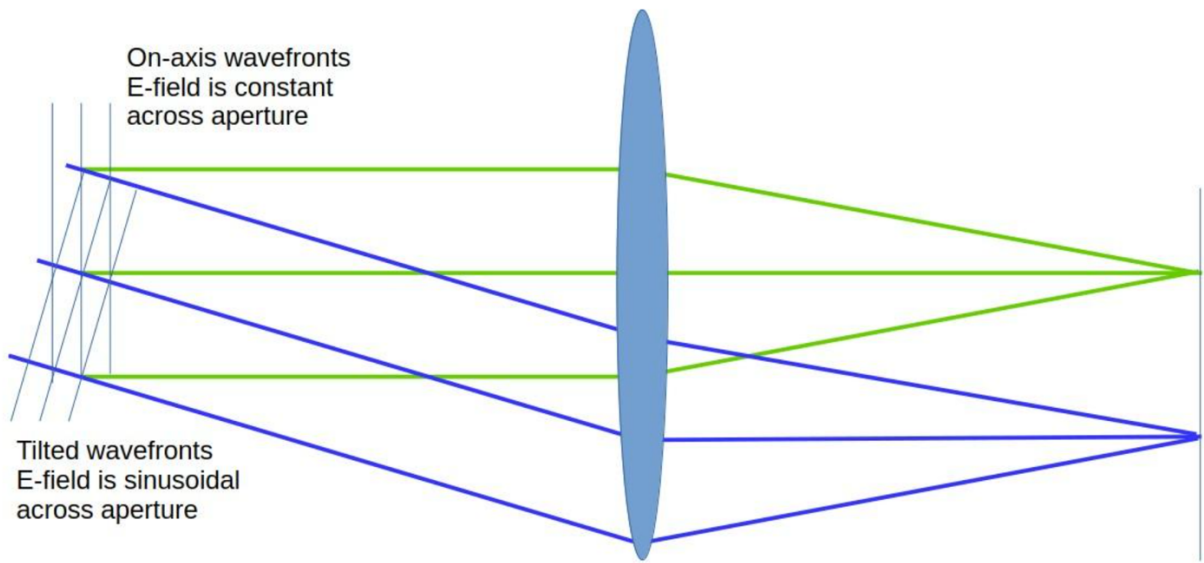

**Figure 3.8** An off-axis object produces a sinusoidal electric field in the aperture, one Fourier component of the image.

To measure interference fringes, the beams must be overlapped in both angle and position. The path lengths from the target to where the beams are combined have to be equal and the electric field vectors parallel. The fringe amplitude is proportional to the cosine of the angle between the field vectors making the measurement relatively insensitive to electric field misalignment. However, the path lengths have to be stable to better than $\lambda/10$ and matched to within the coherence length; high spectral resolution can relax this requirement to tens of microns. The path lengths as well as the angle and positions of the beams will be monitored and stabilized during the observations. With more than a couple of apertures, the simplest way to measure the fringes is shown in Figure 3.9. A lens (or any powered optic) taking an afocal beam as input forms an image in the focal plane. Our targets are unresolved by the individual apertures so the image will be an Airy pattern. Additional beams passing through the lens will form additional Airy patterns. When the beams are parallel, the patterns overlap but because they arrive at the focal plane from different directions, the relative phase of the beams will vary linearly across the image resulting in a sinusoidal fringe across the Airy pattern. The frequency is proportional to the spacing of the beams at the lens. If all pairs of beams have different spacings, the signals can be separated allowing us to measure the amplitude and phase for all the baselines simultaneously.

Additionally, when all the fringes are parallel, a cylindrical lens can be added to the system to compress the image to one pixel, perpendicular to the fringes. Then a diffraction grating can be used to provide spectral resolution on the detector in the dimension perpendicular to the fringe pattern. An important feature of this design is that it is easily expanded. To add a beam, simply increase the diameter and focal length of the lenses and use a detector with more pixels. The size of the combiner grows linearly with the number of telescopes while the number of baselines grows quadratically.

Interferometric combiners using this design with six input beams are currently in operation at CHARA and NPOI. This design can be readily expanded to accommodate more apertures by simply including the additional apertures in the existing row of apertures and increasing the system magnification and number of pixels on the detector to accommodate the extra fringes.

The apertures can also be grouped into subsets to increase the signal to noise per baseline at the expense of detecting fewer baselines. If the apertures are arranged in multiple rows, this combiner will produce multiple fringe patterns, one for each row (subset of apertures). Data from a laboratory demonstration with 15 apertures in each of 5 rows is shown in Figure 3.10. The width of the fringe pattern scales with wavelength resulting in the fountain pattern seen in this figure.

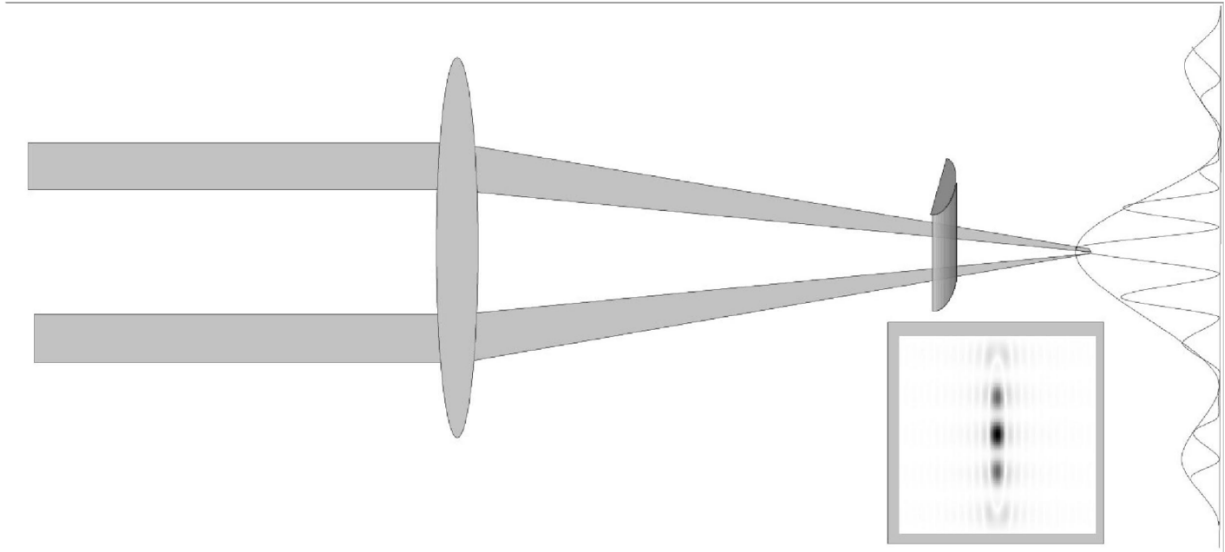

**Figure 3.9** Beam combiner concept. Each of the parallel beams imaged with a single lens will form an Airy pattern at the same place on the focal plane but since they arrive at the focal plane from different directions, they will form images across that Airy pattern. A cylindrical lens compresses the fringe pattern onto a single row of pixels and a diffraction grating (not shown) provides spectral resolution in the perpendicular direction.

Achieving *AeSI*'s primary science goal of creating 30x30 resolution element images of solar-type stars requires sub-nanoradian resolution and baselines on the order of half a kilometer. An N-by-N image contains $N^2$ unknown, independent intensities. This requires measurements on about $N^2/2$ independent baselines since each baseline provides two measurements (amplitude and phase). Therefore, N telescopes provide $N(N-1)/2$ baselines or $(N-1)^2$ measurements (some phases are not measurable as they are indistinguishable from system phase errors). This problem becomes less important as more telescopes are added to the instrument. System designs can also be modified based on observations of more specialized targets.

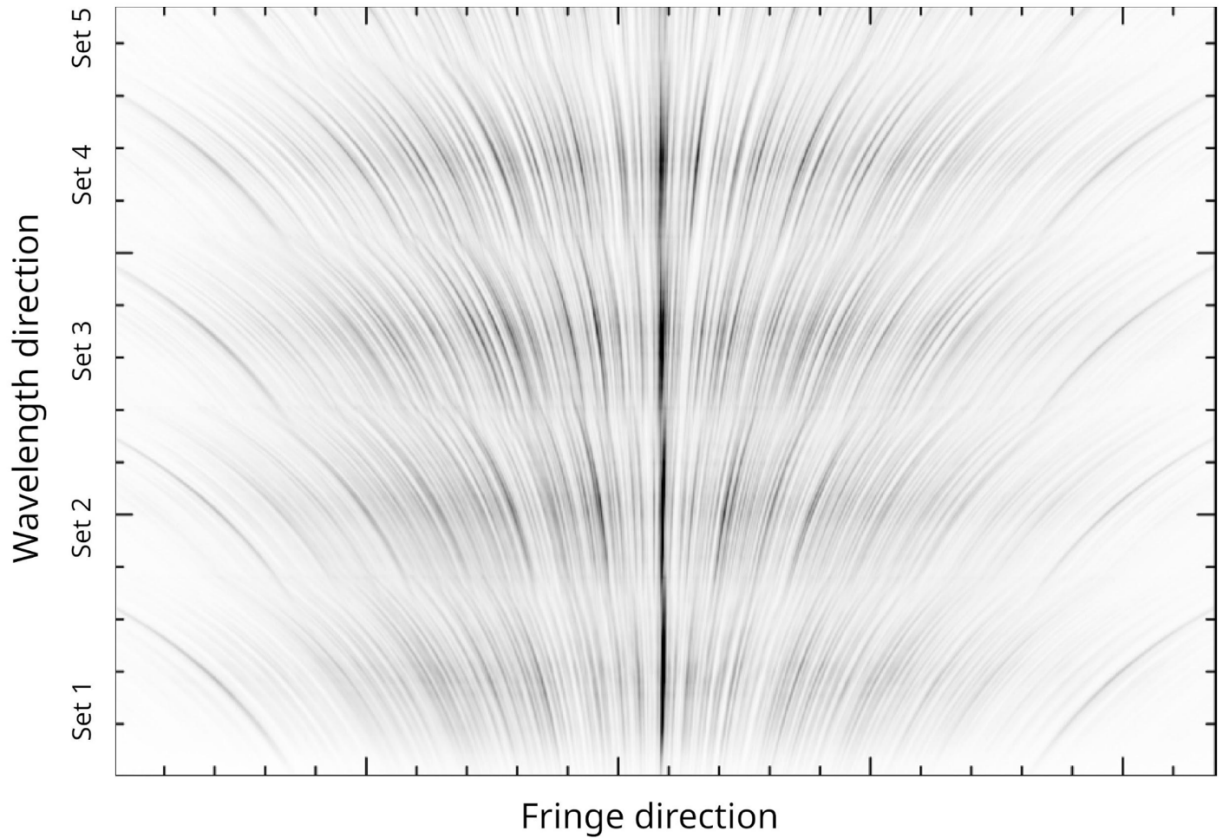

**Figure 3.10** Laboratory data for a beam combiner showing data from 75 apertures. The apertures were divided into 5 sets, each with 15 apertures. Only fringes between apertures in a row are measured. Fringes are in the horizontal direction. The vertical axis, wavelength, repeats once for each set of apertures.

The trade-off in an interferometric system involves balancing the number of telescopes, baselines, and signal-to-noise ratio (SNR). Increasing the number of telescopes enhances image quality by providing more baselines, which improves sensitivity to fine details and reduces sensitivity to image structure changes during an observation. More telescopes also provide better control over systematics and increase the likelihood of detecting fringes on longer baselines. However, with more baselines, fewer photons are available per baseline, necessitating longer integration times. Configuring more rows reduces the total number of baselines but improves SNR per baseline, which is crucial for faint targets. Additionally, resolved targets lose SNR on longer baselines, and stabilizing the fringe pattern becomes more complex with additional rows, requiring careful fringe tracking and combiner design.

The entire system for *AeSI* is modular. It can be built small with additional elements added over time. Our Baseline design starts with 15 primary mirror elements and eventually increases to 30. The primary mirror elements will be mounted on carts, enabling the system to be reconfigured as needed. With 15 primary mirrors, achieving a 30x30 resolution element image requires four reconfigurations.

For observing a target at the zenith, the primary mirror elements are arranged in a circle centered on the beam-combination facility (a.k.a, hub).  With this configuration, all the delays are equal.  For a target off zenith, equal delays can be achieved by stretching the circle into an ellipse with the azimuth of the major axis the same as the azimuth of the target and the hub located at the far focus This is shown in Figure 3.11. The wavefront from a target off zenith reaches one primary mirror before the other. The primaries can be placed asymmetrically around the hub. Per the bottom left panel of Figure 3.11, when A+B=C, the wavefront arrives at the hub from both primaries at the same time and no delay compensation is needed.

As the target moves across the sky, the relative path lengths change. Instead of moving the primary mirror elements while observing, we introduce small delay lines attached to the telescopes. This design has the primaries mounted on carts so they can be moved between observations.  We believe this is a simpler design than using fixed telescopes and adding delay lines that are hundreds of meters long.  This is especially true since the primaries already need to be mounted on carts to match the resolution of the interferometer to the size of the target.  The delay lines are sized to provide about an hour of observing time.

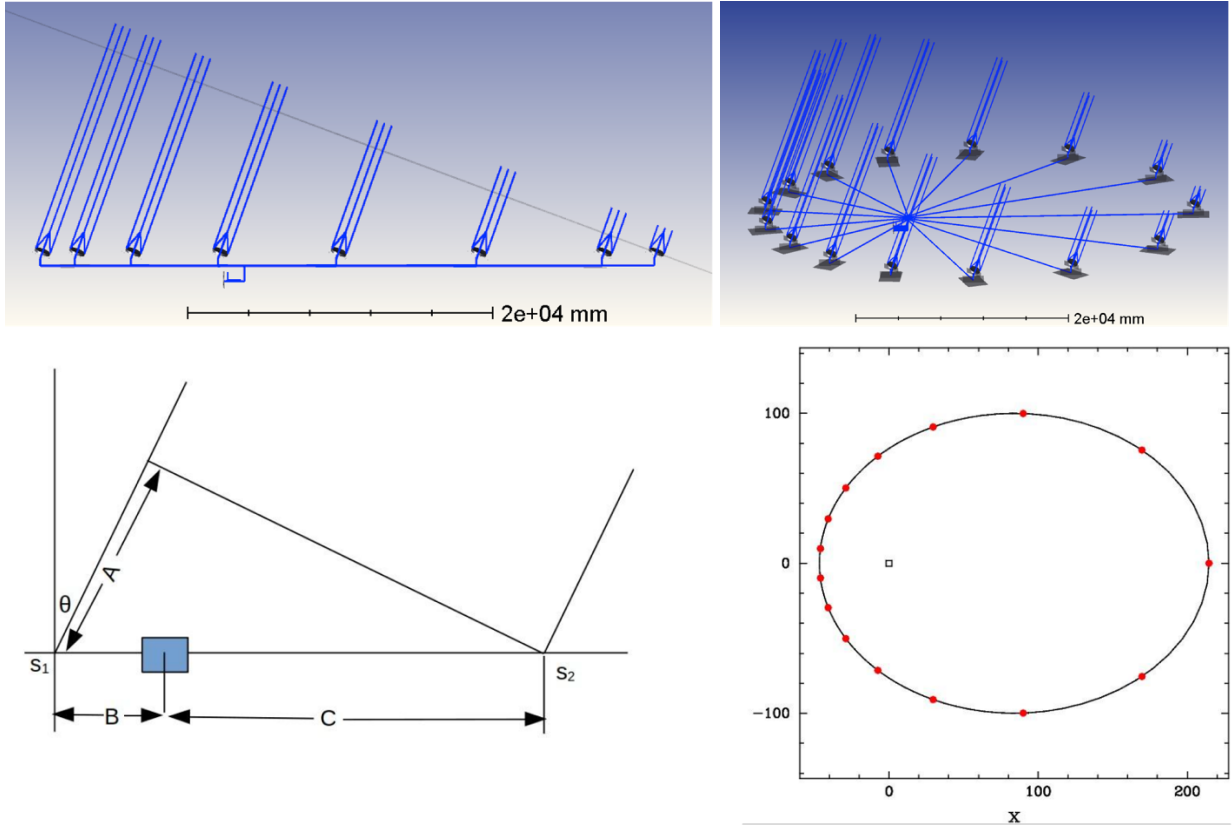

**Figure 3.11** *Top Row:* Equalized optical paths (*left*) and elliptical cart and hub arrangement (*right*) for detecting a wavefront off the zenith by 20 degrees. *Bottom Row left:* The wavefront from a target off the zenith reaches one telescope before the other. The telescopes can be placed asymmetrically around the beam combiner (located at the blue rectangle). When A+B=C, the wavefront arrives at the combiner from both telescopes at the same time and no delay compensation is needed. *Bottom Row right:* The same technique works in two dimensions. The telescopes are placed on an ellipse formed by the intersection of a paraboloid aimed at the target and the surface of the Moon. The ellipse shown is for a 200-meter baseline observing 40 degrees off the zenith.

By adopting a telescope-on-cart design, we are implicitly choosing a structurally compliant system requiring error sensors and active control rather than one that is designed to be long-term stable. This is the natural choice: the stability requirements are too tight to be consistent with a rigid system. The challenge of this design is the need to provide actuators and sensors. The approach to managing this is, while observing, the pointing error signals for all the primary mirror elements can be obtained from starlight with only three cameras and a handful of additional optics. Control can be provided by adding a delay line and tip/tilt control on two mirrors for each telescope. This will also require additional sensors for aligning the system after moving the cart. Pathlength control will be accomplished with a one-dimensional internal metrology system to remove mechanical jitter and delay errors measured on the target to eliminate drift.

With 15-30 telescopes, the system will have a much higher actuator count than seen on a typical satellite mission. The *AeSI* design on the lunar surface is robust to loss of a single station. Additionally, any temporary degradation in image quality due to such a failure can be easily addressed thanks to the redundancy and modularity of the overall system. Most failures would result in the loss of a single station. The system can operate with fewer telescopes although with slower and possibly slightly degraded image quality. This degradation would be temporary; the modular design allows relatively simple repairs by swapping out components either by a robot or by an astronaut from the nearby Artemis base.

### 3.4.2 Optical System Design

The primary mirror elements are placed on carts allowing them to be moved between observations. This compensates for most of the delay difference between stations and allows the array to be scaled to match the resolution to the current target. Additionally, it enables the use of multiple configurations on a single target resulting in better image quality than obtained from a single configuration.

The layout of the primary mirror element optics is shown in Figure 3.12. The first two mirrors, M1 and M2, form a typical afocal telescope while mirror M3 feeds the beam into the delay line. The delay line, mirrors M4 and M5, moves as a unit on a translation stage. The translation stage, measured with laser metrology, is adjusted during the observation to compensate for the changing delay due to Moon rotation and for real-time delay errors due to imperfections in the instrument. Mirrors M6 and M7 redirect the compensated beam to the hub. Two mirrors are needed to prevent electric field rotation when propagating the beam.

Each primary mirror element rotates to track the target during an observation. To maintain alignment, mirrors M3 through M5 are attached to the telescope's elevation axis with mirrors M5 and M6 collocated with the axis. Mirrors M1 through M6 are attached to the azimuthal axis with M6 collocated with that axis.

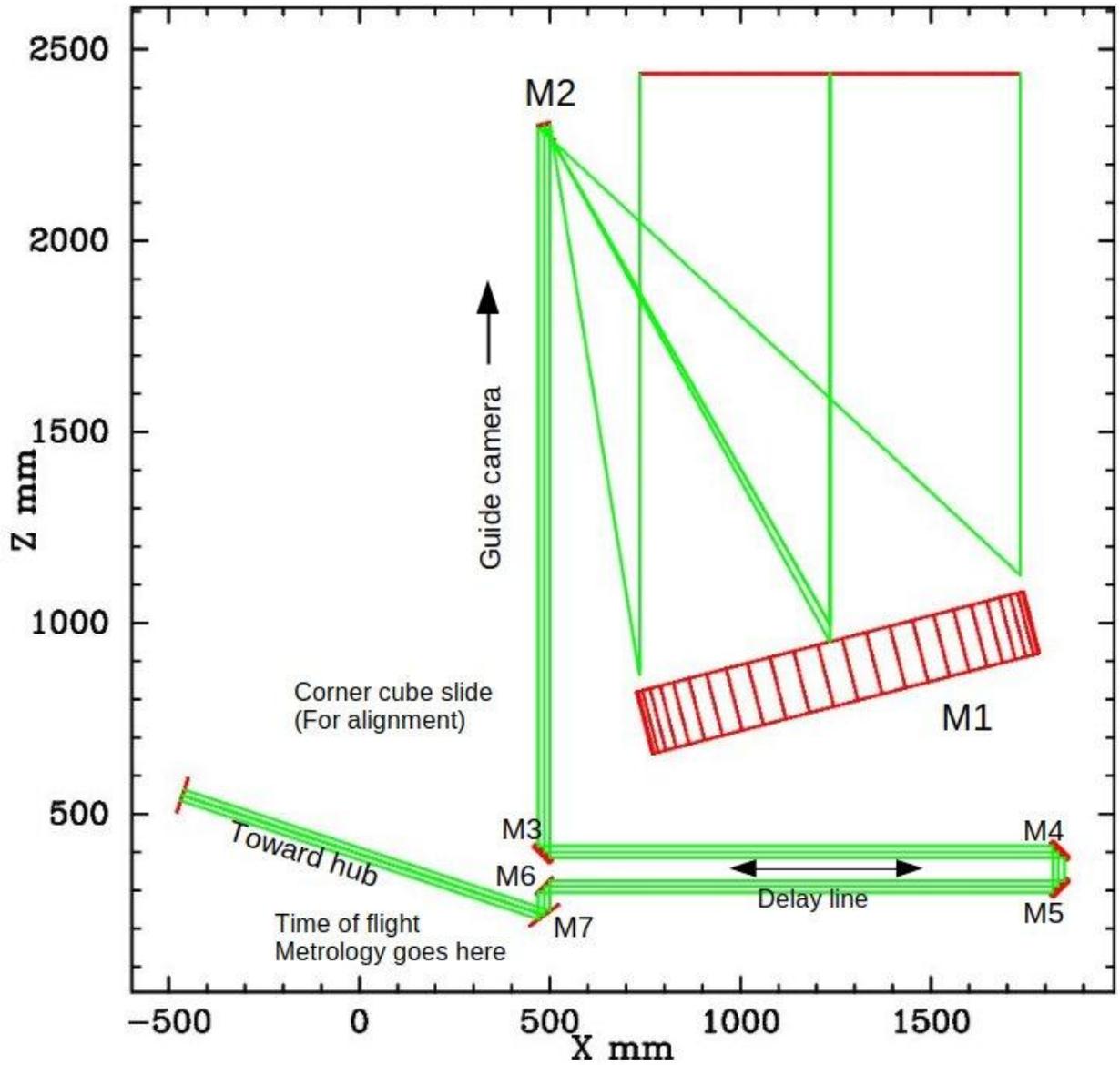

**Figure 3.12** The telescope is on a cart allowing it to be moved between observations. Our current optical design has 7 optical surfaces. M1 and M2 form a traditional, off-axis afocal telescope. Mirror M3 directs the beam into a delay line (M4 and M5). M6 and M7 are needed to transport the beam to the hub.

The hub houses the optics needed to measure the fringe parameters and generate error signals for control. The optical assembly and incoming beams from 15 primary mirror elements are shown in Figure 3.13, while the beam combiner concept is illustrated in Figure 3.9. The hub optics have three primary tasks: arranging the beams into a 1-dimensional, non-redundant pattern, combining those beams, and splitting off the Optical light for the control system. To ensure stability and precision, the entire optical assembly should be pre-aligned and integrated onto a single breadboard before deployment, allowing it to be transported as a solid, pre-configured unit from Earth. This approach eliminates the need for in-situ deployment and alignment, though an Artemis astronaut may verify the alignment upon arrival on the Moon. Accommodating these constraints is particularly challenging given the requirement for an effective focal length of 1.5 km.

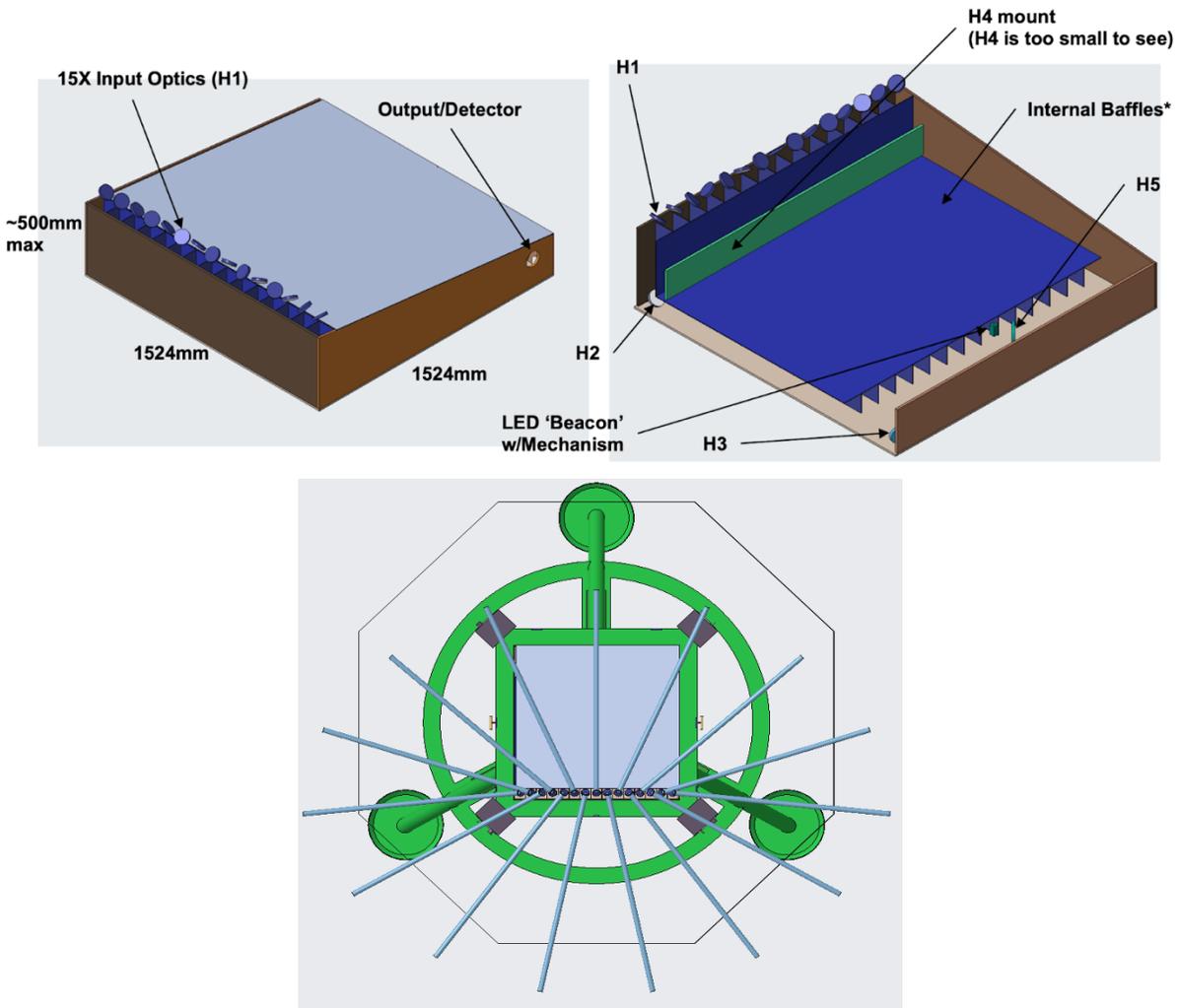

**Figure 3.13** Optical assembly (*top*) and incoming light paths for a 15-element array (*bottom*). The sightlines indicate the stay-out zones for the structure and other connections.

The proposed design is shown in Figures 3.13 and 3.14.  Mirrors $M_7$ (the last mirror on the cart) and $H_1$ form a periscope which preserves the electric field angle.  Mirror $H_2$ deflects the beam to the right, placing all the beams in a plane.  The $H_3$ mirrors rearrange the beams into a non-redundant configuration at $H_4$, where the $H_4$ mirrors form a common image on the field stop just to the left of mirror $H_5$.  Mirror $H_5$ transfers the image to the detector.  The cylindrical optic and diffraction grating described above are not shown in Figure 3.14. The full optical layout is shown in Figure 3.15.

The required magnification is obtained by adding power to all the mirrors.  Between mirrors H2 and H3, the beam diameter is reduced by a factor of ten.  An additional factor of ten occurs between mirrors H3 and H4. With this beam compression, the H4 mirrors all fit onto a 15 mm diameter substrate.  The beam combiner mirror, H5 adds the rest of the magnification.

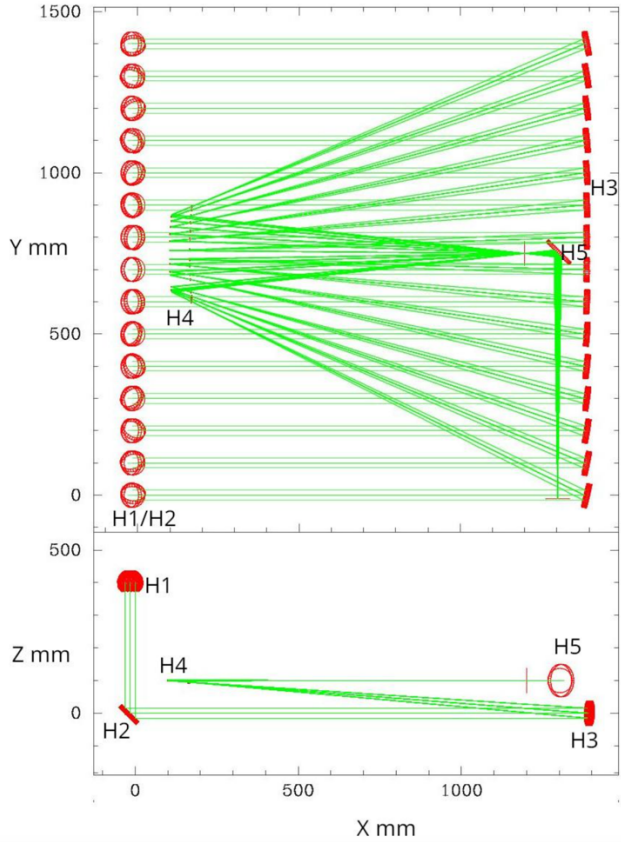
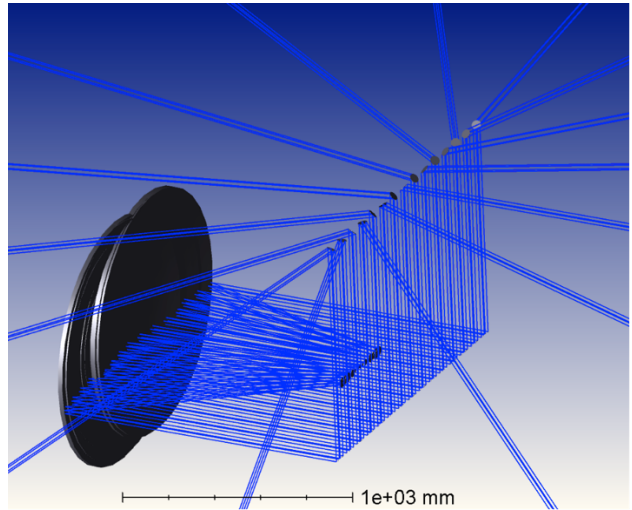

**Figure 3.14** Optical layout of the hub. *Left*: Mirrors H1 and H2 direct the beams down and to the right. Mirrors H3 is responsible for reducing the beam diameter and arranging the beam in an appropriate non-redundant configuration and H4 forms a common image. H5 adds more magnification and directs light onto the detector. The cylindrical lens and diffraction grating are not shown. *Right*: CAD rendering of the top panel schematic, shown from a rotated perspective to illustrate the incoming beams.

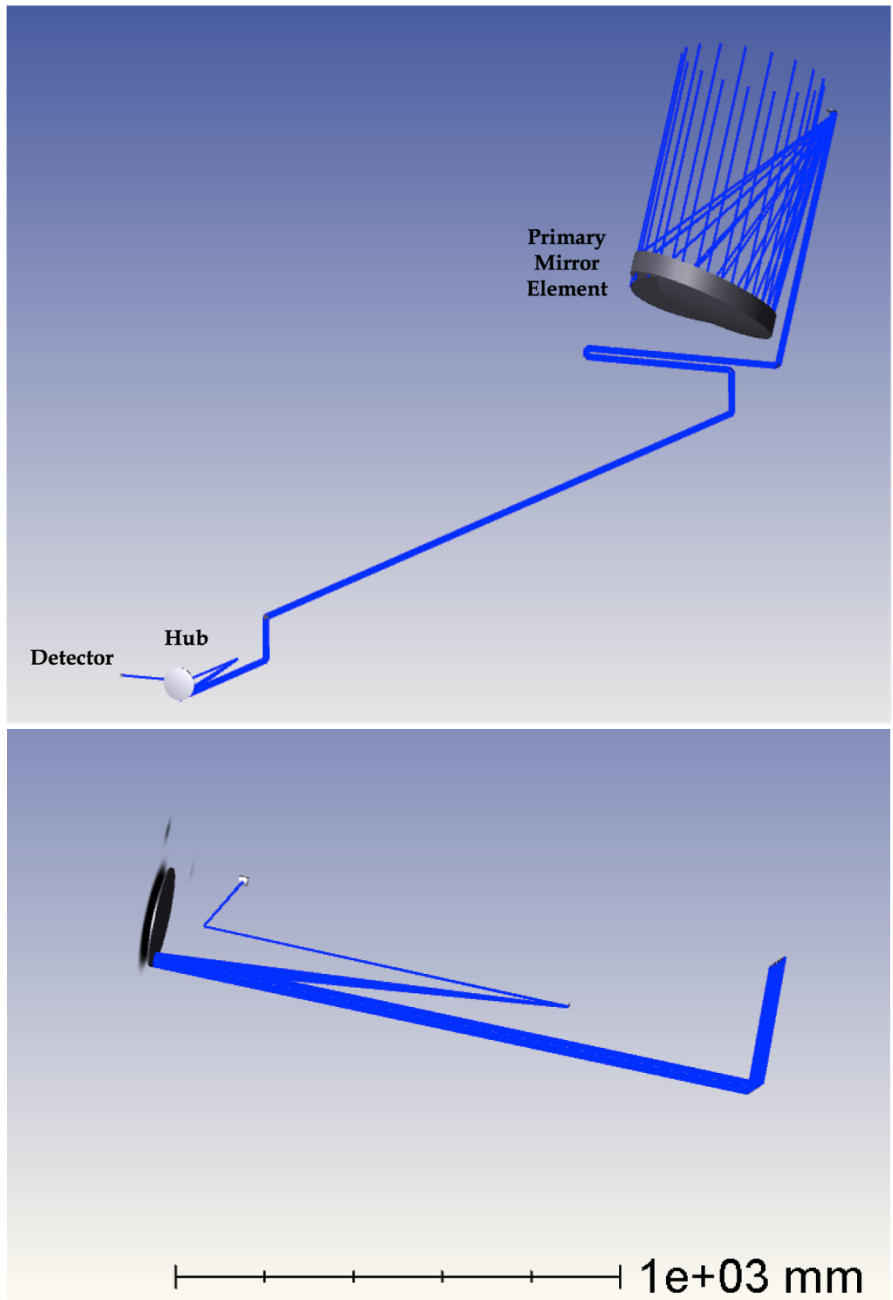

**Figure 3.3.15** *Top*: Optical layout showing the light path from a primary mirror element to the hub and onto the detector. *Bottom*: zoomed in view of the hub optical path.

### 3.4.3 Wavefront Sensing

The concept for the wavefront sensor is shown in Figure 3.16. The light is split three ways: half the light is split off for a fringe tracker, while the other half is split between an angle tracker and a position sensor. The fringe tracker uses the same design as the UV science camera. The angle camera is in the image plane; the position camera is in the pupil. An array of prisms tilt the beams to place them conveniently on the cameras.

All three sensors use Optical light. This is driven by the need for precision on the delay measurement, in particular. At longer wavelengths, targets are less resolved, resulting in higher contrast fringes and better SNR. Also, *AeSI* science cases are expected to have strong continuum signals in the optical, also helping increase the SNR.

The remaining task, separating the optical and UV, is shown in Figure 3.17. Mirror $H_2$ is wedged with the front surface being a dichroic, reflecting the UV light, and the rear surface a mirror for the optical. Both wavelength bands reflect off $H_3$ but are completely separated on $H_4$. Non-common path errors between the optical and UV beams are controlled with this design.

The wavefront control is regulated by three sensors. The delay sensor (fringe tracker) takes half the light and functions as a beam combiner, similar in design to the UV science camera. While its primary role is wavefront control, it also has the capability for Optical-light science, which will be explored in future studies. The remaining light is split between one camera that images the beams for angle tracking and a second camera imaging the pupil for maintaining the proper separations of the beams. Both the image-plane and pupil-plane sensors make use of wedges to separate the images on the detector and make best use of the detector form factors.

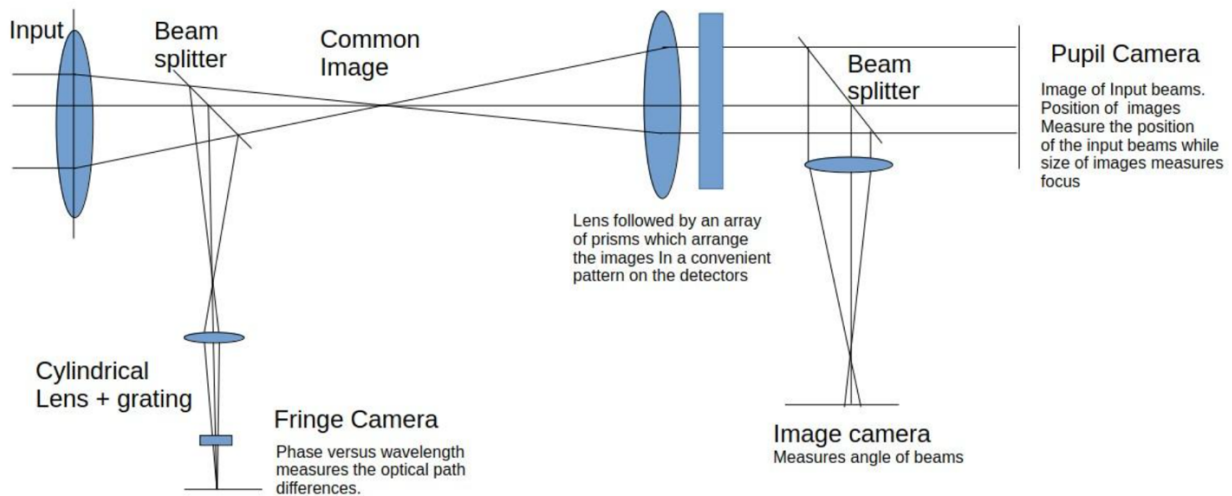

**Figure 3.16** Proposed wavelength sensing. The delay is measured with beam combiner similar to that used in the ultraviolet. The angle and position of the beams are measured with cameras in the image and pupil planes.

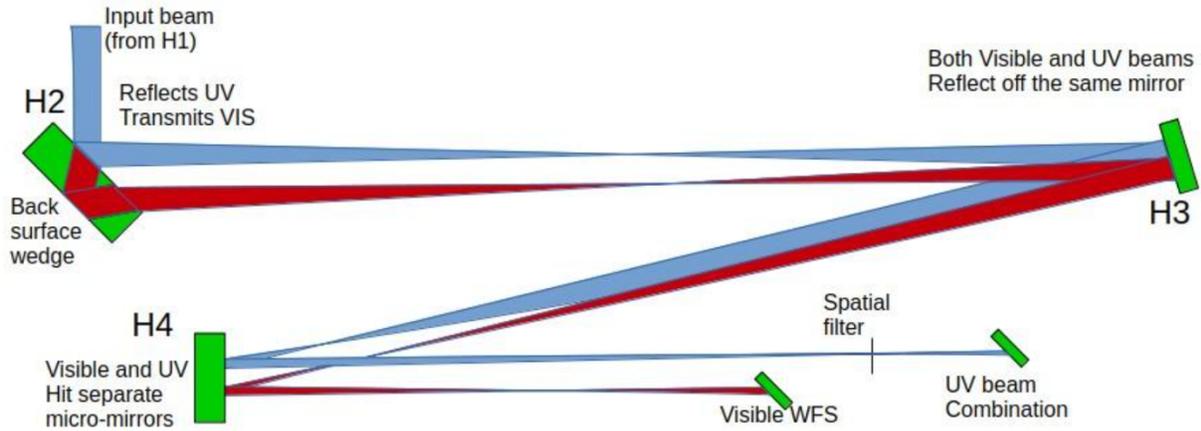

**Figure 3.17** Proposed wavelength separation with a dichroic at H2. The ultraviolet is in blue; visible in red.

### 3.4.4 Coherence Volume on the Moon

For an Earth-based interferometer, the sensitivity is limited by the roiling atmosphere. The atmosphere limits exposures to a few milliseconds and the aperture is limited to a couple tens of centimeters which places fundamental sensitivity limits on any system. Adaptive optics using natural guide stars is also limited to bright targets while laser guide stars have limited utility and become very expensive for an interferometer with several telescopes. Fringe tracking on a bright, nearby guide star would seem to be a natural solution but atmospheric turbulence decorrelates with angle. If the guide star is more than a few arcseconds away from the target, the fringe amplitude is reduced making calibration of the data problematic. However, differential phase measurements between the guide star and target are unbiased allowing precise positions to be determined over an arcminute field in the infrared. These are the fundamental limits of what can be achieved with long-baseline interferometry on Earth.

Faced with these performance constraints, observing from the Moon with *AeSI* becomes an exciting possibility. The absence of a lunar atmosphere eliminates the main limitation to exposure time, aperture and field of view. The delay line and the telescope's two axes are the only parts that move during an observation. The position of the delay line will be controlled and measured with a laser metrology system. Performance of similar systems operating on the Earth's surface achieve nanometer-level jitter, so this should not be a limitation. We must also consider runout in the telescope bearings which will cause the telescope position to slowly rotate about its mean position. Over short times, this will cause the delay to vary linearly with time in which case the visibility amplitude is reduced by:

$$1 - |V| = \left(\frac{\pi^2}{6}\right)\left(\frac{\Delta D}{\lambda}\right)^2 \tag{1}$$

Keeping the visibility reduction to 5%, the drift must be less than ΔD=0.09 μm during an exposure. For instance, this impact can be seen in the quality of bearings needed for the telescopes. Total runout in high-quality bearings is on the order of 2 microns typically over angles of 90 degrees. For *AeSI*, we are interested in small angles – during a one-hour exposure the Moon rotates by half a degree. This should keep the delay drift to an acceptable level.

The exposure time can also be limited by knowledge of the pointing model. The delay naturally changes during an observation due to the rotation of the Moon. To see fringes, we need to model this change and correct for it. Our Baseline design involves telescopes mounted on carts that are moved between observations. Every time a cart is moved, we lose knowledge of the baseline (distance between the telescopes). If our estimate of the baseline is off by *b*, the delay will be wrong by:

$$\Delta D = b \sin(\omega(t - t0)) \tag{2}$$

where the angular velocity of the Moon, $\omega = 3.36 \times 10^{-6}$ radians/s. A 100 second exposure requires baseline knowledge better than 0.25 mm. An error as large as one centimeter after moving a cart will allow a one second exposure. After measuring four bright stars surrounding the target, baseline knowledge can be better than one micron. The length of an exposure will be limited by other unknowns: e.g. the thermal drifts in the instrument or settling of the cart legs into the regolith after moving the telescope, etc.

### 3.4.5 Sensitivity of *AeSI*

The fringe amplitude signal to noise is given by:

$$SNR = \frac{\sqrt{2N^2 V^2}}{(n_T N + P\sigma_R^2)} \tag{3}$$

where N is the number of detected photons per telescope, V is the target visibility amplitude, $n_T$ is the number of telescopes, P is the number of detector pixels and $\sigma_R^2$ the read noise variance. In this discussion below, we assume no read noise.

Throughput is summarized in Table 3.1. The system described in Section 3.4.2 has 13 reflections. The mirror reflectivity, 0.85, is low by Optical light standards but has been achieved in the UV with overcoated aluminum between 120 and 200 nm. We assume similar reflectivity at longer wavelengths. Higher reflectivity silver mirrors can be used for Optical light after the wavelengths are separated. The size of the spatial filter/field stop is a free parameter, trading throughput for lower scattered light. Here, it is set at 2.44λ/D. The Optical light system has a beam splitter to pick off some light to control beam angles and positions. Combined, these predict a throughput of 6.6% for the UV. Since half the Optical light is used for wavefront sensing, the throughput is only 3.5%.

Table 3.1 Throughput

| SYMBOL | UV | VIS | DESCRIPTION |
|---|---|---|---|
| $n_R$ | 13 | 10 | Number of Al mirrors |
| $n_S$ | 0 | 3 | Number of Silver mirrors |
| $\eta_A$ | 0.8 | 0.85 | Reflectivity of Al |
| $\eta_S$ | | 0.98 | Reflectivity of Silver |
| D | 0.8 | 0.8 | Dichroic |
| S | 0.8 | 0.8 | Spatial filter |
| Q | 0.85 | 0.9 | Detector QE |
| B | 1.00 | 0.5 | Beam splitter |
| $\eta$ | 0.066 | 0.035 | Total |

Interferometric imaging differs from focal plane imaging in that every photon in the image contributes noise to all image pixels[8] not just the pixel where the photon is detected. As a result, image quality is determined both by the number of photons and by the structure of the image. Thus, we define dynamic range as the integral of the signal over the image divided by the noise floor. In frequency space that is equivalent to:

$$D = \sqrt{\frac{n_M}{\sigma_V^2 + \sigma_\phi^2}} \qquad (4)$$

where $n_M$ is the number of visibility measurements. This is only a guide; good imaging also requires good u,v coverage. If the signal is evenly distributed over $n_{pix}$ pixels, the signal to noise in each of those pixels will be SNR = $D/n_{pix}$. For a single observation with $n_T$ telescopes

$$Nt = \frac{2(snr\ n_{pix})^2}{n_T - 1} \qquad (5)$$

where N is the number of photons per aperture and $n_T$ is the number of telescopes and t is the length of the observation. The results are shown in two figures. Sensitivity for UV spectral line imaging is shown in Figure 3.18 and broad-band imaging at Optical wavelengths is in Figure 3.19. These plots are for a signal to noise of 5.

---

[8] Throughout this section, pixel is used to mean the size of a resolution element, not how finely the image is oversampled.

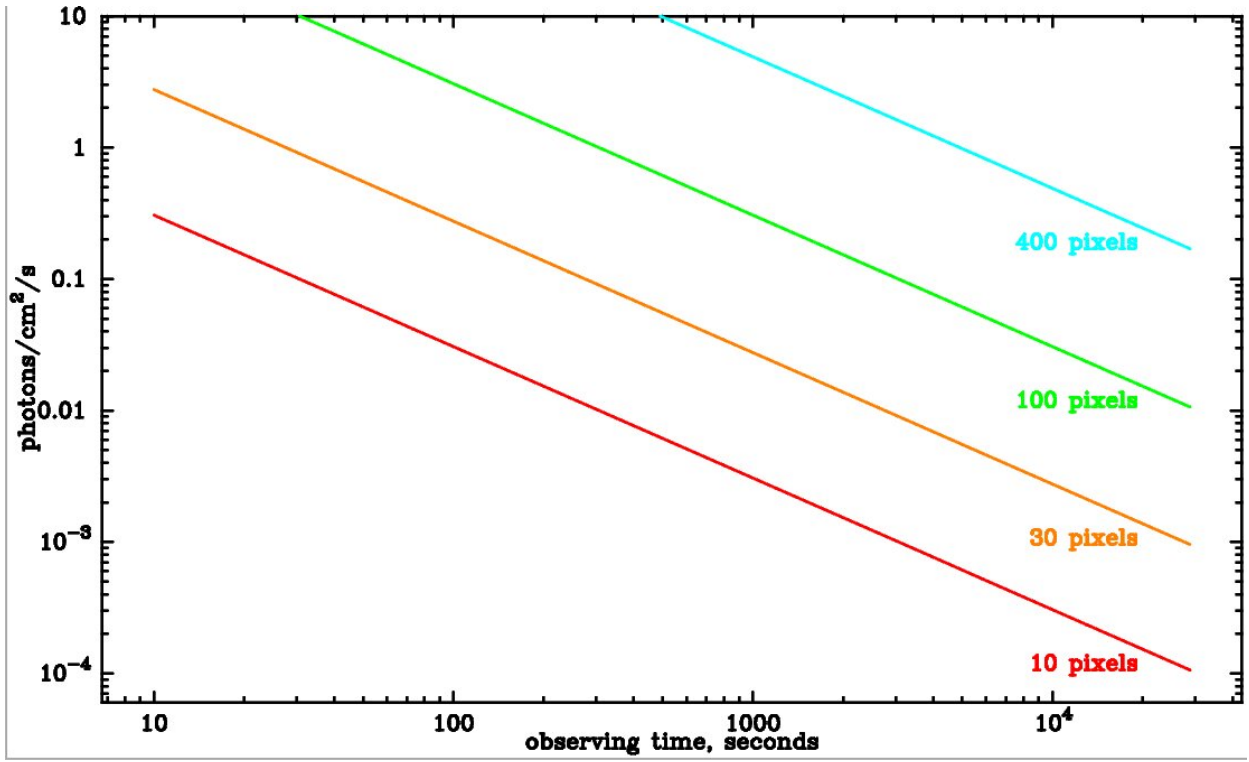

**Figure 3.18** *AeSI* sensitivity for UV line imaging assuming the flux is distributed uniformly over the number of pixels with a SNR=5 in those pixels.

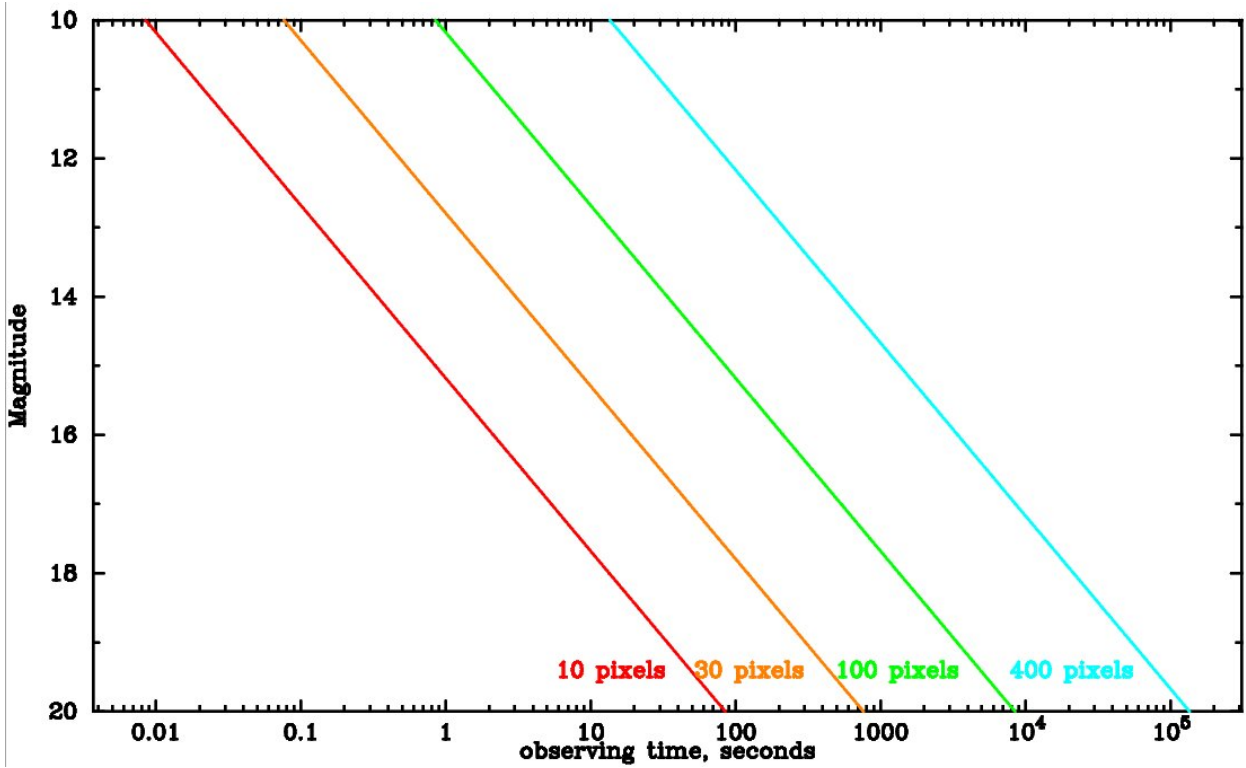

**Figure 3.19** *AeSI* sensitivity for broad-band Optical light imaging. This assumes the flux is evenly divided between the specified number of pixels with an SNR=5 in those pixels.

### 3.4.6 Scattered Light Control

The baseline *AeSI* design is to observe during the lunar day. This allows us to use solar panels for power without the need for extravagantly large batteries. But for daytime observations to be possible, we must control scattered sunlight. A detailed scattering model is a high priority for a future Phase II study. For now, we argue that scattered light is not a significant problem by presenting a simple, order of magnitude estimate. To avoid a direct path for sunlight to reach the detector, we have to avoid pointing at or too near to the sun. This may seem obvious but is a bit of a challenge near the South Lunar Pole because the sun is always near the horizon and therefore also near a potential light path between the cart and the hub. This orientation needs to be avoided when placing the telescopes for an observation, which will result in a small negative effect on *AeSI*'s u,v coverage. A single scattering of sunlight onto the detector can be avoided due to the instrument's very small field of view. The area at the cart focused onto the field stop is on the order of $2.44 \lambda L/D$ = 30 mm for a cart at a distance L=500 meters and a beam diameter of D=35 mm at a wavelength $\lambda$=0.8 µm. With a 50 mm diameter mirror, the detector can see only the sky and the upstream mirror surfaces.

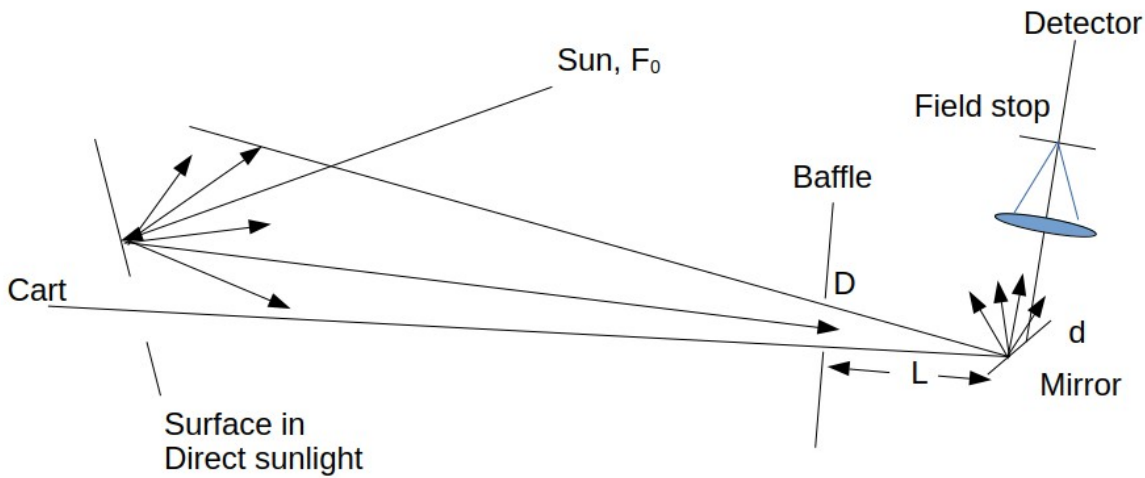

**Figure 3.20** Scattering geometry. Two scatterings are needed for sun light to reach the detector. First by a surface that can be seen by a mirror, then by the surface of that mirror into the field of view.

The first significant term requires sunlight to be scattered twice to reach the detector as shown in Figure 3.20. First sunlight illuminates a surface nearly in line with the cart. Some of the light scattered from this surface will pass through a baffle at the hub and reach mirror $H_1$. Light scattering off this mirror will be parallel to the starlight beam from the cart and pass through the field stop. The scattered flux reaching the detector can be approximated as:

$$F_S = F_0 \: \eta_S \: \Omega_B \left(\frac{\pi d^2}{4}\right) \eta_M \: \Omega_F \tag{6}$$

Here, $F_0$ is the solar flux incident on the surface, $\eta_S$ is the albedo of that surface, $\Omega_B = \pi(D/L)^2$ is the solid angle of the baffle, d is the diameter of the mirror, $\eta_M$ the fraction of light scattered by the mirror surface, and $\Omega_F = (\pi/4)(2.44\lambda/d)^2$ is the solid angle of the field stop.

The solar photon flux is shown in Figure 3.21. Scattering from the mirror is represented by:

$$\eta_M = \eta_0 + 1 - e^{-\left(\frac{2\pi r}{\lambda}\right)^2} \tag{7}$$

Where $\eta_0$ is a surface contamination term and r is the mirror surface microroughness. The adopted values are shown in Table 3.2. The result is shown in Figure 3.22.

A control bandpass of 450 to 900 nm will see approximately 150 scattered solar photons/second. This corresponds to a magnitude V=19.4 star, which is orders of magnitude lower than that needed for stellar surface imaging.

**Table 3.2** Values for equations 6 and 7

| | | |
|---|---|---|
| $\eta_S$ | 0.15 | surface emissivity |
| D | 0.05 m | Baffle diameter = diameter of mirror |
| L | 1.0 m | Baffle length |
| d | 35 mm | Diameter of beam |
| $\eta_0$ | 10-4 | Mirror contamination |
| r | 0.5 nm | Microroughness |
| $\lambda$ | 0.5 μm | Design wavelength for field stop |

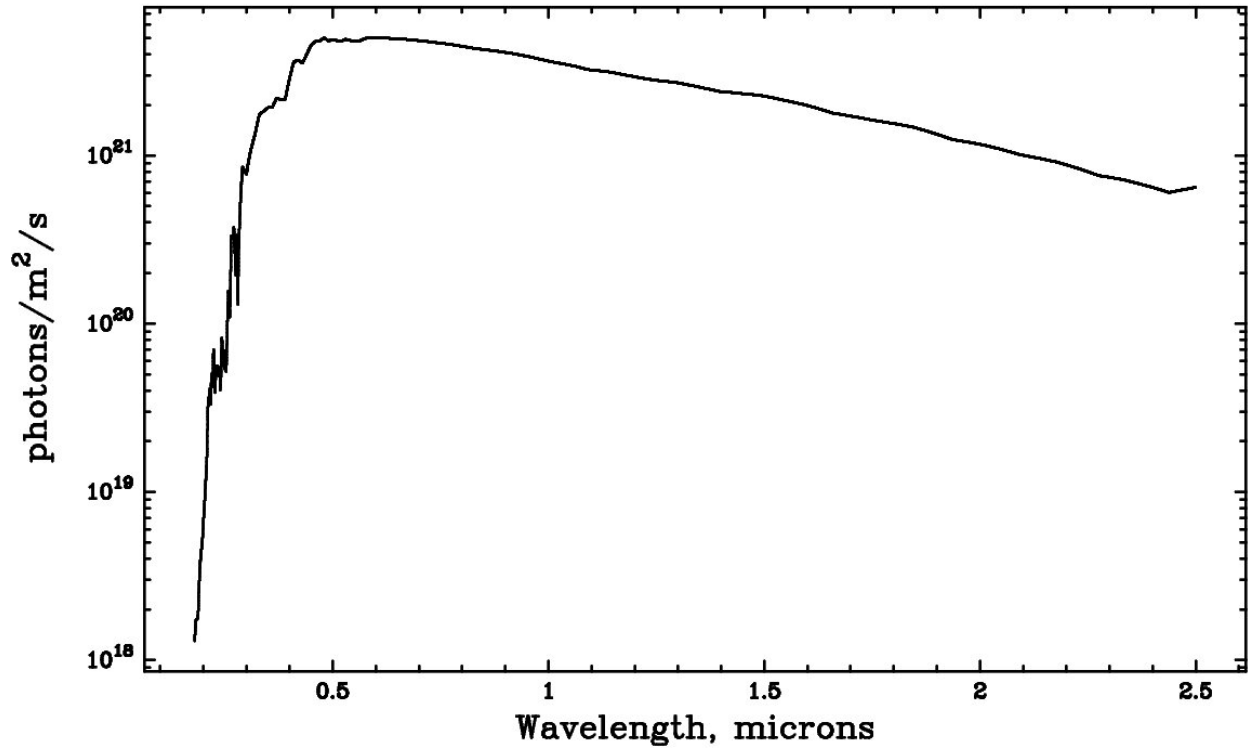

**Figure 3.21** Solar photon flux.

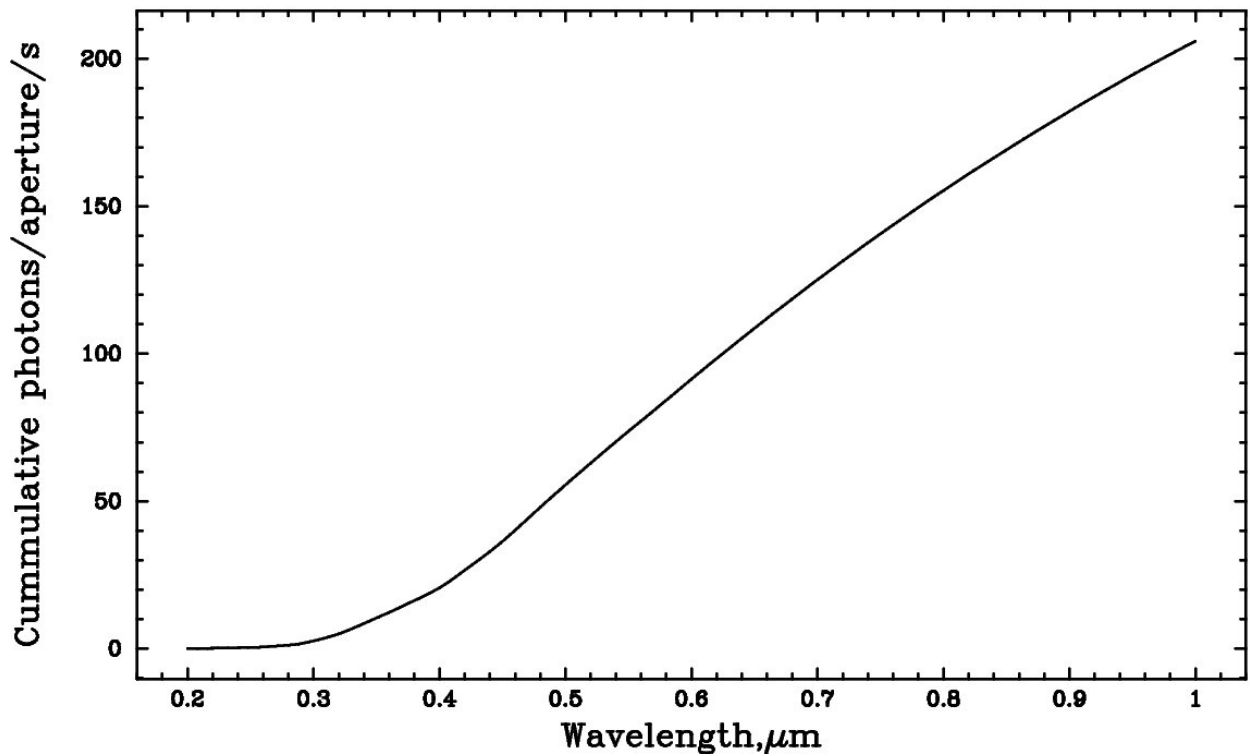

**Figure 3.22** Cumulative scattered sunlight reaching the detector.

### 3.4.7 Detectors for *AeSI*

The choice of detectors and the optimization of the beam combiner and observing modes are areas that need to receive significant work in a Phase II study. The detector proposed by the GSFC IDL for both the Optical and UV bands is the Teledyne CIS-115 detector. It has a 7 μm pixel pitch with a 2000x1504 pixel array and is well-matched to *AeSI*'s observing requirements with full frame rates as high as 7.5 frames/sec. The dark current is presently less than 1 e-/s at 210K which can be rendered negligible at lower temperatures. The quantum efficiency (QE) is 90% for wavelengths between 400 and 700 nm, dropping to 40% at 900 nm. The QE at shorter wavelengths is equally satisfactory.

The main concern with this detector technology is the $\sigma_R$=5e$^-$ read noise. The boundary between photon-noise dominated and read-noise dominated is at:

$$n_P \sigma_R^2 = \eta \frac{\pi d^2}{4} n_t t N \qquad (8)$$

where $n_P$ is the number of pixels, $\eta$ the system throughput, d, the diameter of the telescope, $n_T$, the number of telescopes, t is the length of the exposure and N the photon flux. This boundary is shown in Figure 3.23 for UV spectral line observations and in Figure 3.24 for the Optical continuum (4500 - 9000 A). We need read noise at the 1 to 2 electron level, which should be possible with an electron multiplying charge coupled device (EMCCD) or a good complementary metal-oxide semiconductor (CMOS) detector.

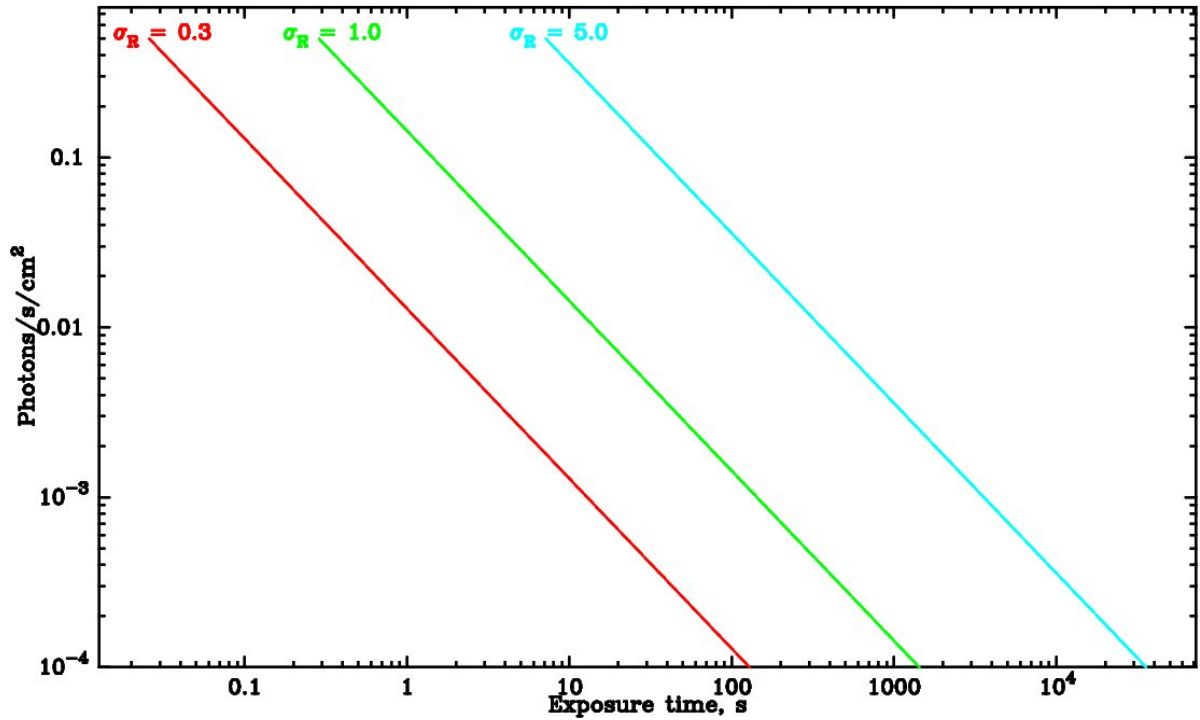

**Figure 3.23** The effect of read noise on UV line imaging. The lines show where the detector read noise, $\sigma_R$, equals photon noise. Since we assumed the system throughput is independent of wavelength through the UV, it is valid for any spectral line.

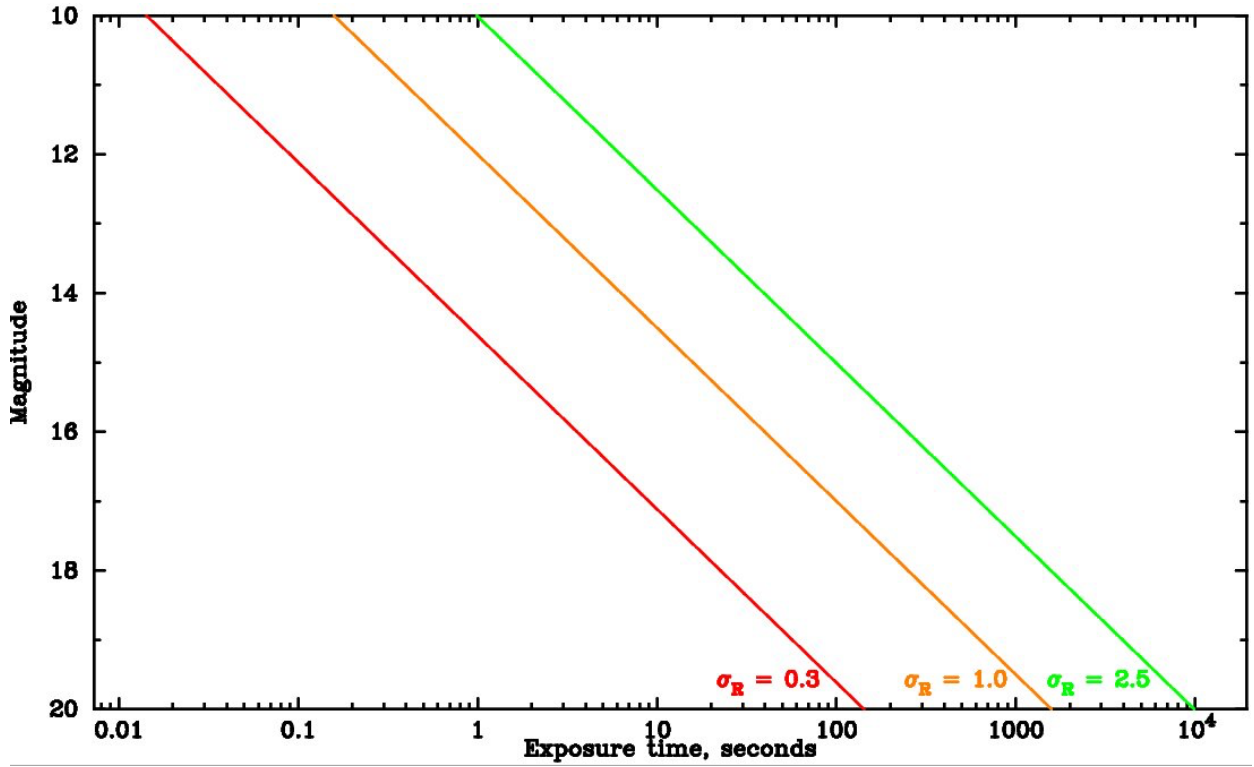

**Figure 3.24** The effect of read noise on sensitivity for continuum, Optical-light observations. The lines indicate the boundary where photon and read noise are equal.

### 3.4.8 Control While Observing

*AeSI*'s layout represents a large system with 11 optics in each of 15 beams before the light enters the beam combiner and wavefront sensors. There are 4 optics in continuous motion in each beam while observing which all need control signals and actuators. The rest of the optics can be considered quasi-static, only two tip/tilt mirrors to correct slow drifts are required.

The telescope has two degrees of freedom for tracking the target across the sky as the Moon rotates. The error signal for this will be a guide camera attached to the optical telescope assembly. Feedback is directly to the telescope axes, so no additional mechanisms are needed.

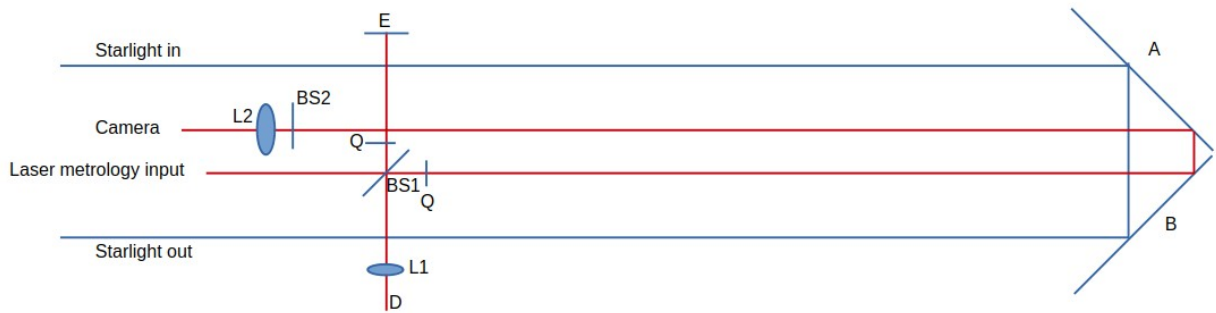

**Figure 3.25** Delay line concept. A dihedral (mirrors A and B) is used to offset the output starlight beam from the input (shown in blue). It translates on a stage parallel to the input beam to compensate for the changing delay which is measured with a heterodyne metrology system (red). The reference polarization is reflected by a beam splitter, BS, reflects from mirror E and is detected at D. The other polarization follows the starlight beam through the delay line, reflects from the second beam splitter BS and mixes with the reference at the detector D. This second beam splitter siphons off some of the metrology light for a tip/tilt sensor consisting of a lens and a camera. The rotation of the dihedral about its apex does not need to be controlled but a tilt into or out of the figure does.

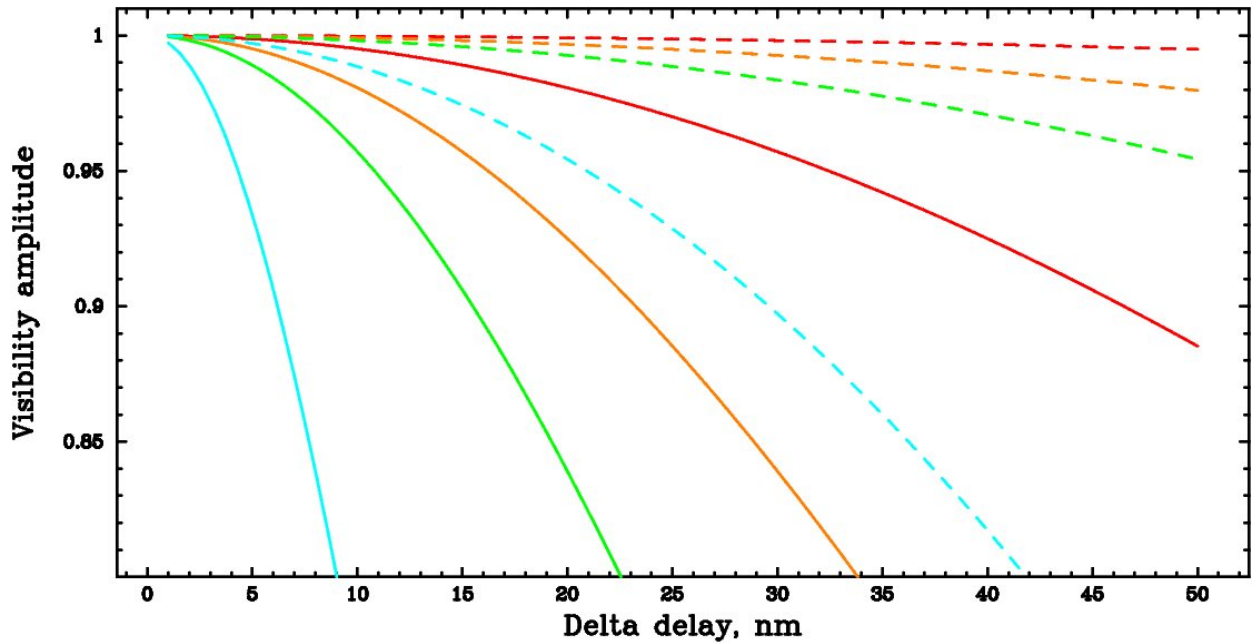

**Figure 3.26** Fringe amplitude reduction as a function of delay error. The dashed lines are for a linear drift through the exposure while the solid lines are for Gaussian variations (jitter). From red to blue, the lines represent 900, 450, 300, and 120 nanometer wavelengths.

During an hour-long observation, the delay can change by 2 m which can be corrected with a delay line having 1 m of physical travel (double pass) as in Figure 3.25. The delay line needs to track the fringe smoothly and at the correct rate. Delay errors on time scales slower than the detector exposure time can be corrected with post-processing software. Delay errors during the exposure smear the fringe reducing the fringe amplitude. These are shown in Figure 3.26. A delay error of 4 nm reduces the measured visibility amplitude and hence SNR by 5%. A delay drift of 20 nm during an exposure results in a similar reduction. Delay lines with this level of performance are in use on the Earth today.

A typical delay line design is shown in Figure 3.25. The dihedral optic is mounted on a flexure stage mounted on a carriage which rolls on precision rails. A motor drives the carriage to achieve the required delay rate while the delay is monitored with a laser metrology system. Carriage errors are corrected at the flexure stage and the dihedral tilt is controlled. Single axis control is sufficient with the error signal derived by focusing some of the laser metrology light onto a camera.

The type of metrology will depend on which systems are space-qualified first. A heterodyne metrology system is shown in Figure 3.25. The laser is preconditioned to give its orthogonal linear polarizations different frequencies. One of the frequencies (the reference) is reflected off a polarizing beam splitter (BS1), and a mirror, E, to a detector. The second polarization makes a double-pass through the delay line by being reflected off the second beam splitter BS2. The return beam is deflected into the detector at D. Quarter wave plates, Q, prevent crosstalk between the polarizations.

The phase of the beat frequency between the reference and unknown is proportional to the delay modulo a half wavelength. Some light is transmitted to the camera and is used to monitor the angle of the beam by monitoring the position of its image on a camera. Mirrors A and B are on a cart. If the mirrors rotate while moving, the angle of the output beam changes. The change is detected by the camera and corrected with appropriate actuators on the cart.

Finally, we need to correct for slow drifts in the angle and position of the beams. These are measured by the wavefront sensor and can be corrected with tip/tilt actuators on any two mirrors. The best choice is probably M3 and H1 because these mirrors must be aligned every time the telescope cart is moved.

### 3.4.9 Realignment after Reconfiguring the Array

There are three sets of alignments that need to be implemented once *AeSI* is deployed. They are: stabilization of a cart after a move, alignment of the cart to the hub, and fiducial position of the cart relative to the hub to estimate path lengths.

The cart leveling and stabilization will be accomplished with retractable feet (for moves). The tolerance for leveling on the order of a degree and can be accomplished with a bubble-level. The orientation of the cart is even looser since the instrument can be rotated on the cart.

The alignment of M7 with H1 can be achieved with a camera-beacon system. First, slide a camera into the beam before M7 and a beacon at H1 then adjust M7 to center the image of the beacon. Next, reverse the arrangement placing a camera in the beam after H1 and a beacon at M7 using the image position to adjust H1.

Finally, to determine the relative positions of the cart and hub we can use a time-of-flight laser ranging system to determine the distance between the cart and hub. For the relative azimuth and elevation, a camera-target system can be employed. A beacon on the cart can be imaged with a small telescope at the hub. An accuracy of 1 microradian is a translation error of 1 mm at 1 kilometer. A 5-inch diameter telescope has an airy pattern diameter of 10 microradians at 500 nm wavelength. Centering the image to one tenth of its size is straightforward. With the approximate baseline, searching for the correct delay is accomplished in less than a minute. Making measurements before and after a move eliminates long-term stability issues. Once the fringes are measured, higher-accuracy positions can be determined from the delay rate so performance should not degrade even after several moves. There will also be shutters and multiple dust covers for each beam train.

### 3.4.10 Thermal Control for *AeSI*

Driving Requirements and Assumptions

The thermal model is strongly affected by *AeSI*'s location at the Lunar South Pole, which minimizes variation in solar loading in sunlit conditions and makes the zenith an ideal direction for radiator pointing. *AeSI* must be capable of (a) operating in sunlight while meeting all operational temperature requirements, and (b) surviving for up to 15 days during lunar night. While there is a desire to operate in darkness, this is held for future Phase II study and not addressed in this Baseline design. Table 3.3 indicates how the requirements drive thermal subsystem design features.

**Table 3.3**

| Requirement | Drives |
| --- | --- |
| **Meet operational temperature requirements in sunlight** | Minimum radiator size; make-up heater power for intermittently operating components (e.g., laser range finder, comm system, batteries). |
| **Survive in darkness for up to 15 days** | Survival heater power sizing; battery capacity. |
| **Battery minimum allowable discharge temperature of -20 °C** | Limits maximum battery discharge capacity to 80% (i.e., depth of discharge is calculated against 80% of the total battery capacity). Also drives overall battery sizing. |

While operating, all electronics are on, the laser range finder is assumed to be active (which drives radiator sizing), and the batteries are charging. In survival (night time) mode, all the science components are off, power requirements for communication and control electronics are minimal, and the batteries are in discharge mode.

Table 3.4 lists the assumptions made when modeling the *AeSI* thermal subsystem during this IDL study. It is expected that future study phases will revisit these assumptions. For example, greater power margins may be required, and we expect to impose temperature stability requirements on some components, and finally mechanisms may require thermal control. High thermal-conductivity material is chosen for large radiators to ameliorate the need for heat pipes or other devices to distribute the heat load.

Operational and survival power and temperature requirements are detailed in Tables 3.5(a) and (b) at the hub and carts respectively. The hub electronics box has a 10% power mode, which is used in Survival mode. The hub battery discharges when the solar arrays are not illuminated.

**Table 3.4.** Assumptions invoked when modeling the *AeSI* thermal subsystem

| Thermal Subsystem Assumptions |
|---|
| There is no solar loading on the radiators. |
| Reduction in radiator efficiency to account for lunar dust and other inefficiencies. |
| Most radiators are 100-mil Al 1100-F, which was chosen for its relatively high thermal conductivity. Annealed Pyrolytic Graphite (APG), a sealed aluminum component with embedded graphite sheets to improve in-plane thermal conductivity, is used for high-power units, such as the Hub and Cart electronics boxes. |
| 25% power margin is applied to component dissipations. (This is less than the recommended pre-Phase A growth allowance per GSFC Gold Rules.) |
| 30% of the Hub Ka-band transmit power is modeled as waste heat, and it is captured in the Ka-band Comm electronics box. |
| +/-5 °C temperature margins are applied to the temperature limits per Goddard Gold Rules. |
| The batteries dissipate ~ 5% of the total power load. |
| Thermostat deadband of +/-3 °C. |
| *AeSI* experiences no variation of optical properties due to coating degradation. |
| Transmitting transients are not modeled. The thermal system is sized to accommodate steady-state power for transmission. This is reasonable for transmission lasting >2 hrs. |
| No stability or gradient requirements are levied on any components. |
| No thermal control is modeled for mechanisms or optical components. |

**Table 3.5(a)** Hub power and temperature requirements

| COMPONENT | OPERATIONAL POWER (W), NO MARGIN | SURVIVAL POWER (W), NO MARGIN | OPERATIONAL TEMPERATURE (C) | SURVIVAL TEMPERATURE (C) |
|---|---|---|---|---|
| Fringe Camera / Angle Camera / Pupil Camera | 23 | 0 | -10 to +40 | -55 |
| Wavefront Sensing and Control System | 16.5 | 0 | -10 to +40 | -55 |
| UV Fringe Camera | 15 | 0 | -10 to +40 | -55 |
| Laser Ranging System | 25 | 0 | -10 to +40 | -55 |
| Hub Electronics Box | 237.2 | 23.72 | -10 to +40 | -55 |
| Hub Comm Electronics Ka-Band* | 60 | 0 | -10 to +40 | -40 |
| Hub Comm S-Band 1 Transmit | 32 | 0 | -10 to +40 | -40 |
| Hub Comm S-Band 2 Transmit | 32 | 0 | -10 to +40 | -40 |
| Hub Comm S-Band 1 Receiver | 8 | 8 | -10 to +40 | -40 |
| Hub Comm S-Band 2 Receiver | 8 | 8 | -10 to +40 | -40 |
| Hub Battery | 0 | 17.5 | -20 to +20 | -20 |
| Hub Star Tracker CH 1 | 1.4 | 0 | -30 to +55 | -50 |
| Hub Star Tracker CH2 | 1.4 | 0 | -30 to +55 | -50 |
| Hub Star Tracker DPU | 8.78 | 0 | -30 to +55 | -45 |

* 5% of 350 W

**Table 3.5(b)** Cart power and temperature Requirements

| COMPONENT | OPERATIONAL POWER (W), NO MARGIN | SURVIVAL POWER (W), NO MARGIN | OPERATIONAL TEMPERATURE (C) | SURVIVAL TEMPERATURE (C) |
|---|---|---|---|---|
| Cart Electronics Box | 140.9 | 14.09 | -10 to +40 | -55 |
| Cart Comm Electronics S-Band Transmitter | 5 | 0 | -10 to +40 | -40 |
| Cart Comm Electronics S-Band Receiver | 8 | 8 | -10 to +40 | -40 |
| Cart Battery | 0 | 5.5 | -20 to +20 | -15 |

Thermal Design Features and Model Predictions

A passive thermal design was developed for *AeSI* with zenith-facing radiators to dissipate power from the components listed in Tables 3.5 (a) and (b) and notional mechanical accommodation on a hub bench or cart structure, neither of which is represented in detail. Future modeling can address cases with power turn-down in various subsystems.

The modeled components of *AeSI* and their respective salient characteristics are shown in Figures 3.27 (hub) and 3.28 (cart). Most of the radiators are modeled as separate components with no cross-coupling, but in a few cases components with common temperature ranges (e.g., the hub cameras and wavefront sensing and control system, shown as "camera suite" and depicted in yellow, pink and blue in Fig. 3.27) were accommodated together to explore possible radiator area and power savings. The radiators are sized to hot power rejection needs with thermostatically controlled heaters to maintain temperatures above the survival limits shown in Tables 3.5(a) and (b). The hub and cart electronic boxes are power-hungry and require the largest radiators. In future work, we will model the effects of shallow angle illumination of the radiators and take steps to mitigate the excess environmental backloading, allowing for the fact that the carts will tip and tilt to some extent in response to the lunar topography.

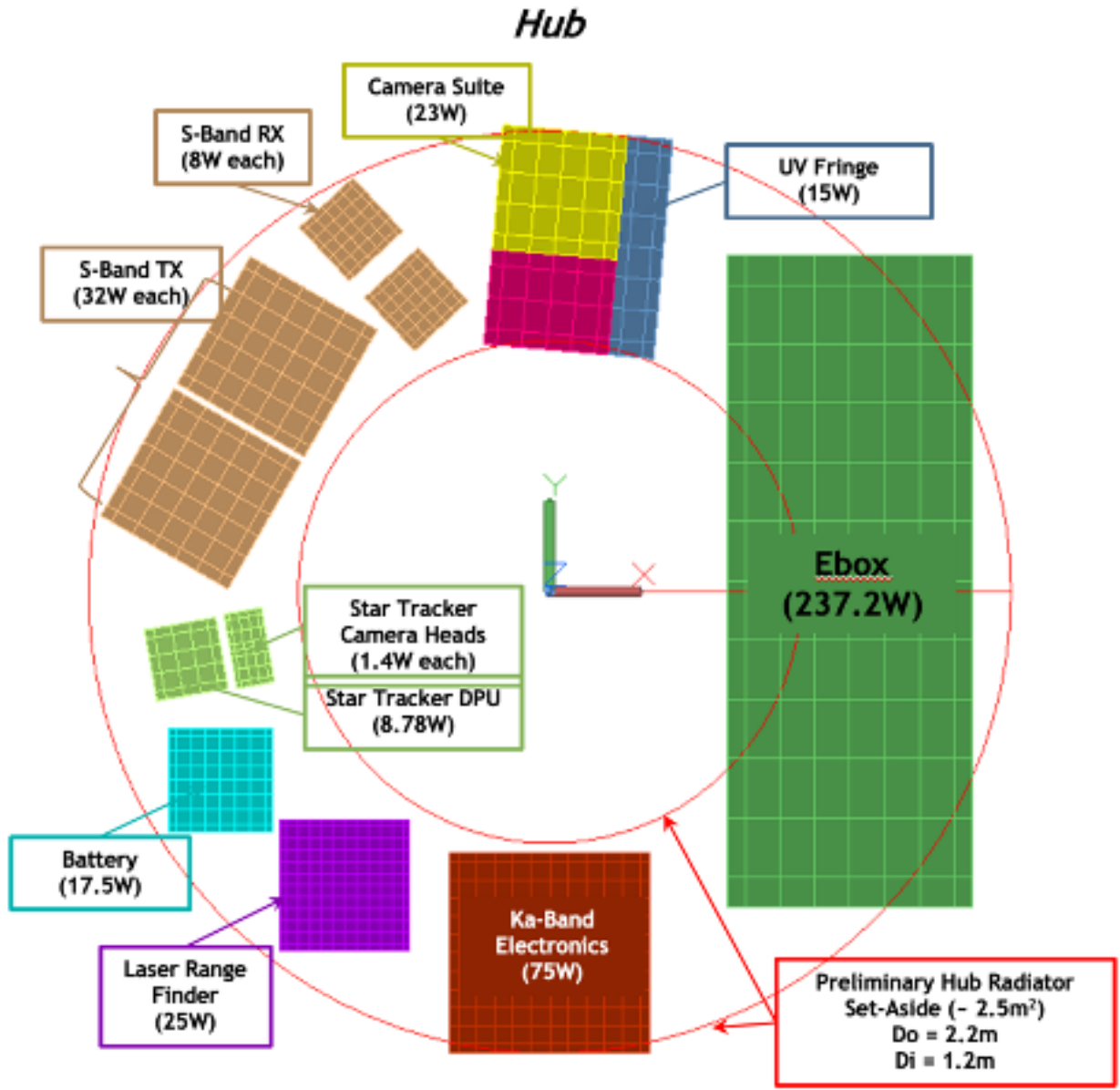

**Figure 3.27** Sizes and notional placement of the Hub radiators within an annulus reserved for this purpose in the *AeSI* mechanical model. Dimensions of the annulus are shown. The area was arbitrarily chosen to be annular. The electronics box radiator could be shaped to fit within the annulus, if desired.

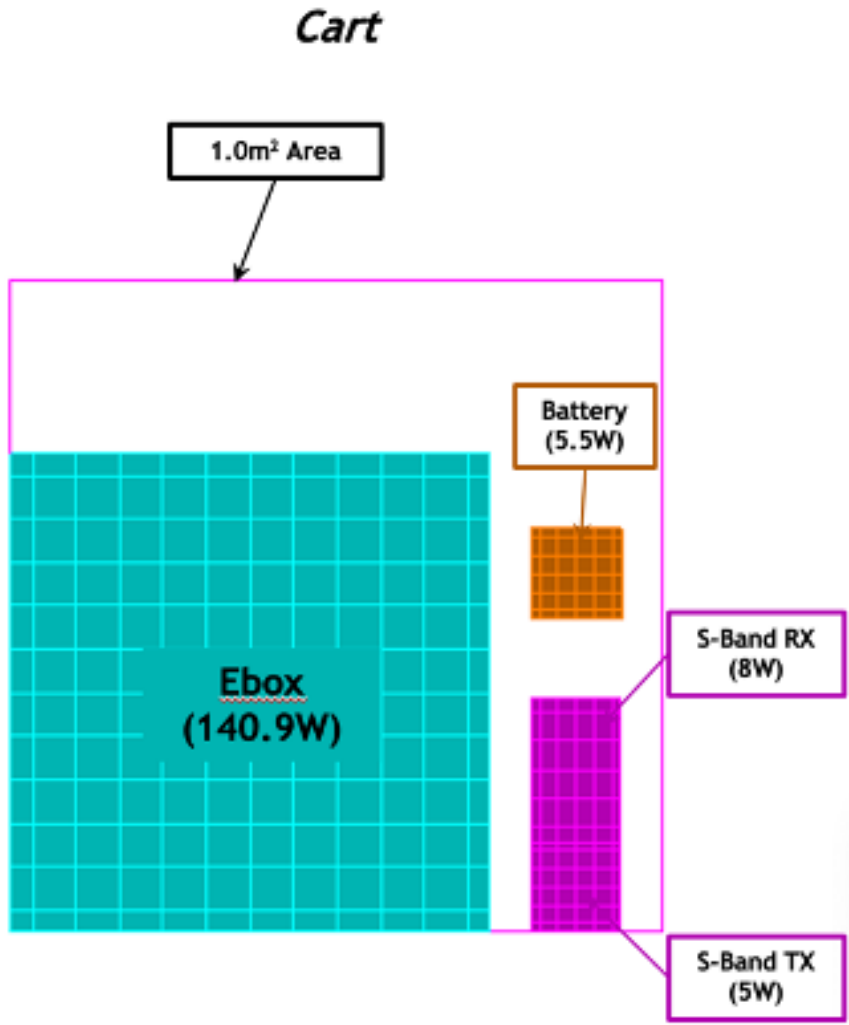

**Figure 3.28** Sizes and notional placement of the Cart radiators within a 1 m2 area reserved for this purpose in the *AeSI* mechanical model.

Tables 3.6 and 3.7 detail the *AeSI* model temperature and heater power results, respectively. Temperature limits, power dissipations, heater power consumption, and heater duty cycle are included in the appropriate tables on a per-unit basis. In this Phase I study, we analyzed two configurations: Operational Mode and Survival Mode. Both cases require all temperature limits to be maintained with adequate margin and any heaters (operational or survival) must remain at or below the 70% duty cycle requirement. For the Baseline design, only one unit shows a temperature limit violation, the Hub Star Tracker Data Processing Unit (DPU; highlighted in red in Table 3.6). This issue can be addressed in future work with attentiveness to radiator gradients, adjustment to radiator area, and adjustments to the survival heater thermostat deadband.

**Table 3.6** *AeSI* temperatures relative to their adopted limits.

| Component | Limits (°C) | | | | *AeSI* Operational Predictions (°C) | | | *AeSI* Survival Predictions (°C) | | |
|---|---|---|---|---|---|---|---|---|---|---|
| | Surv Low | Op Low | Op High | Surv High | Min | Avg | Max | Min | Avg | Max |
| HUB_FRINGE_CAM | -55 | -10 | 40 | 50 | 5.8 | 5.8 | 5.8 | -50.2 | -48.2 | -46.9 |
| HUB_WAVE | -55 | -10 | 40 | 50 | 5.9 | 5.9 | 5.9 | -50.4 | -48.2 | -47.0 |
| HUB_UV_FRINGE_CAM | -55 | -10 | 40 | 50 | 5.9 | 5.9 | 5.9 | -50.0 | -48.1 | -47.0 |
| HUB_LR | -55 | -10 | 40 | 50 | 17.9 | 17.9 | 17.9 | -50.2 | -48.4 | -46.4 |
| HUB_EBOX | -55 | -10 | 40 | 50 | 15.9 | 15.9 | 15.9 | -50.3 | -48.7 | -47.1 |
| HUB_KATWTA | -55 | -10 | 40 | 50 | 17.2 | 17.2 | 17.2 | -35.3 | -33.5 | -31.6 |
| HUB_SBAND_1_TX | -40 | -10 | 40 | 50 | 3.2 | 3.2 | 3.2 | -35.3 | -33.5 | -31.9 |
| HUB_SBAND_2_TX | -40 | -10 | 40 | 50 | 3.2 | 3.2 | 3.2 | -35.3 | -33.5 | -31.9 |
| HUB_SBAND_1_RX | -40 | -10 | 40 | 50 | 3.7 | 3.7 | 3.7 | 3.7 | 3.7 | 3.7 |
| HUB_SBAND_2_RX | -40 | -10 | 40 | 50 | 3.7 | 3.7 | 3.7 | 3.7 | 3.7 | 3.7 |
| HUB_BATTERY | -20 | -20 | 30 | 50 | -15.6 | -13.5 | -11.6 | 9.3 | 9.3 | 9.3 |
| HUB_ST_CH_1 | -55 | -30 | 55 | 65 | -7.7 | -7.7 | -7.7 | -45.1 | -43.5 | -42.0 |
| HUB_ST_CH_2 | -55 | -30 | 55 | 65 | -7.7 | -7.7 | -7.7 | -45.1 | -43.5 | -42.0 |
| HUB_ST_DPU | -45 | -30 | 55 | 65 | 26.8 | 26.8 | 26.8 | **-45.3** | -43.5 | -41.9 |
| CART_EBOX | -55 | -10 | 40 | 50 | 17.5 | 17.5 | 17.5 | -50.1 | -48.5 | -47.0 |
| CART_SBAND_TX | -40 | -10 | 40 | 50 | 17.8 | 17.8 | 17.8 | -16.6 | -16.6 | -16.6 |
| CART_SBAND_RX | -40 | -10 | 40 | 50 | 18.3 | 18.3 | 18.3 | -9.7 | -9.7 | -9.7 |
| CART_BATTERY | -20 | -20 | 30 | 50 | -15.7 | -13.2 | -11.5 | 9.5 | 9.5 | 9.5 |

Table 3.7 shows that the *AeSI* heater **power consumption** predictions are 13W Operational (10W hub, 3W per cart) and 196W Survival (161W hub, 35W per cart). It is important to note that the power consumption predictions are raw temperature predictions and are not representative of heater sizing, nor do they include additional margin. In practice, the heaters will be sized to allow for margin and for the expected duty cycles. None of the heaters saturate with operation at 100% duty cycle. Although several of the heaters are predicted to have duty cycles >70%, this is not concerning at an early design stage, as the radiators and heaters have not yet been optimized.

**Table 3.7** *AeSI* heater power consumption relative to allocated power per unit.

| Component | Available Power (W) | | Heater Power Consumption | | | |
|---|---|---|---|---|---|---|
| | | | *AeSI* Operational Mode | | *AeSI* Survival Mode | |
| | Operational | Survival | Power (W) | Duty Cycle (%) | Power (W) | Duty Cycle (%) |
| Hub Battery Power Dissipation | 0 | 17.5 | 10.01 | **71.5%** | 0.00 | 0.0% |
| Hub E-box Power Dissipation | 296.5 | 29.65 | 0.00 | 0.0% | 58.30 | 50.7% |
| Hub Fringe Camera Power Dissipation | 18.75 | 0 | 0.00 | 0.0% | 8.49 | **70.7%** |
| Hub Ka, TWTA, Dissipation | 75 | 0 | 0.00 | 0.0% | 31.32 | 46.7% |
| Hub Laser Range Finder Power | 31.25 | 0 | 0.00 | 0.0% | 14.20 | 19.5% |
| Hub S-Band Receiver #1 Power Dissipation | 8 | 8 | 0.00 | 0.0% | 0.00 | 0.0% |
| Hub S-Band Transmitter #1 Power Dissipation | 32 | 0 | 0.00 | 0.0% | 16.01 | **79.2%** |
| Hub S-Band Receiver #2 Power Dissipation | 8 | 8 | 0.00 | 0.0% | 0.00 | 0.0% |
| Hub S-Band Transmitter #2 Power Dissipation | 32 | 0 | 0.00 | 0.0% | 16.01 | **79.2%** |
| Hub Star Tracker Camera Head 1 | 1.75 | 0 | 0.00 | 0.0% | 0.86 | **78.6%** |
| Hub Star Tracker Camera Head 2 | 1.75 | 0 | 0.00 | 0.0% | 0.86 | **78.6%** |
| Hub Star Tracker DPU Power Dissipation | 10.975 | 0 | 0.00 | 0.0% | 3.14 | 69.8% |
| Hub UV Fringe Camera Power Dissipation | 18.75 | 0 | 0.00 | 0.0% | 5.62 | **77.0%** |
| Hub Wavefront Power Dissipation | 20.625 | 0 | 0.00 | 0.0% | 6.47 | **80.9%** |
| Cart Battery Power Dissipation | 0 | 5.5 | 3.05 | 67.7% | 0.00 | 0.0% |
| Cart EBox Power Dissipation | 176.125 | 17.6125 | 0.00 | 0.0% | 34.52 | **71.9%** |
| Cart S-Band Receiver Power Dissipation | 10 | 10 | 0.00 | 0.0% | 0.00 | 0.0% |
| Cart S-Band Transmitter Power Dissipation | 6.25 | 0 | 0.00 | 0.0% | 0.00 | 0.0% |
| *Hub Survival Power Consumption Total* | -- | -- | *10.01* | *N/A* | *161.28* | *N/A* |
| *Cart Survival Power Consumption Total* | -- | -- | *3.05* | *N/A* | *34.52* | *N/A* |

Steady-state radiator temperature maps in the Operational and Survival Modes are shown in Figures 3.29 and 3.30, respectively. Large temperature gradients arise in high power-density regions, but the heat load distribution is notional and not yet well known. The hub and cart electronics box radiators have been baselined with APG included to improve heat distribution. Any intolerable gradients, such as those seen above the avionics boxes, can be mitigated in future design work, when the locations of heat sources will be better known. If necessary, APG doublers and heat pipes can be added to the design (at the expense of mass and complexity) to improve the heat distribution and reduce the temperature gradients. Figure 3.30 shows that all radiators reach a steady-state temperature that satisfies the specified survival limits.

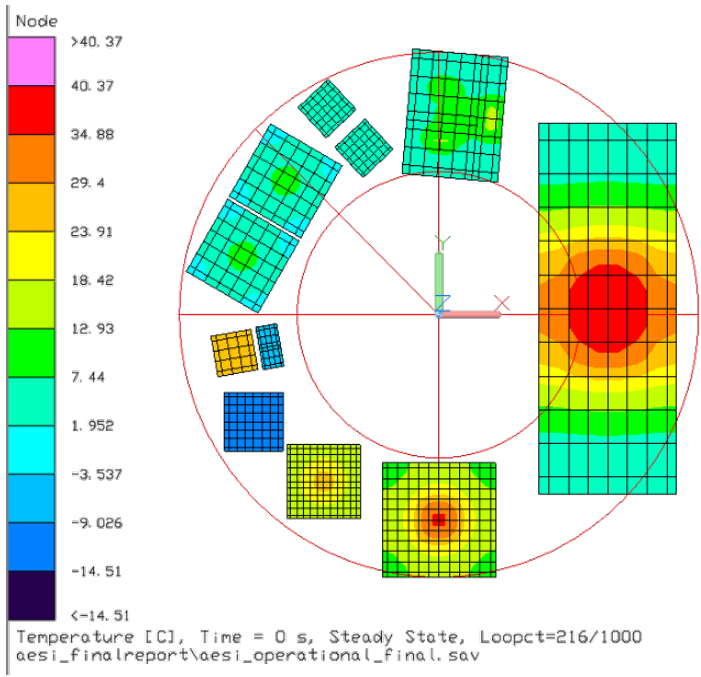

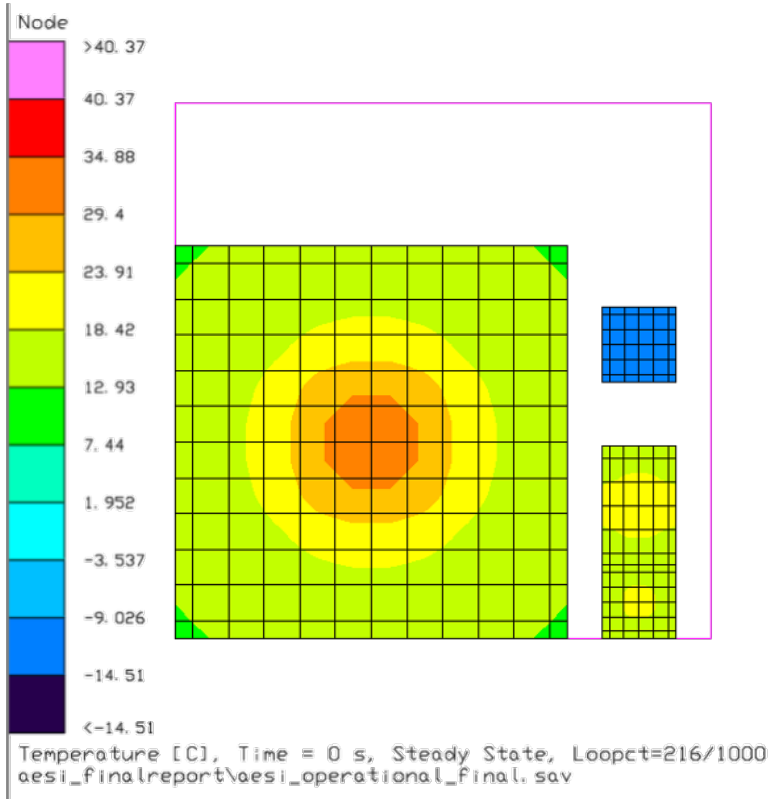

**Figure 3.29** *AeSI* Operational Mode steady-state radiator temperature map. (top) Hub and (bottom) Cart. The temperature scale is in degrees Celsius. See Figures 3.3.3-1 and 3.3.3-2 for labels indicating the components served by each of the radiators.

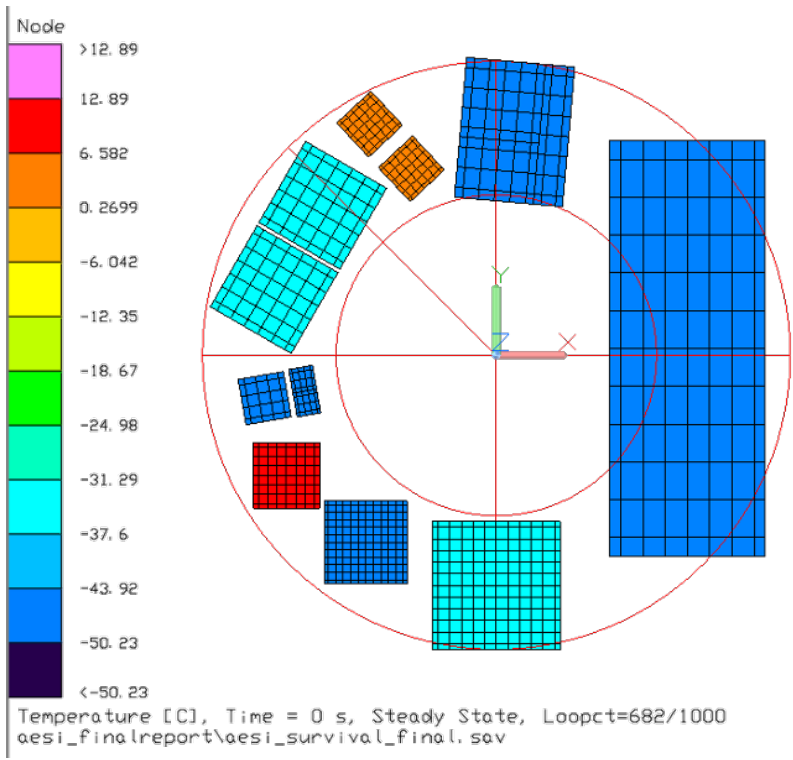
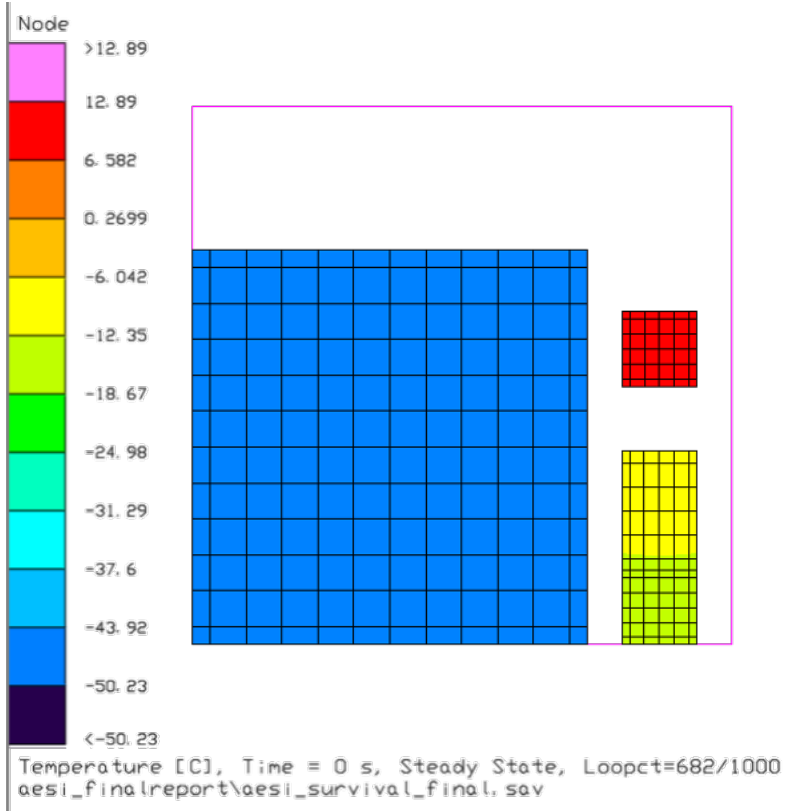

**Figure 3.30** *AeSI* Survival Mode steady-state radiator temperature map. (top) Hub and (bottom) Cart. The temperature scale is in degrees Celsius. See Figures 3.3.3-1 and 3.3.3-2 for labels indicating the components served by each of the radiators.

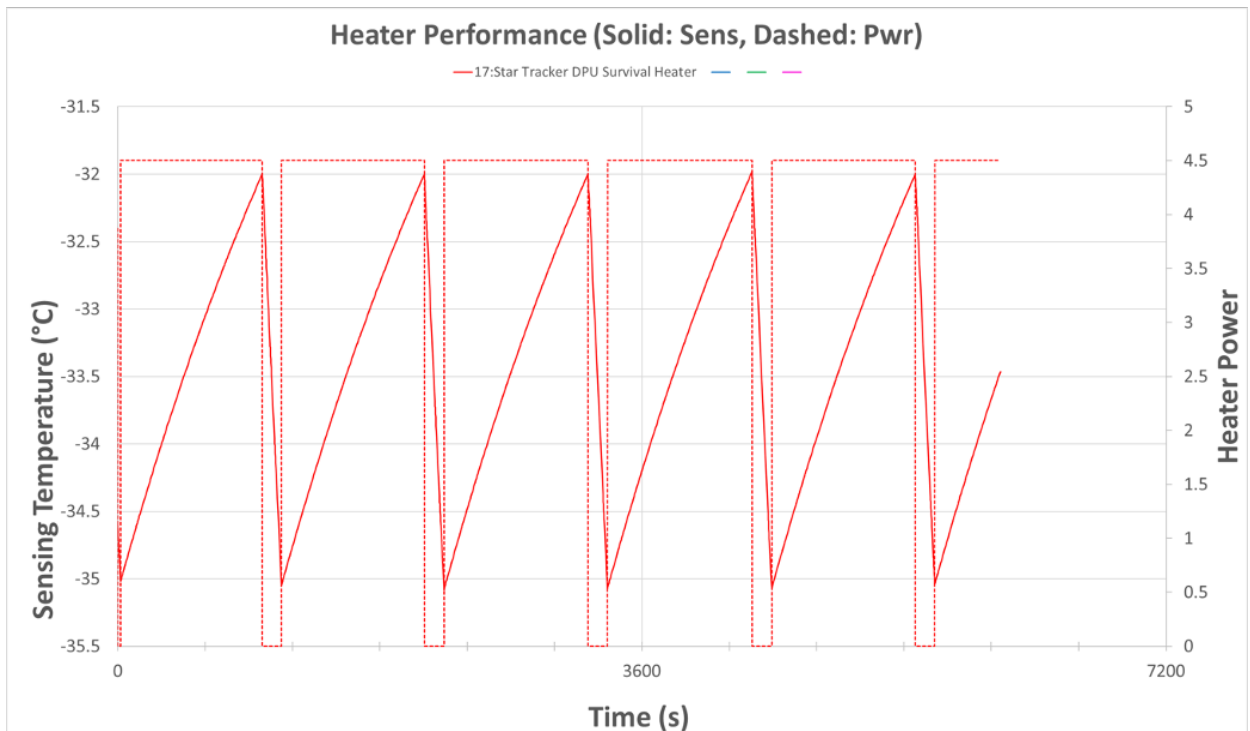

**Figure 3.31** Temperature (solid curve) and heater power (dotted curve) exhibit periodic behavior in Survival Mode, indicating positive heater control at the DPU Star Tracker.

Figure 3.31 shows the predicted Survival-Mode variation in temperature and power for the star tracker DPU, a component with a relatively high duty cycle. Heating and cooling oscillate regularly at ~18-minute intervals. This behavior indicates that the heater has positive control and it does not saturate. Temperature cycling continues throughout the 14-day Survival Mode. The star tracker DPU has a relatively small radiator, which is expected to operate more efficiently than assumed. If factored into future modeling, a higher radiator efficiency will reduce the duty cycle.

The thermal subsystem employs high Technology Readiness Level (TRL) hardware like that depicted in Figure 3.32. Items like those shown have flown successfully on many missions. With high conductance turndown ratios and radiators operating at colder temperatures, we can reduce the heater power requirements while enabling survival through the long lunar night. Such situations call for lower-TRL high-capacity heat switches, which we consider mission-enhancing (rather than mission-enabling) technology. Arrays of commercial off-the-shelf (COTS) thermal switches (~5 – 12 W maximum) can be employed, but less mature Passive Thermal Control Heat Switches (Figure 3.33), which make or break thermal contact in response to temperature changes, are designed to work for high turndown ratio. Examples include two-phase switches comprising a vapor and bellows, and Shape Memory Alloys, which change shape in response to temperature changes. Table 3.8 shows the potential heater power savings as a function of increasing turndown ratio at three representative (low-power/small area and high power/large area) hub locations. Figure 3.34 indicates that the potential power savings are substantial at the high-power components, such as electronics boxes. Note, however, that mass, complexity, or other challenges may outweigh the power savings.

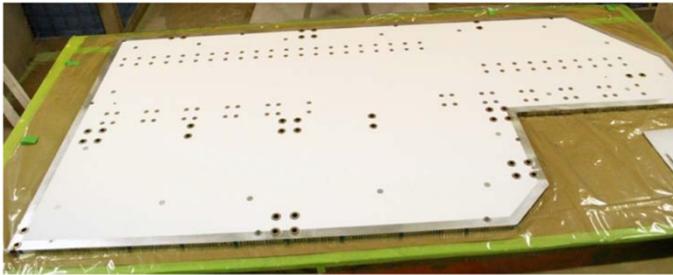
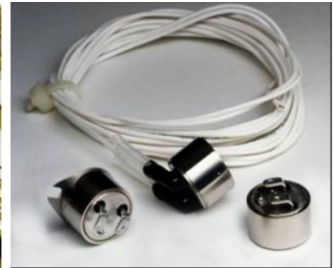
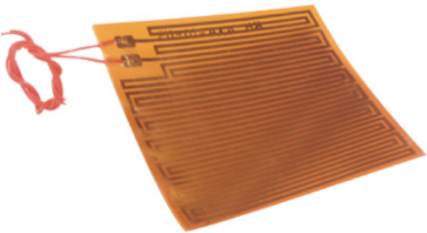
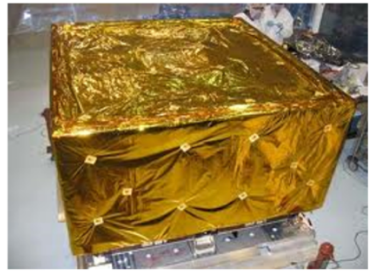
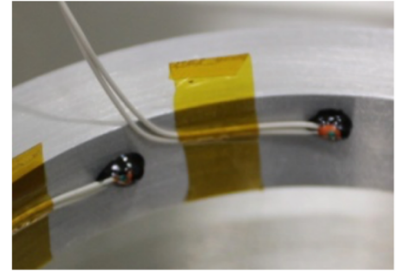

**Figure 3.32** *AeSI* will employ standard thermal subsystem hardware, including items like those shown here.

**Table 3.8** Potential heater power savings as a function of increasing conductance turndown ratio

| Component | Unit Power (W) | A (m²) | Heater Power Consumption (W) | Heater Power Size (w/ 70% DC) | Turndown Ratio |
|---|---|---|---|---|---|
| **Hub** **Fringe Camera / Angle Camera / Pupil Camera** | 28.75 | 0.088 | 9.6 | 13.7 | 2 to 1 |
| | | | 9.2 | 13.2 | 3 to 1 |
| | | | 8.6 | 12.3 | 5 to 1 |
| | | | 7.7 | 11.1 | 7 to 1 |
| | | | 7.4 | 10.5 | 10 to 1 |
| | | | 5.9 | 8.4 | 20 to 1 |
| **Hub** **Wavefront Sensing and Control System** | 20.625 | 0.063 | 7.5 | 10.8 | 2 to 1 |
| | | | 7.3 | 10.4 | 3 to 1 |
| | | | 6.9 | 9.8 | 5 to 1 |
| | | | 6.3 | 9.0 | 7 to 1 |
| | | | 6.1 | 8.7 | 10 to 1 |
| | | | 5.0 | 7.1 | 20 to 1 |
| **Hub E-Box** | 296.5 | 0.906 | 59.9 | 85.6 | 2 to 1 |
| | | | 50.4 | 72.0 | 3 to 1 |
| | | | 39.0 | 55.7 | 5 to 1 |
| | | | 31.6 | 45.1 | 7 to 1 |
| | | | 25.9 | 37.0 | 10 to 1 |
| | | | 16.1 | 23.0 | 20 to 1 |

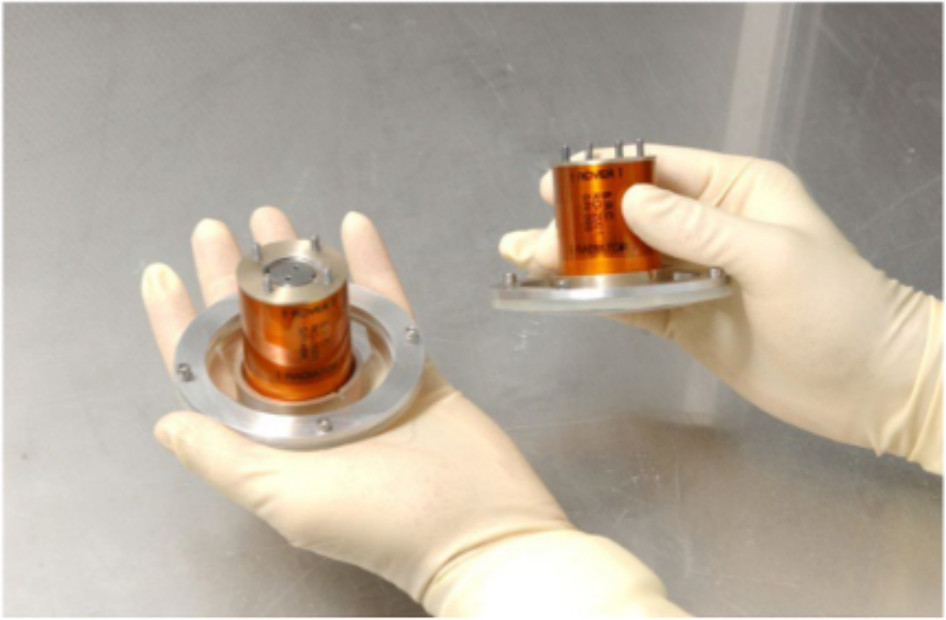

Passive Thermal Control Heat Switch. Shown here are heat switches with dust cover shields.

**Figure 3.33** Passive Thermal Control Heat Switches could reduce power demand where the turndown ratio is 50:1.

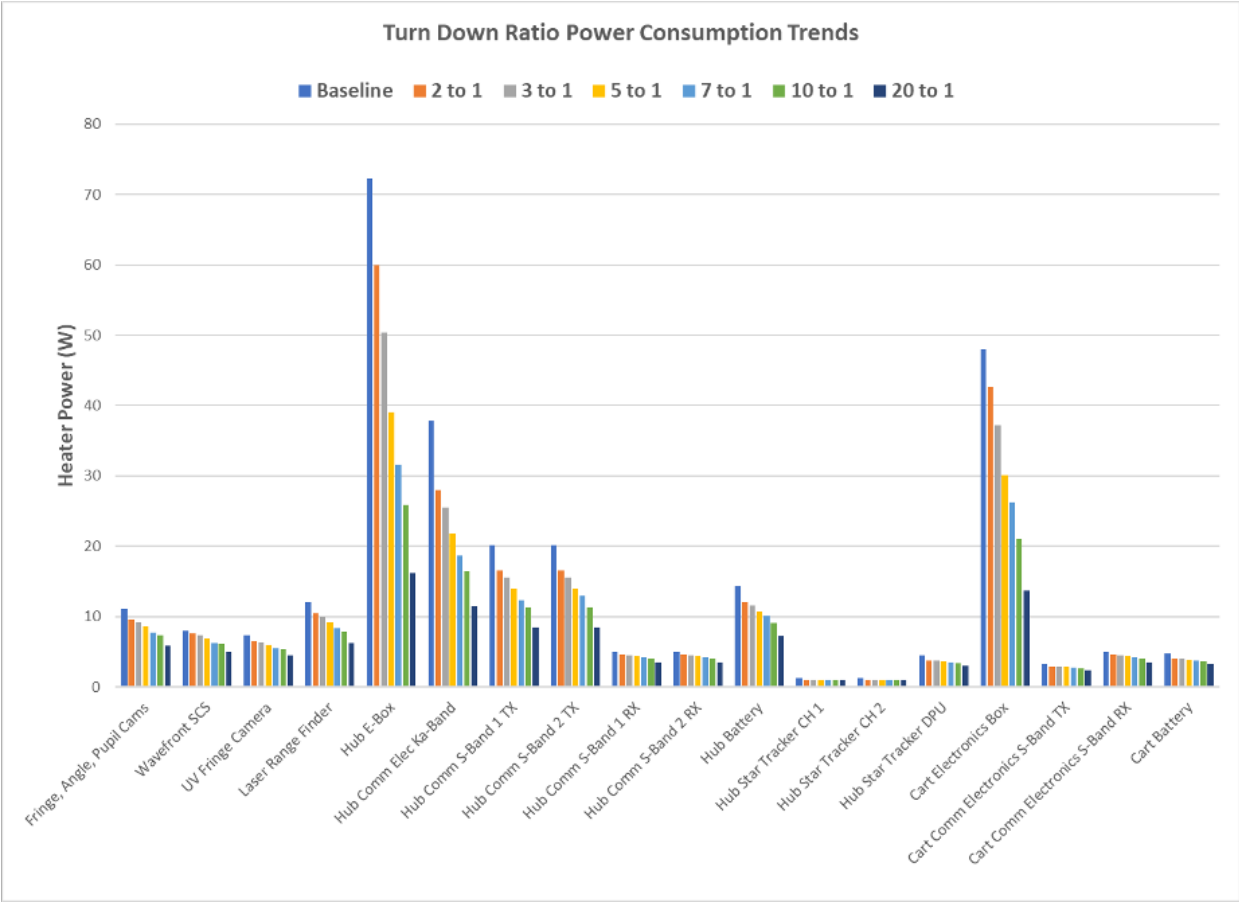

**Figure 3.34** Potential heater power savings as a function of increasing conductance turndown ratio.

## 3.4.11 Hub Mechanical Design

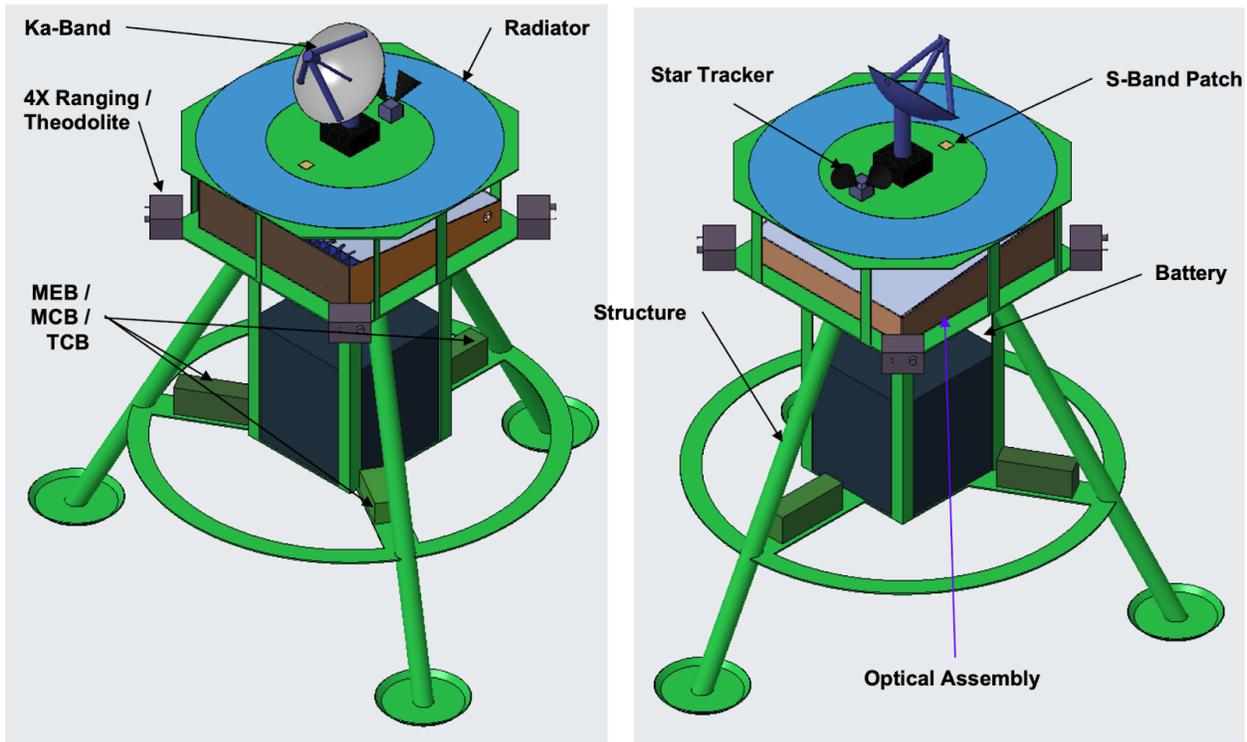

**Figure 3.35** CAD modeling of *AeSI* hub components.

The Baseline design of the hub for *AeSI* (Figure 35) requires several essential components and systems to ensure its functionality and protection. First, the hub must be protected from dust during both transport and potential post-installation dust events. To achieve this, deployable closeouts and mechanisms, not currently shown in the CAD models, will be incorporated. The hub must be located within a stellar reference system, utilizing a star tracker for precise positioning, as shown in the CAD.

To manage the movement of the hub, a system to measure primary mirror element (cart) positions and guide the carts' movements will be implemented. This system includes a Laser Ranging/Theodolite, also shown in the CAD, and is essential for precise alignment and positioning. The optical systems of the hub must include both UV and Optical interferometry capabilities, with an optical bench provided in the CAD. The input optical train must accommodate both UV (shown in the CAD) and Optical/wavefront sensing, although the latter is not depicted in the current CAD.

Internal baffling will be used as needed, as shown in the CAD, to manage beam position and pathlength. The laser ranging/theodolite system will again be used to ensure proper alignment between the cart and the hub. Cameras will be employed to monitor the system, and communication with the LCRNS (described below in Section 3.5.2) and carts will be facilitated via the appropriate communication systems. The hub will be equipped with a Ka-band dish and an S-band patch and dipole, as indicated in the CAD.

For power generation, solar panels and a battery system will be incorporated, as shown in Figure 36 without the support structure. The hub must also manage heat generation, with a 2.6m² deployable radiator and cover (shown in the CAD, with the cover not depicted) to dissipate heat to the environment when necessary and insulate from the environment at other times. Additionally, a seismometer will be included to detect moonquakes, though this is not currently shown in the CAD.

A rough initial mass estimate for the hub (that reflects no optimization and does not include all components) suggests a total around 2700 kg,and is broken down by component in Table 3.9. Future work to refine this estimate and trade studies to reduce the mass are outlined in Sections 4.4 – 4.6. Mass estimates for the primary mirror elements are excluded due to the Phase I study not studying rover options.

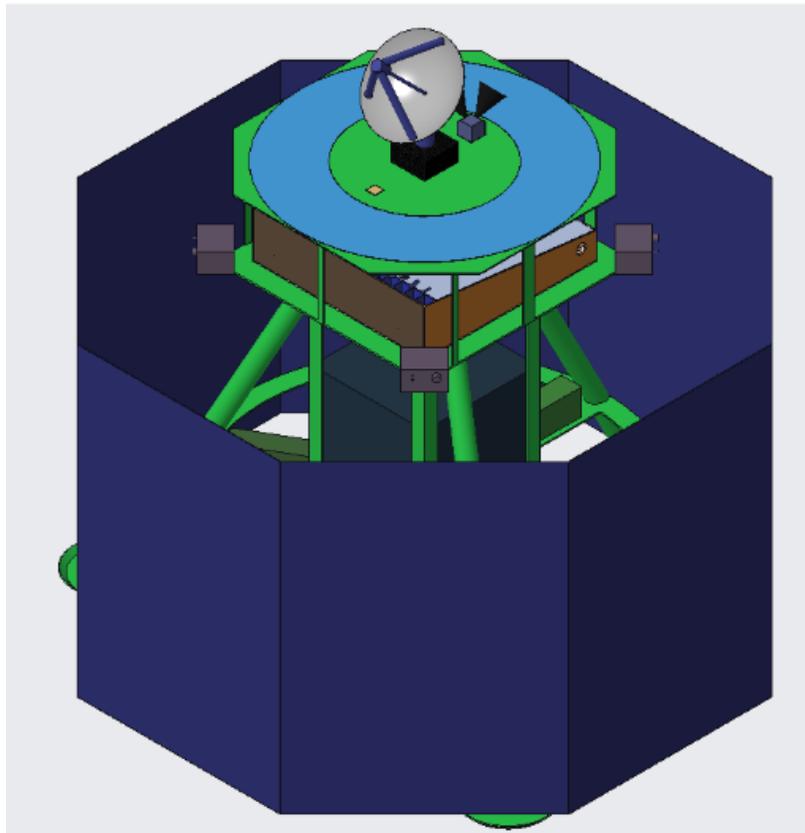

**Figure 3.36** CAD design showing the solar panel configuration (dark blue) for the Hub.

**Table 3.9** Rough mass estimates for hub components

| Description | Material | Mass (kg) |
| --- | --- | --- |
| Optics (total) | Fused Silica | 9 |
| Optical Bench | Al-6061 | 183 |
| Solar Arrays | Al-6061 structure | 422 |
| Battery | | 1060 |
| Structure | Al-6061 | 902 |
| Ranging Assy | | 5.6 |
| Ka-Antenna | | 30 |
| Star Tracker | | 1.3 |
| Radiator | Al-6061 | 22 |

### 3.4.12 Summary of Design Elements for Hub Carts

In Figures 3.37 and 3.38 we present, in a schematic sense, the design of the Hub and Carts produced by our Phase I study. It does not show the actual layout but does identify all the components included in this initial design. These figures are meant to conceptualize the architecture and highlight the complexity of each component rather than show the physical layout of the hub and carts. These can thus be considered a high-level summary of the component designs at the end of Phase I.

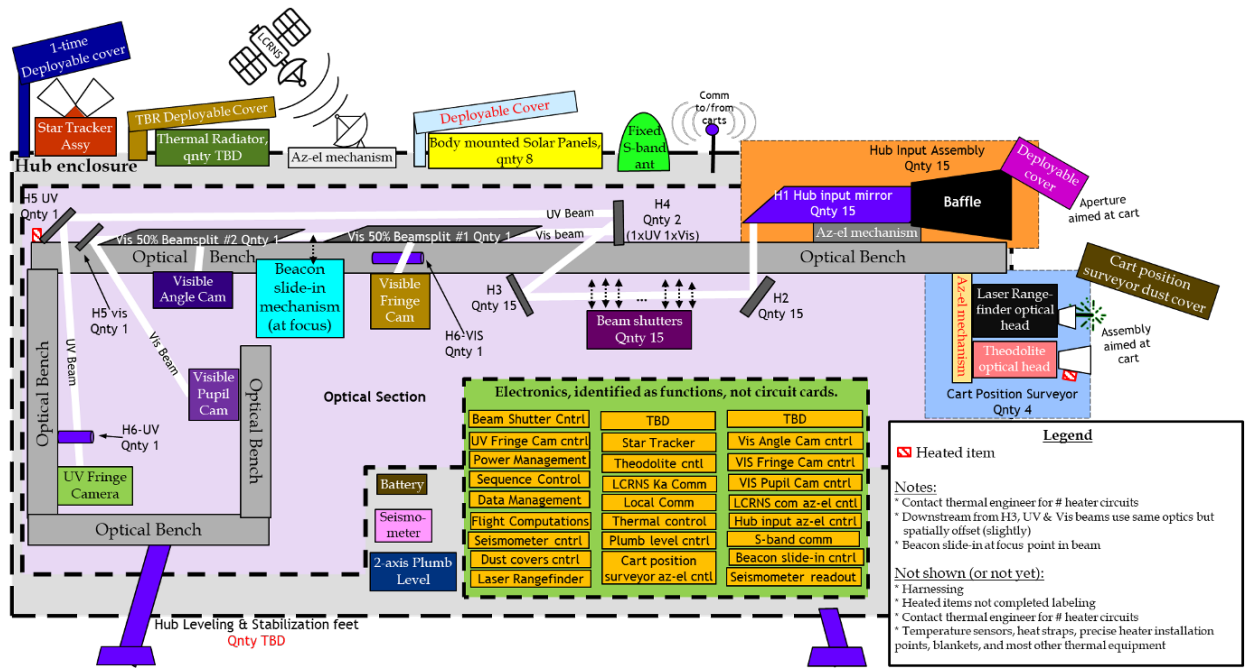

**Figure 3.37** Architecture diagram of the hub.

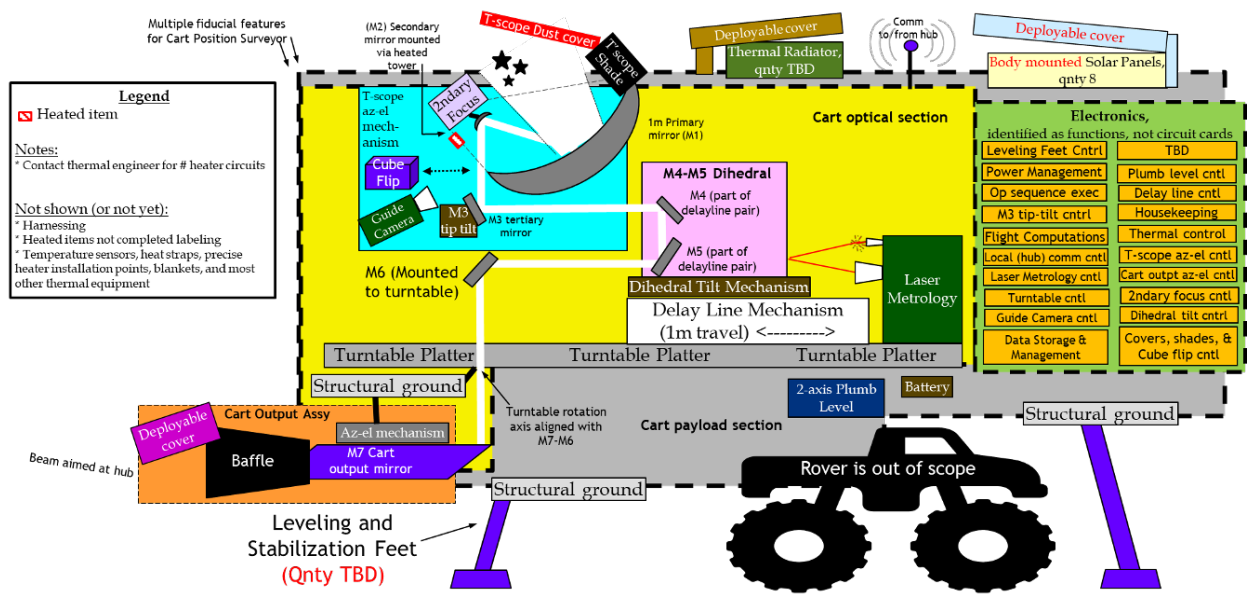

**Figure 3.38** Architecture diagram of the cart.

## 3.5 Areas for Lunar Infrastructure Enhancement

The baseline architectural concept for *AeSI*, designed in the IDC study, adopts a highly localized approach where each primary mirror element (cart) and hub is self-contained, minimizing interfaces with Lunar infrastructure (Figure 3.39). Examples include, solar power being collected and managed locally at each array element and communications between the hub and carts being point to point and thus not requiring the Lunar infrastructure. Thus, the IDC solution is in fact, an existence proof; a solution to provide *AeSI* with support services (power, communication, and pointing, navigation, and timing (PNT)) has not yet been fully considered. The Baseline design presented in this report is only one solution and not an optimal solution. An optimal solution requires a trade study to explore the solution space.

The Baseline designed *AeSI* architecture does require the Lunar infrastructure in four areas: 1) Pre-Launch support; 2) Deployment support to land, unpack, and distribute the *AeSI* telescope nodes on the Lunar surface; 3) communications between the hub and Earth; 4) PNT to support the primary mirror elements.

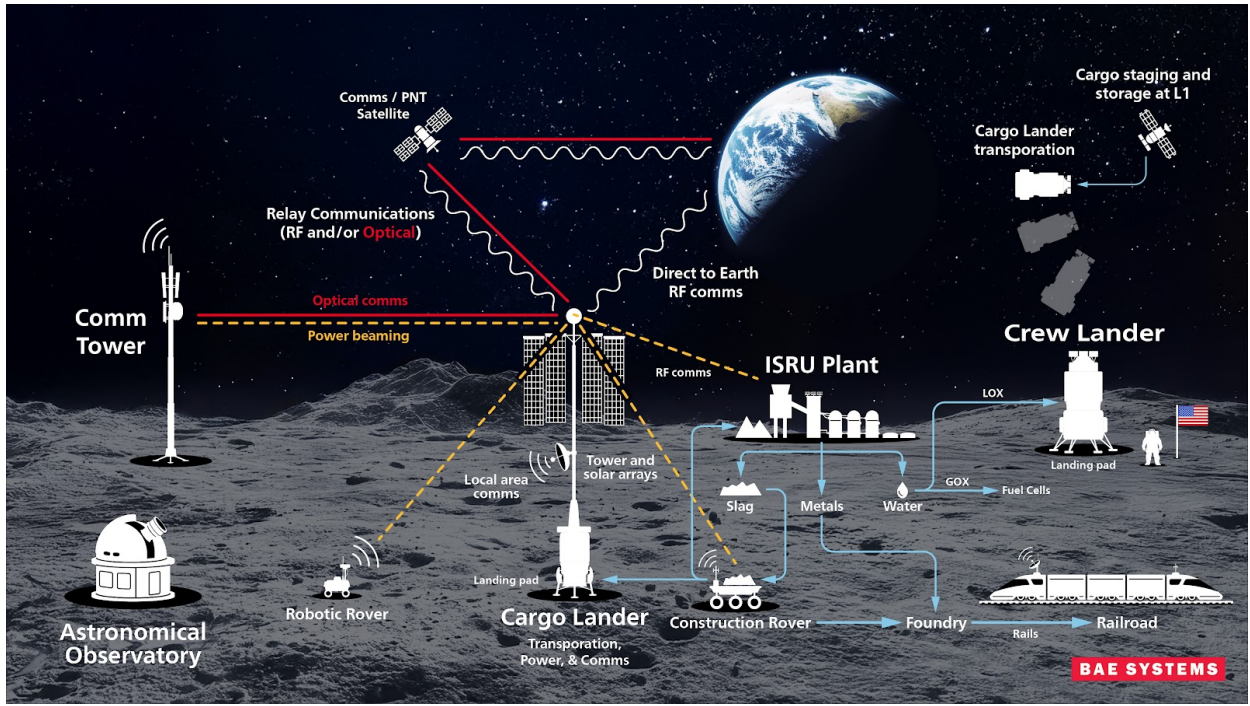

**Figure 3.39** Conceptual sketch of the potential lunar infrastructure under evaluation highlighting the power, PNT, mobility, and communications interfaces. The Astronomical Observatory (lower left) would leverage the different parts of the lunar infrastructure. (image credit: BAE Systems)

## 3.5.1 Power

Power is the primary challenge for operating instrumentation on the Moon, where assets must endure extreme environmental conditions. Lunar surface temperatures fluctuate drastically, ranging from approximately 35 K to 400 K, over prolonged (>10 Earth day) day and night cycles, combined with highly variable heat loads (from about 5 W to 10 kW). Ensuring systems can survive and function through these extremes is essential for sustained surface operations. Potential power sources include solar arrays and batteries (such as regenerative fuel cells), fission surface reactors, solar towers on nearby ridges (though current technologies are not optimized for low-horizon sunlight at the lunar poles), and power beaming.

In April 2024, the NASA Space Technology Mission Directorate (STMD) identified and ranked 187 areas of technology requiring further development[9], placing surviving and operating through the lunar night as <u>the top</u> priority and high-power energy generation on Moon and Mars surfaces as the second. Specifically mentioned developments include new power, thermal management, actuation technologies, and fission surface power systems.

---

[9] https://techport.nasa.gov/strategy

For the IDC study, driving requirements were chosen to assess feasibility with currently available technology. The most important driver for the Baseline design is to simply survive lunar night, assuming no science operations nor communications. For the Baseline design, TRL 9 solar and battery cells were considered, and a 5-year mission length was assumed. Longest lunar night was assumed to be 15 days.

The survival load distribution for the Hub is dominated by thermal power, as components such as electronic assemblies must not freeze, meaning they must retain enough power to turn on in the morning (Figure 3.40). The baseline Hub design assumes a total of 8 solar arrays (Figure 3.41, *left*). To survive a 15-day lunar night, and assuming a temperature of 135 C, a 28% efficiency, 95% packing factor, and Sun pointing, this sets the required area of each array to be 4.5 m$^2$ and power to be 1217 W (Figure 3.42). The mass per side is 9 kg. Assuming a lithium-ion battery with a max depth discharge of 80% and minimum operating temperature of -20C, the battery size must be large: mass = 1806 kg, volume = 1298 l, and capacity 220 kWh (Figure 3.42). For reference, this is equivalent in size to almost 2 vending machines. For each cart, the solar array design makes the same assumptions as the Hub, including having 8 arrays per cart (Figure 3.41, *right*). The array area per side is 1.4 m$^2$, power per side is 379 W, and mass per side is 2.8 kg. The battery on each cart needs to be 438 kg, with a volume of 315 l, and capacity of 53.4 kWh.

In terms of Science Operational time, the sizing of the described electrical power system would allow for science operations during lunar day and shorter (2-3 day) lunar nights (Figure 3.43).

In terms of safety during launch, most rockets have large tanks of hydrazine propellant. As the battery is a different energy source, preliminary conversations with safety experts suggest this might be okay. Current safety rules, however, prohibit humans to be on board with a battery this size, so cargo-only launches should be considered. Further complications arise from the current Crewed Space Vehicle Battery Safety Requirements document[10] that prohibits astronauts from handling or approaching batteries of this size. A proposed trade would be to utilize robotic options rather than astronauts for setup, maintenance, or servicing.

---

[10] https://ntrs.nasa.gov/citations/20150000860

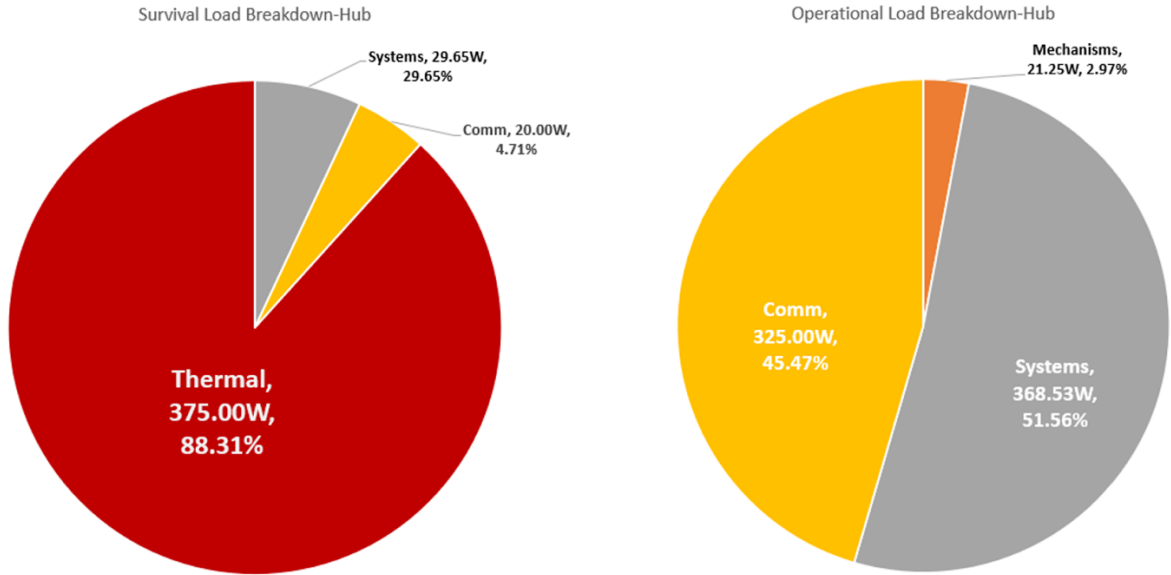

**Figure 3.40** Power load distribution for the Hub when in survival mode (left) and while operating (right).

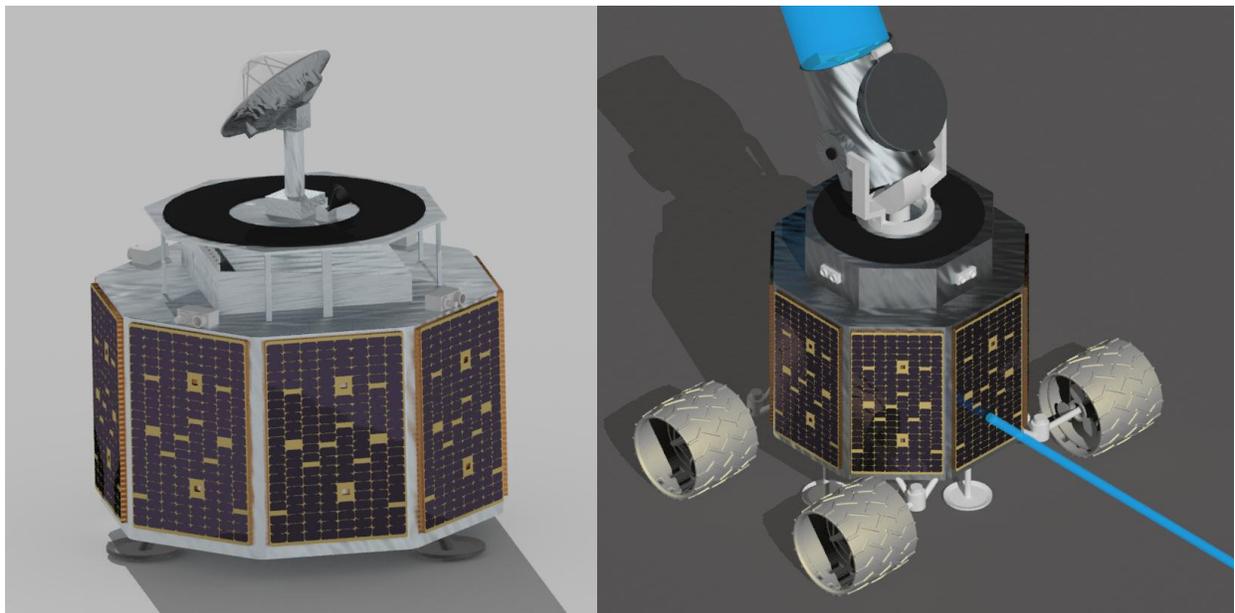

**Figure 3.41** Artist rendering of the hub (left) and one of the carts (right), depicting the solar panel configuration for both elements of *AeSI*. (image credit: Britt Griswold)

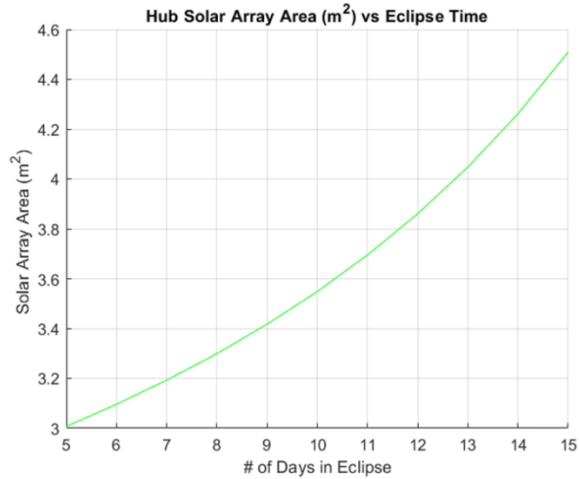
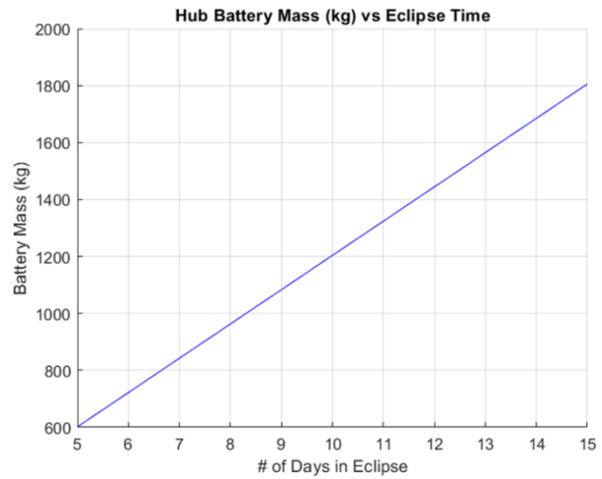
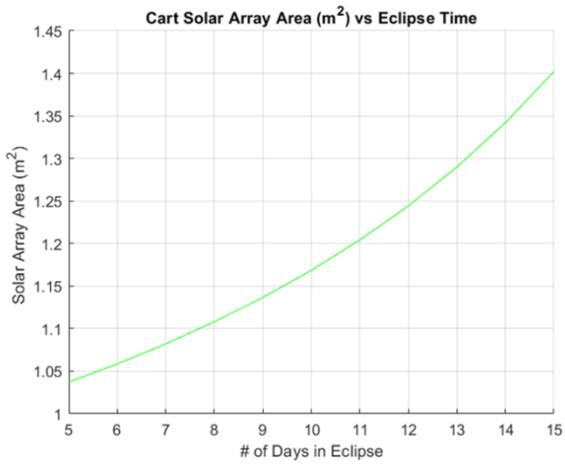
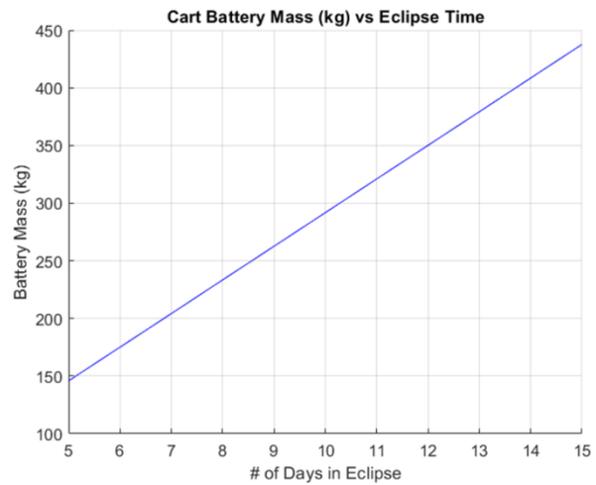

**Figure 3.42** Dependence of solar array area (*left column*) and battery mass (*right column*) on the duration of the longest lunar night for the Hub (*top row*) and Cart (*bottom row*).

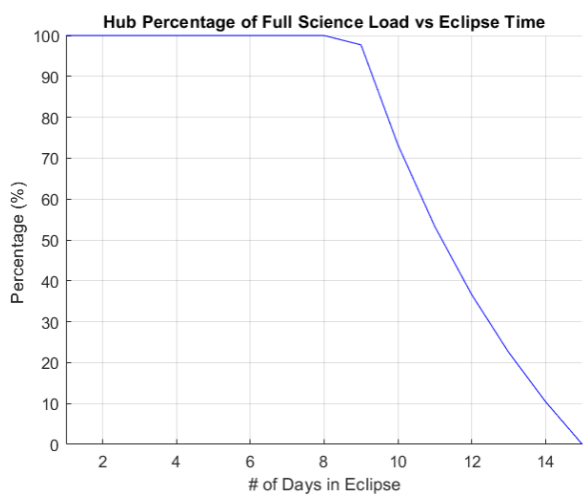
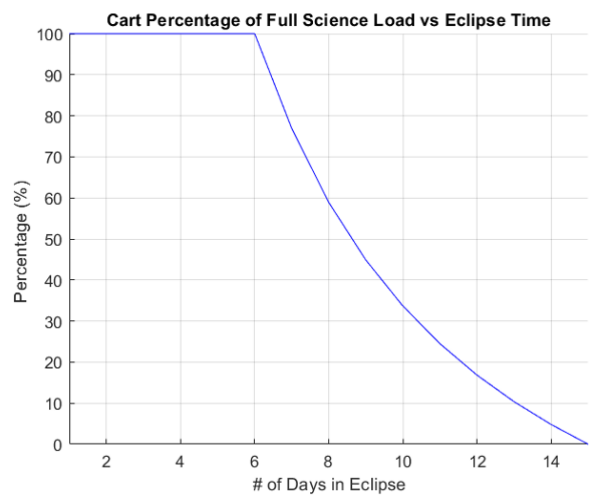

**Figure 3.43** Science operational time assuming an electrical power system required to survive 15 days of lunar night.

The main conclusion drawn from the IDC study on using TRL 9 solar and battery cells is that technology development is needed to improve feasibility. The required battery size is very large - refining the site selection is the primary method of reducing power system size as the longest eclipse drives the battery size, so a trade in this area could be explored. A detailed electronics architecture development, including sizing of electronics boxes based on functionality, would also help refine the power requirements for both hub and cart architecture. And finally, a more detailed trade could be done on single vs. multiple battery design. Regarding other technology development, cryo-tolerant electronics & batteries will allow for lower survival temperatures and less heater power overall; their usage and potential technology maturation/implementation of these technologies should be investigated. Alternative power sources and storage should also be studied in detail. Future work should include investigating potential collaborations with Glenn Research Center on fission power[11], researching potential implementation of fuel cells, and performing a detailed assessment of power transmission options, including cable transmission (at potentially superconducting temperatures) and power beaming.

### 3.5.2 Communications

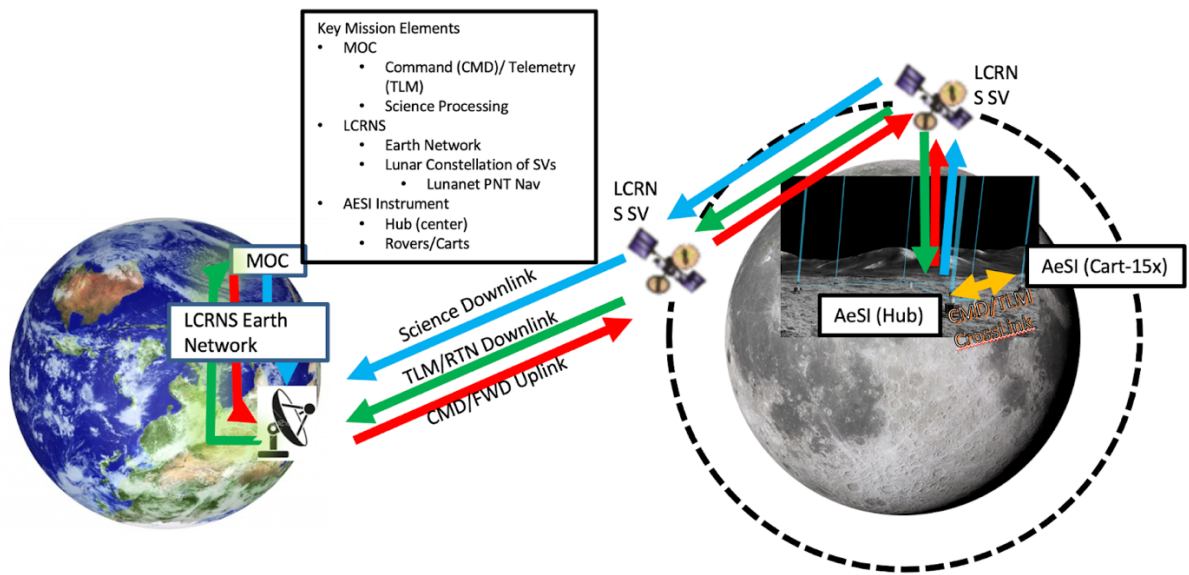

**Figure 3.44** Notional *AeSI* mission architecture.

The notional *AeSI* mission architecture assumes that the hub is responsible for receiving commands from the Mission Operations Center (MOC), parsing and distributing the commands to the carts, receiving data back from the carts, and sending data to Earth for subsequent analysis (Figure 3.44).

---

[11] https://www.nasa.gov/centers-and-facilities/glenn/nasas-fission-surface-power-project-energizes-lunar-exploration/

For assessing the command and communication requirements for *AeSI*, the Baseline design assumes that either NASA's planned Lunar Communications Relay and Navigation Systems (LCRNS) or Artemis Enabled Architectures will be used to receive data. Specific requirements that were considered include: 1) delivering 43 GB per day from the Lunar Surface to LCRNS system to Earth, 2) receiving commanding data from Earth with a 2-kbps link periodically, and 3) providing data between the Hub and Cart at 200 bps (which must be sent to all carts simultaneously).

The communications requirements place constraints on allowed Lunar site selection in that the site must have a low slope (0-5 degrees), a maximum lunar night duration of 15 days, and low Earth visibility. Three sites were identified that satisfy these needs: Behind Connecting Ridge: -88.8730, -136.6792 (No Earth Visibility), Peak of Connecting Ridge: -88.8730, -137.6793, Near Rim of Shackleton Crater: -89.5730, -135.6794 (Figure 3.45).

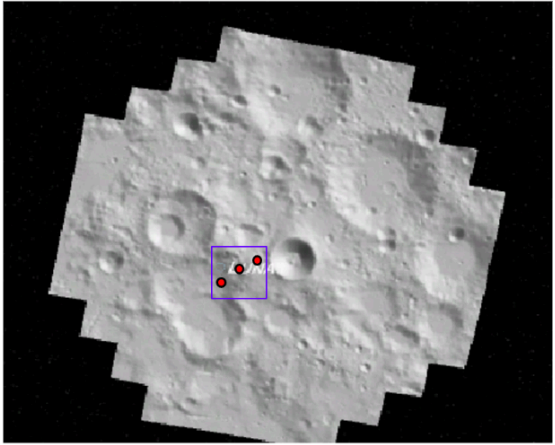
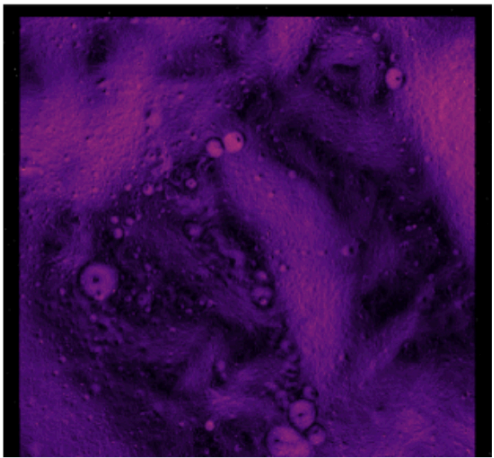
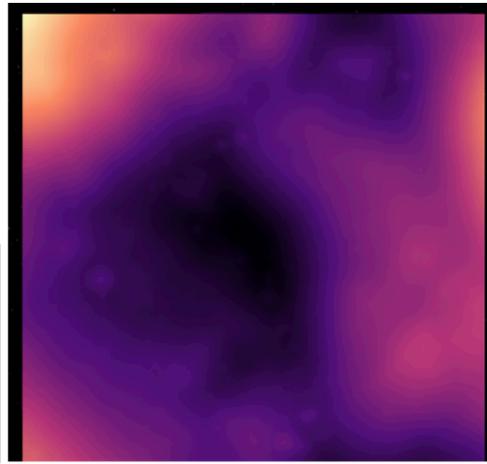
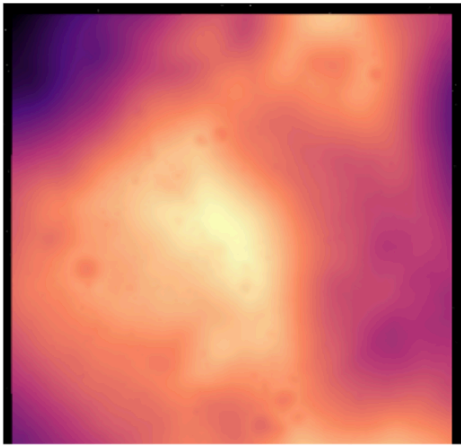
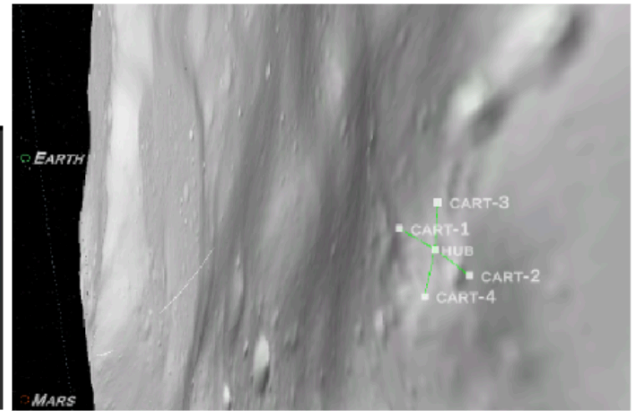

**Figure 3.45** Lunar terrain maps of identified sites that meet the communications requirements.

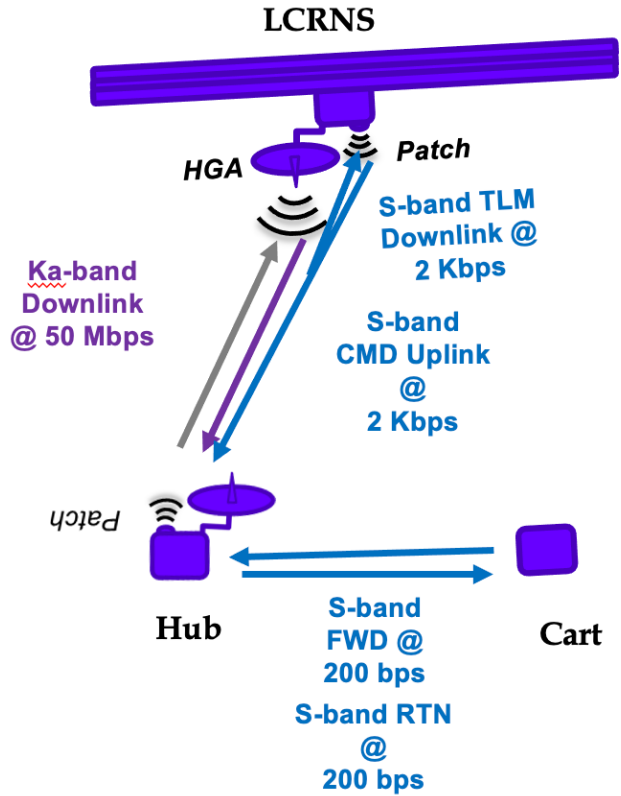

**Figure 3.46** RF ConOps design.

In support of the Science Operations Phase, the identified IDC communications solution is to use RF ConOps. For this, in the Hub, a Ka-band will be used to achieve the high amount of primary science data volume: Ka-band (27 GHz) at 50 Mbps with a 2-axis gimbaled 0.5 meter antenna will be sent to the LCRNS Constellation over 2 hours of contact per day. S-bands will be used between the Hub and LCRNS via 2 kbps command and telemetry and between the Hub and Carts at 200 bps forward and return (Figure 3.46).

Mass, power, and size summaries are given in Table 3.10. The current design requires too much power and has been identified as an area for improvement. Future work includes investigating space-qualified milliwatt transmitters for hub and cart communications (e.g., Endurosat S-Band Radio) and refining communications ConOps for both hub-cart and hub-LCRNS communications. A lower-power, more network utilization solution may be possible to reduce overall comm power: an S/Ka antenna at 5 Mbps and 2 Mbps could provide a closed design, for example. However, this may use a lot of LCRNS time; a trade study needs to be performed for higher data rate/higher transmit power Ka usage vs. lower power, lower data rate S/Ka usage (but more LCRNS time) as well as investigating resources afforded by the Artemis Base. An additional area for future work is to improve the frequency plan for the S-Band links, specifically to mitigate radio frequency interference (RFI).

**Table 3.10** Mass, power, and size summaries for communications design.

| AeSI SWAP | Quantity | Mass (kg) | | DC Comm Power | | | Dimensions | | |
|---|---|---|---|---|---|---|---|---|---|
| Component | | Unit | Total | Peak | Operations Average | Safehold Average | Length (mm) | Width(mm) | Height(mm) |
| Antenna (Ka-band) | 1 | 3 | 3 | 0 | 0 | 0 | 500 | 200 | 500 |
| S Band Patch antenna | 1 | 0.4 | 0.4 | 0 | 0 | 0 | 80 | 80 | 5 |
| Antenna (S-band) dipole | 2 | 0.1 | 0.2 | 0 | 0 | 0 | 25.4 | 25.4 | 127 |
| Comm Electronics (Ka-band) | 1 | 4.7 | 4.7 | 45 | 3.75 | 0 | 152 | 66 | 206 |
| Ka TWTA (30 W waste heat captured in Comm Electronics) | 1 | 1.6 | 1.6 | 130 | 2.6 | 0 | 380 | 80 | 55 |
| Diplexer | 3 | 0.5 | 1.5 | 0 | 0 | 0 | 170 | 140 | 40 |
| Comm Electronics (S-band Transmitter) | 2 | 0.3 | 0.6 | 5 | 0.05 | 0 | 200 | 100 | 100 |
| Comm Electronics (S-band Receiver part of Transponder) | 2 | 0 | 0 | 8 | 8 | 8 | 0 | 0 | 0 |
| Comm Electronics (S-band Transmitter) Cart | 15 | 0.3 | 4.5 | 5 | 0.05 | 0 | 200 | 100 | 100 |
| Comm Electronics (S-band Receiver part of Transponder) Car | 15 | 0 | 0 | 8 | 8 | 8 | 0 | 0 | 0 |
| Antenna (S-band) dipole Cart | 15 | 0.1 | 1.5 | 0 | 0 | 0 | 25.4 | 25.4 | 127 |
| Diplexer Cart | 15 | 0.5 | 7.5 | 0 | 0 | 0 | 170 | 140 | 40 |

### 3.5.3 Pointing, Navigation, and Timing

The position and orientation of the carts relative to the hub is critical to facilitate interferometric alignment. As described in Section 3.4.9, alignment and measurement of the relative orientation of the telescope nodes with respect to the hub can be accommodated using onboard telescopes and lasers. Depending upon the nature of the lunar infrastructure, alternate alignment methodologies may be available that reduce the complexity inherent in the Baseline design. For example, a lunar beacon network may provide the carts and hub with sufficient positional knowledge. Alternatively, if a rail system is used then the positions of the telescope nodes can be effectively defined, likely reducing the time to bring the array into phase alignment. In addition to positional requirements, the time needed to configure and align the *AeSI* network is a consideration in the trade.

The PNT requirements for position accuracy, attitude accuracy, and timing accuracy are discussed further in Section 3.4.9.

# 4   Baseline Design Evaluation

## 4.1   Feasibility of Baseline Design

Over the course of the Phase I Study, the *AeSI* team worked with NASA/GSFC's IDC to turn our initial mission concept into a design with solid engineering behind it. The goal was to create a Baseline design that demonstrated the feasibility of this concept and to identify areas needing further study and technology development to enable an optimal design. This collaboration resulted in a credible Baseline design that initially features 15 primary mirrors and an array outer diameter of 1 km but is expandable to 30 primary mirrors and even larger array sizes, in one or more staged deployments. Importantly, this study demonstrates that a **lunar-based interferometer can be largely self-sufficient**, without relying heavily on undeveloped lunar infrastructure, proving its feasibility as a near-term concept. Since we did not identify any critical roadblocks or fundamental obstacles in this Baseline design, it serves as a logical foundation for further study into optimization strategies that rely more heavily on planned lunar infrastructure. This includes trade studies on potential improvements, which are intended to be pursued in follow-up investigations, such as a NIAC Phase II investigation. We discuss these below in Sections 4.4, 4.5, and 4.6.

This collaboration concluded that the Baseline design architecture and the required initial resources are acceptable in that no requirements were identified without a conceptual implementation.  Each item in the Baseline design has been demonstrated on the Earth and is TRL 6 or higher except for the delay line mechanism, which is still considered lower TRL, yet having a clear path to implementation.

The IDC collaboration agreed that *AeSI* has a compelling motivation that justifies ≥$1B, including a strong science case, a design that employs the lunar economy (e.g., transportation to and on the lunar surface and Artemis-enhanced deployment and servicing) and helps justify the establishment and operation of the Artemis bases, as they enable visionary science far beyond the lunar surface itself.  It takes advantage of the goodwill around Artemis and offers straightforward opportunities for international partnerships and the involvement of astronauts and/or their robotic assistants.

## 4.2   Enabled Science with the Baseline Design

The Baseline design of *AeSI* enables groundbreaking high spatial- and temporal-resolution imaging across a wide range of astrophysical targets (Section 2). Imaging in both the UV and optical, from 1200 - 6600 Å, *AeSI* can resolve fine details on stellar surfaces, probe magnetically driven activity such as starspots and plages, and capture convective patterns across different stellar types.

Figure 4.1 illustrates a color-magnitude diagram of potential science targets within 20 parsecs,

highlighting the solar-type stars where *AeSI's* imaging capabilities can resolve stellar surfaces. Considering the full sky and assuming a maximum baseline of 1 km and a central wavelength of 2000 Å, 462 stars achieve at least 10 pixels across their surface (N(pixels) > 10), covering main-sequence stars from spectral types A to M. As resolution requirements increase, the number of available targets decreases: 143 stars meet the N(pixels) > 20 threshold, while 63 stars meet N(pixels) > 30 (Table 4.1).

Table 4.1 highlights examples of objects available to study for select stellar science cases discussed in the text. We curated the exoplanet host star sample from NASA Exoplanet Archive[12]. Supergiants were selected from catalogs built from GAIA (Messineo & Brown 2019), which we filtered to retrieve stars with parallax between 0.3 and 0.5 milliarcseconds (~2 - 3.5 kpc) in order to ensure ~30 resolution elements across the target's disk while also maximizing *AeSI*'s SNR. In addition, as an example of interacting binaries, we investigated several catalogues of symbiotic stars (Belczynski et al. 2000, Luna et al. 2013) and found over several dozens of potential targets to study within *AeSI*'s capabilities, including the "inter-binary" regions, and close circum-binary regions for targets within few pc, with sizes of few milliseconds to arcseconds. We also explored nearby AGN targets, containing BELRs, and extended wind outflows, resolvable with 0.1 milli arcsec; Depending on the mirror/detector sensitivity, and the location of the *AeSI*, we expect at least a dozen of targets to be available for imaging (e.g., up to 10x10 pixels across the studied regions).

When restricting the target availability to the Southern sky (all negative latitudes; Figure 4.1, *right*), the total number of targets with N(pixels) > 10 is cut almost in half: the number of solar type targets with drops to 229 and known exoplanet host stars and supergiants drop to 8 and 153, respectively.

---

[12] https://exoplanetarchive.ipac.caltech.edu/ (as of December 2024)

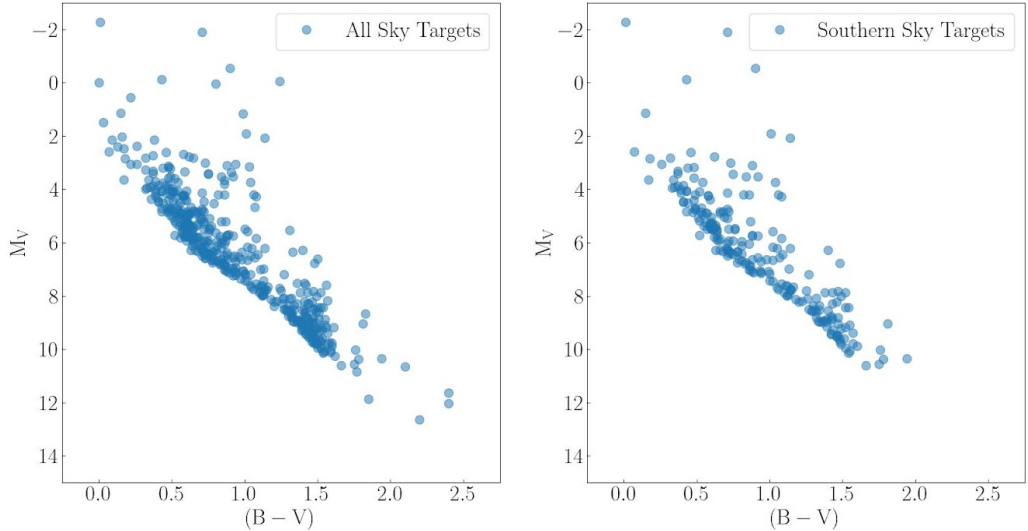

**Figure 4.1** Color-magnitude diagram of the 462 solar-type stars available for surface imaging in the 20-parsec sample. The left panel shows the all-sky sample for *AeSI* at equatorial latitudes, and the right panel shows the sample for a south polar location on the lunar surface.

**Table 4.1** Potential target sample yield for select stellar *AeSI* science investigations for all latitudes. "N(pix)" refers to the number of resolution elements across the stellar disk assuming a wavelength of 2000 Å and a baseline of 1,000 m.

| SCIENCE CASE | #OBJECTS WITH N(PIX)>10 | #OBJECTS WITH N(PIX)>20 | # OF OBJECTS WITH N(PIX) >30 |
|---|---|---|---|
| **Solar-type stars** | 460 | 143 | 63 |
| **Known exoplanet host stars** | 17 | 6 | 2 |
| **Supergiants** | 280 | 280 | 273 |

## 4.3 The Role of Site Selection in Mission Feasibility

Final site selection for *AeSI* is closely tied to the locations of Artemis habitats, as these bases will provide critical logistical support. However, since the Artemis base site has not yet been chosen, this adds complexity to the selection process. *AeSI* site planning must account for multiple potential Artemis locations and remain flexible to integrate with the eventual base site. Within the viable regions, a suitable location must be identified that meets both scientific and operational requirements. Key considerations include:

- **Terrain Suitability:** The selected site must feature a flat enough area to accommodate a 1–2 km diameter array while also having a nearby elevated location suitable for a remote solar or nuclear power station.
- **Accessibility:** The site must be reachable from the Artemis landing zone, allowing for efficient transport of equipment, personnel, and maintenance resources.

- **Lunar Darkness Pattern:** The site must be analyzed for its illumination conditions, including periods of extended darkness, which impact power generation, thermal control, and observational windows.

Next Steps in Site Selection

To refine the selection process, several key tasks remain:

- **Refining Site Selection Criteria:** Science and operational constraints—including line-of-sight (LOS) requirements, surface slope, and sunlight duration—must be solidified and used to assess potential sites.
- **Developing a Design Reference and Key Requirements:** Using the refined site selection criteria, a design reference will be established to define infrastructure needs. This includes considerations such as landing pad locations, staging areas, delivery logistics, personnel support, emergency services, and long-term maintenance.
- **Analyzing Site Options for Illumination and Topography:** Potential sites must be evaluated to ensure they meet the illumination and surface conditions required for *AeSI's* operations.
- **Power Systems and Thermal Analysis:** More detailed studies are needed to assess power generation feasibility and thermal management strategies for different candidate sites (see Sections 4.5 and 4.6 for additional trades in these systems).
- **Determining Necessary Site Preparation:** The level of site preparation required—such as compaction, leveling, and boulder removal—must be identified to ensure *AeSI* can be deployed successfully.

Because Artemis base site selection is still uncertain, *AeSI's* planning must remain adaptable. As more details emerge, an optimal site can be selected that balances scientific goals with practical deployment and operational constraints.

### 4.3.1 Impact of Site Selection on Science Goals

Target availability is significantly affected by *AeSI's* location. A site near the lunar equator provides access to nearly the entire sky, maximizing the number of observable targets. In contrast, if *AeSI* is placed at a polar location and constrained to observing outwards from the south pole, the accessible target sample shrinks considerably. This effect is shown in Figure 4.1, where the number of resolvable stars is reduced by half in the polar configuration compared to an equatorial placement. Figure 4.2 illustrates this distribution with a histogram of available targets by ecliptic latitude, highlighting the loss of targets at high latitudes when observing from the Lunar South Pole. Figure 4.3 further visualizes the impact on target selection with a scatterplot showing available targets by latitude and surface resolution, emphasizing the significant reduction in accessible targets under a polar constraint. While a polar site offers advantages such as thermal stability and reduced dust accumulation, it imposes strict observational limitations that must be carefully considered when optimizing *AeSI's* science return. Additionally, local terrain features such as crater walls and elevated ridges can further restrict the field of regard, limiting access to low-elevation targets and necessitating careful site selection to minimize obstructions.

Further analysis is needed to quantify the trade-offs between site selection and science yield, ensuring that the Baseline design continues to support its broad scientific objectives even under operational constraints.

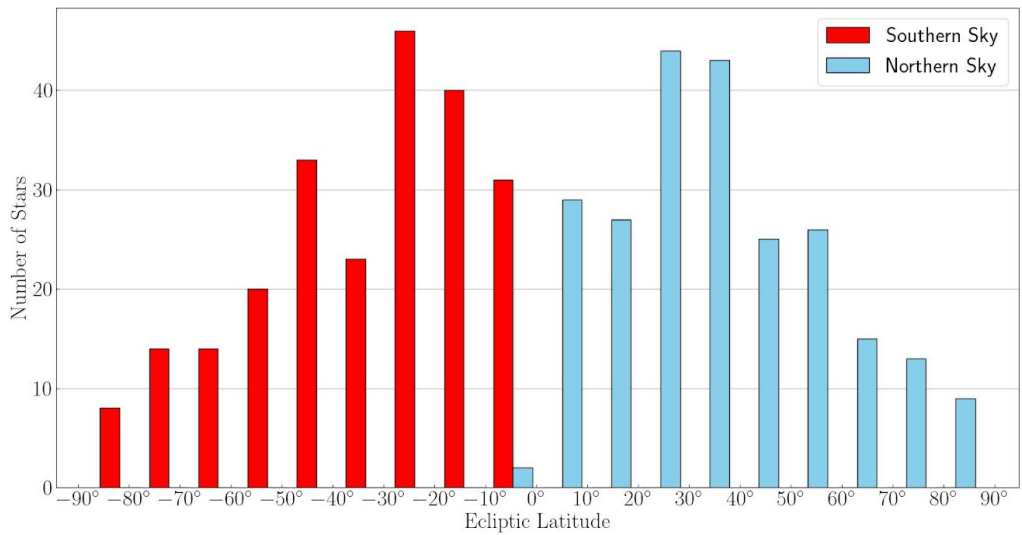

**Figure 4.2** Dispersion of targets described in Section 4.2 by latitude.

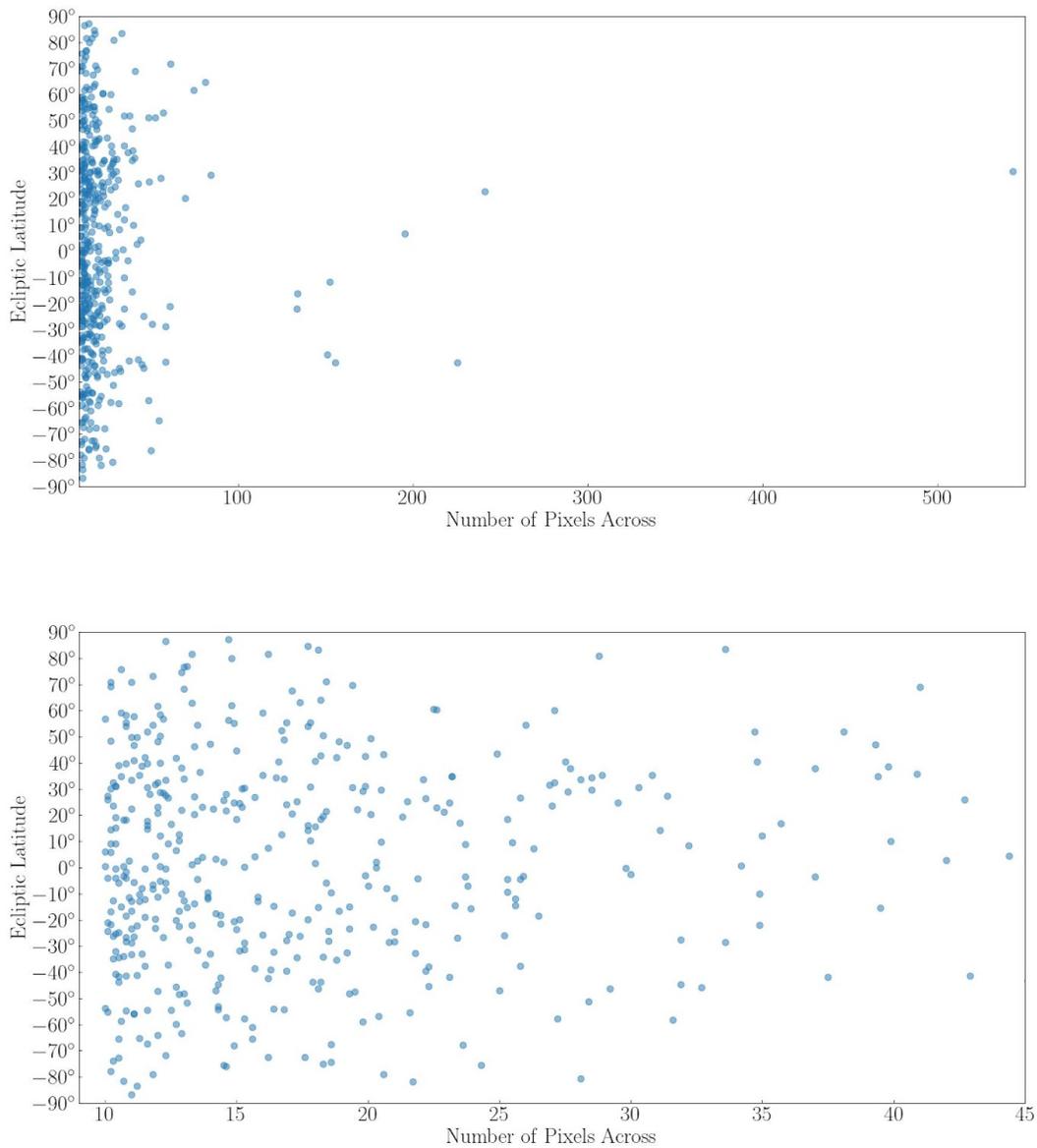

**Figure 4.3** Dispersion of available targets by latitude and surface resolution (number of pixels; bottom panel restricts x-axis to less than 45 pixels across), as described in Section 4.2. Targets are relatively evenly distributed across all latitudes; however, if *AeSI* is positioned at the Lunar South Pole, the field of regard is significantly reduced, limiting access to more than half of the targets. Placing *AeSI* away from the south pole increases target availability.

## 4.4 Future Work and Trades in Mission Implementation

While this Phase I study establishes the feasibility of *AeSI's* Baseline design, several critical areas require further exploration to refine mission implementation. These include deployment strategies, dust mitigation, cost considerations, array configuration, and operational requirements. Addressing these factors will be essential to ensuring the success and optimization of the mission.

**Launch and Deployment Considerations**

Strategic planning for development, launch, and deployment remains a key area of study. One major question is how mass constraints influence mission architecture, particularly in the context of Starship's landing capacity on the Moon. The viability of inexpensive rovers at this payload mass and the potential need for astronaut or robotic servicing must also be considered. Any new strategic capabilities identified should be incorporated into the overall mission concept.

The selection of an appropriate launch vehicle is another important factor. Options under consideration include:
- **Starship** – high payload capacity, reusable architecture
- **New Glenn** – competitive commercial option with lunar capability
- **SLS** – government-led, heavy-lift option with proven lunar mission support

Once delivered to the lunar surface, *AeSI* components must be transported from the landing site to the interferometer array location. The feasibility of cost-effective lunar transport solutions remains unproven, and further research is needed to assess:
- Methods for transferring large, strain-sensitive components onto the lunar surface
- Storage and handling requirements during transport
- Challenges specific to deploying the array elements, particularly those that are not self-contained on individual carts

**Dust Mitigation and Contamination Control**

Lunar dust presents a challenge to long-term operation. As rovers traverse the surface, they will inevitably disturb the regolith, raising concerns about contamination and potential damage to optical surfaces. A more detailed assessment is needed to determine how quickly dust settles and whether actuation of dust covers could inadvertently redistribute it.

This study did not explicitly address contamination and dust mitigation strategies, but future work should include:
- Investigation of active and passive dust mitigation technologies
- Opportunities for technological maturation in dust-resistant coatings and mechanical shielding
- Design modifications that integrate dust mitigation strategies from the outset

**Configuration Trades: Flexible vs. Constrained Array**

*AeSI's* interferometric array could be deployed in a flexible (rover-based) or constrained (rail-based) configuration, each with trade-offs:
- **Rover-Based Design**: Offers adaptability in U-V plane coverage but may introduce stability concerns, require more infrastructure, and demand longer reconfiguration times.
- **Rail-Based Design**: Provides increased stability but reduces flexibility and requires more extensive site preparation, including the transportation of heavy rails.

Future work should conduct a detailed trade study evaluating mass, infrastructure requirements, and operational efficiency for both configurations.

**Operational Planning and Autonomy**

The interferometer must maintain precise alignment, either autonomously or with human oversight. The Baseline design assumes alignment via a guide camera, but the level of human-in-the-loop (HITL) involvement remains an open question. Any required HITL operations must be carefully weighed against computational demands and potential science time loss. Future work should:
- Confirm autonomous alignment capability and associated computational power needs
- Define all necessary HITL elements and assess their impact on operational efficiency

**Cost Considerations**

Understanding the total cost to complete the mission is essential for assessing feasibility and securing funding. Future efforts should focus on:
- Developing a detailed cost model for the full mission lifecycle
- Evaluating whether the scientific return justifies the projected expenses
- Identifying potential funding sources capable of supporting a mission at this scale

## 4.5  Leveraging Lunar Infrastructure for Mission Optimization

The Baseline design demonstrates that relatively self-sufficient communications, PNT, and power solutions are feasible. While we have developed a technical solution that can be refined, the most effective way to fully optimize the system is by leveraging existing and planned lunar infrastructure. The Baseline design results in an exceptionally large overall mass for a surface payload when considered as a fleet, though individual components remain within the range of other landed assets. This mass is primarily driven by the large batteries required on each rover and the hub, as well as design choices related to communications and PNT. In the following sections, we outline strategies for improving the Baseline design and explore the trade space necessary to optimize the system using lunar infrastructure.

**Communications and Data Handling**

The communication trade space includes three primary options: (1) direct *AeSI*-to-Earth communication, (2) *AeSI*-to-Artemis base relay to Earth, and (3) *AeSI*-to-orbital relay to Earth. The key selection criteria for this trade study include data volume, data rate, accessibility, and power consumption.

The baseline communication design, which relays data from the hub to LCRNS before transmission to Earth, results in excessive power consumption. Future efforts will focus on optimizing communication cadence between the hub and carts and exploring low-power transmitters. Additionally, trade studies are needed to evaluate S- and Ka-band usage, frequency planning, and potential lower-power alternatives.

The feasibility of relying on Artemis- and LCRNS-based infrastructure remains uncertain, as their deployment and technological maturity may not progress as anticipated. To ensure flexibility, we will explore multiple options for orbital relay, landing site connectivity, and transport logistics. Designing for greater autonomy and further assessing the *AeSI*-to-Artemis base communication trade space will be critical for optimizing system performance.

**Pointing, Navigation, and Timing**

The Baseline design for *AeSI* assumes a local PNT system, but a trade study is needed to assess the benefits of leveraging lunar PNT infrastructure. Key factors in this trade include accuracy, complexity, and range. A local system requires onboard hardware such as atomic clocks and inertial measurement units (IMUs), adding mass and increasing power consumption, whereas lunar PNT infrastructure could offload these functions to external satellites or ground stations. Additionally, a local system increases complexity by requiring internal redundancy and precise synchronization, which becomes more challenging when coordinating multiple mobile platforms. Without external PNT signals, long-term accuracy may degrade due to drift, requiring frequent recalibration—an issue for an interferometric system needing nanometer-scale precision. While a local os system offers independence, leveraging lunar infrastructure would reduce mass, power demands, and system complexity, improving overall efficiency.

**Power**

The Baseline design for *AeSI* relies on solar-powered batteries, which must be large enough to ensure survivability through the coldest, longest lunar night. Optimizing site selection—such as positioning near peaks of eternal light or placing solar arrays on a nearby hill to maximize sunlight exposure—could reduce these power requirements, while refining the electronic architecture would help define precise power specifications. Additionally, assessments need to be made to ensure the batteries meet astronaut safety requirements, with autonomous solutions where human interaction is not possible. Solar array dimensions also play a key role in system envelope constraints, warranting further refinement. To enhance resilience, cryo-tolerant electronics and low-temperature battery technologies could be explored.

Beyond solar, alternative power sources must be considered, including fission power, fuel cells, and power transmission via cables or beaming. The full trade space includes both local and centralized implementations of:

- Solar
- Nuclear (fission)
- Beamed power
- Artemis Base-supplied power

Selection criteria for this trade space include array configuration, system complexity, operational considerations for day and night cycles, lunar night survival, power availability, mass constraints, and overall project scale. Conducting these trades will be critical for determining the most efficient and reliable power solution for long-term lunar operations.

## 4.6 Future Work and Trades in Instrument Engineering and Architecture

The baseline instrument design for *AeSI* demonstrates both promise and feasibility, but several key elements require further refinement to ensure reliable operation on the Moon. Next steps include validating the overall architecture, particularly for accessory cameras, rangefinders, and other components not assessed in detail during the Phase I study, as well as ensuring adequate survival heating. Additionally, a detailed electrical architecture must be developed for both the *AeSI* hub and carts based on required functionality, which will inform updates to power and thermal analyses.

The following subsections outline the current design status, necessary advancements, and key considerations for developing a fully functional and sustainable instrument.

**Optical**

An end-to-end detailed optical design was not completed in Phase I. The optical model should refine light paths for both the UV and optical channels, with a focus on thermal limits and stability. Future steps include developing mounts, laser metrology systems, and analyzing stray light paths from the Sun and Earth, particularly for mirrors M1, M7, and H1. Optical elements require thermal trade studies and stray light baffle solutions, including dust mitigation measures. Design verification is also needed for the UV/optical dichroic (H5) and the elliptical mirror array (H4) to ensure performance at targeted wavelengths.

Further work is required on performance error budgeting and validation, as UV wavelengths impose extremely small allowable errors. Additional studies are needed to assess potential vignetting, occlusion, and stray light effects.

**Mechanisms**

A full mechanism definition for the hub and carts was outside of scope in this Phase I study, and engineering resource estimates (mass, power, volume) given for the Baseline design are approximate. Precision and motion requirements need updating, with specific heritage considerations for cryogenics. In subsequent studies we will define mechanism trades for non-cryo implementations, if more heater power becomes available. Dust covers require further definition and mechanism design. Additional development is needed on the M4-M5 delay line mechanisms. Addressing mechanical vibrations and evaluating potential vibration mitigation strategies will also be necessary.

**Detectors**

A detailed design of the focal plane arrays (FPAs) for the UV and Optical channels were not achievable in Phase I due to the current level of concept maturity. The design assumes 2 μm pixel pitch detectors, but current technology supports only 7 μm. We will assess whether optical adjustments or detector development can achieve the desired resolution.

**Thermal**

A final cart and hub point design was beyond the scope of Phase I, so a complete thermal control system was not developed. Future work must define mechanism geometries, duty cycles, power needs, and survival temperatures to guide thermal control strategies for each mechanism. Operational temperature requirements for optical components and structural elements must also be established to inform appropriate thermal control methods.

Heater power savings need to be assessed, and efficient thermal transport methods must be explored, particularly for large distances between components and radiators. Radiator designs should be revisited to improve efficiency, and once a detailed thermal point design is achieved for surviving the longest lunar night, system performance should be evaluated for operations through shorter lunar nights. Exploring all possible avenues for thermal turndown could yield significant power savings, with a high return on investment.

Among these considerations, mechanism modeling is the most critical, as it has the greatest potential to drive power requirements well beyond initial estimates. Beyond mechanisms, the Baseline design presents a high-power solution as an existence proof, but further analyses may identify opportunities for power savings, potentially at the cost of integrating new technologies.

**Mechanical**

Mechanical mass and volume estimates for the Baseline design are preliminary and currently very large. Future work should focus on refining the mechanical model by incorporating optical mounts, mechanisms, and structural elements, including the solar array assembly. The optical model also requires refinement to resolve interferences, integrate a wavefront-sensing optical channel, and conduct a trade study on materials for the optical bench.

Careful attention is needed for connections between the lower structure (e.g., e-boxes and battery) and the upper structure (e.g., radiators and star trackers), ensuring they are routed effectively between optical paths. Further studies should explore options for stray light baffles and optimize material selection across the system (e.g., aluminum vs. composites vs. other alternatives). Additionally, launch vehicle compatibility and rover + hub leveling and stabilization analysis must be completed.

**Rover Design**

A rover design and its interfaces with the payload were not studied in Phase I. Once a rover vendor is selected, payload accommodations must be integrated into the vendor's design, with subsystem updates to reflect the new rover configuration. A key consideration is determining how to effectively distribute and carry loads of this magnitude on the regolith. Further development is required to mature the functional concepts for mechanisms outlined in the IDC study, including the creation of a CAD model for the delay line mechanism, which was only represented schematically in the Phase I study. Stability mechanisms for the optics cart may need to be restructured into a separate cart that establishes a stable platform for the optics cart to roll onto. Additional information is also necessary to accurately partition electronics functions for practical packaging estimates and to determine the appropriate solar panel sizes. Further guidance is needed on the electronic support required for the cart drive system to ensure seamless integration and operation.

**Other**

Other features that were not addressed in the Phase I study that should be considered for future development include:
- Electrostatic dust mitigation
- More input capability in the hub (beyond 15 inputs)
- Moonquake nulling system (operate successfully through a moonquake)
- Optimized safety vs power savings tradeoff during non-operational periods (e.g. uptime cadence to be able to receive a command)
- Redundancy and optimized reliability

# 5 Technology Development Roadmap

The *AeSI* instrument relies on a Baseline design that incorporates technology with a Technology Readiness Level (TRL) of 6 or higher, having been developed and demonstrated in Earth's environment. However, operating on the Moon presents new challenges, including extreme temperature gradients, lunar dust, moonquakes, and our design requires the need for unprecedented stability and increased scale. This roadmap outlines the critical advancements necessary to transition the existing technology into a lunar-ready system. The primary objective is to demonstrate that the proposed design is not only feasible on paper but also viable in practice. To achieve this, we must build and test key components in relevant environmental conditions, increasing their TRL and ensuring their performance and reliability under lunar conditions. This Phase I Study identified the following technology developments as the most important needed to fully enable *AeSI's* success on the Moon:

**Space qualified delay lines**

Delay lines are well understood and in active use across a broad range of Earth-based optical interferometers, so the challenge here is simply to qualify them for use in space and on the lunar surface and ensure they will work in a vacuum and survive the range of temperatures expected to be encountered on-site. This is expected to be straight-forward and GSFC Engineers have already begun to work on it.

**UV detectors with high quantum efficiency, High UV reflectivity optical coatings, and Innovative delay-line designs that may require fewer mirrors/reflections**

The biggest challenge facing the *AeSI* development is the low sensitivity in the UV due to the numerous reflections in the currently-designed optical train. Three parallel approaches are being pursued to address this and ameliorate excessively long exposure times. The two most feasible approaches are to develop UV detectors with significantly higher Quantum Efficiency and mirror coatings with substantially higher UV reflectivity. Both of those are also required for optimal designs for other UV space-telescopes, especially the segmented-mirror Habitable Worlds Observatory (HWO). A third approach, which may be difficult to achieve, is to find an innovative delay-line design which uses fewer than the traditional designs used on Earth-based interferometers. We have already reduced the required length of the delay lines by planning to use innovative elliptical configurations of the mirror array when observing off the zenith to take out most of the optical path length difference for those cases.

**Surface-based fission power source**

While a full trade study has not yet been conducted, we believe there is a need to move beyond the co-located solar cells plus batteries power system used in our Baseline design, as that significantly limits the amount of time that *AeSI* can spend observing during dark time. Although we expect the interferometer will be able to work in an optical sense during both bright and dark time, being able to operate in both greatly increases the productivity of the observatory and provides more time when scattered light from the nearby lunar landscape does not limit performance. One option that partially addresses this would be to find a nearby hill with longer duration sun exposure than the interferometer site (such sites are available near candidate Artemis landing sites) and send power by cable or wirelessly down to the observatory which would be located in areas with shorter-duration sun exposure. Although this approach would improve the time over which *AeSI* could observe, it does not enable full-time observations. In order to enable continuous observations, both to facilitate optimal asteroseismology studies as well as double the observing efficiency, a better solution is to take advantage of a nuclear fission power source, such as the small fission reactor proposed by the Glenn Research Center (GRC) for Artemis[13]. That would likely be placed outside of the array itself (perhaps over a nearby hill) to minimize any issues due to radiation from the reactor. Power would be transmitted, as for the remote solar array case, via cables or wireless transmission, to the observatory.

---

[13] https://techport.nasa.gov/projects/105671

# 6 Conclusions

The ultimate goal of this effort is to assess the feasibility of the *Artemis-enabled Stellar Imager*, a lunar-based UV/optical interferometer, and determine whether it offers advantages over a free-flying version. In this Phase I study, the *AeSI* team, working with NASA GSFC's IDC, firmly established the feasibility of building and operating a reconfigurable, dispersed aperture telescope on the lunar surface by leveraging the infrastructure being developed to support the Artemis program. Our study confirmed that a lunar-based UV/optical interferometer can be competitive—and in some ways superior—to a free-flying version, provided that its deployment, operations, and servicing take advantage of expected Artemis infrastructure, including robotic and/or human support. However, this study did not fully optimize the design to maximize these benefits. Rather than being a standalone point study, this work represents a logical progression from the 2005 Stellar Imager concept—originally envisioned as a free-flying system—to evaluating the technical challenges and benefits of placing it on the Moon, with the next step being to fully realize the Artemis-enabled implementation. Real progress was made in understanding the feasibility of this approach, and we have not identified any fundamental flaws in the concept. With a clearer understanding of the challenges and opportunities, the next phase is to push forward, optimizing the design and rigorously determining whether a lunar-based interferometer is ultimately the superior solution.

## 6.1 Comparison to Free-Flyer

Our Phase I study demonstrated that a lunar-based UV/optical interferometer can rival—and in some aspects outperform—a free-flying version, provided it fully integrates with Artemis infrastructure. *AeSI* eliminates the need for ultra-precision formation-flying of numerous spacecraft over long distances, a significant challenge in realizing a free-flying interferometer. Additionally, the lunar-based design allows for easier maintenance and potential upgrades compared to a free-flyer operating near the Sun-Earth L2 point. The challenges of operating in a lunar environment—including dust, extreme temperature variations, and moonquakes—can be addressed with existing technology or with clear development pathways.

### 6.1.1 Advantages of *AeSI* Over a Free-Flyer

One of the most significant advantages of *AeSI* is that it removes the need for formation flying of ~31 separate spacecraft, a key technical hurdle in the original *SI* concept. Instead, the Moon's surface provides a stable platform for the interferometer, and any transient disturbances, such as lunar seismic activity, can be detected with onboard seismometers and later removed from the data without compromising science return.

The Artemis program plays a pivotal role in making *AeSI* a competitive alternative to a free-flyer. Even in its Baseline design, *AeSI* benefits from Artemis in several key ways:
- **Transportation and Deployment**: Artemis infrastructure provides launch, transit to

lunar orbit, and landing capabilities, eliminating concerns about total mass and volume. The planned availability of Starship (with New Glenn and SLS as alternatives) ensures sufficient payload capacity for delivering *AeSI* to the lunar surface. Additionally, general-purpose rovers under Artemis could facilitate the transport of *AeSI* components from the landing site to the interferometer's deployment location.
- **Servicing and Upgradability**: Unlike a free-flying observatory, which would require highly complex robotic servicing missions, *AeSI* could be maintained and upgraded by either human or robotic systems. This includes dust mitigation, replacing failed components using modular designs, and expanding the observatory's capabilities over time by deploying additional mirror elements and improved detectors.

### 6.1.2 Optimizing *AeSI* Through Artemis Integration

To fully demonstrate the superiority of *AeSI* over a free-flyer, further integration with Artemis infrastructure is necessary. Our Baseline design demonstrated technical feasibility, but the next step is to incorporate additional Artemis-enabled capabilities, such as:
- **Power Alternatives**: Transitioning from solar-powered batteries to alternative power solutions—such as power beaming from a central Artemis power station or a surface nuclear fission reactor—could reduce mass and provide continuous, reliable energy for operations.
- **Hardline Communications**: Establishing a direct communication link between the *AeSI* hub and the Artemis base would improve data transmission reliability compared to relay-based solutions.
- **Lunar PNT (Positioning, Navigation, and Timing) Services**: Utilizing lunar PNT infrastructure would enhance positioning accuracy and further simplify operational constraints.

By relying more on Artemis-provided infrastructure, the mass requirements for *AeSI* could be significantly reduced, further improving its feasibility.

### 6.1.3 Key Challenges and Consideration

While *AeSI* has many advantages, there are some areas where the free-flyer currently holds an edge, particularly in potential science yield. Because the lunar version must be co-located with the Artemis base to take advantage of the infrastructure, its sky coverage is constrained—primarily looking downward from the south pole—whereas a free-flyer has greater flexibility in targeting.

Additionally, some aspects of *AeSI* require further development to match or exceed *SI*'s capabilities. These include:
- **High-Reflectivity UV Mirror Coatings**: To optimize performance, advancements in UV reflectivity are needed.
- **High-Quantum-Efficiency UV Detectors**: Improving detector efficiency would enhance

scientific return.
- **Dust Mitigation Strategies**: While maintenance is possible, a more detailed plan for active and passive dust mitigation is necessary.
- **Power Solutions**: The high battery masses in the Baseline design indicate the need to explore alternative thermal management solutions. A compelling option is a nuclear fission power source, such as those under development at Glenn Research Center for Artemis missions[14]. Such a system would provide continuous power and enable nighttime operations.
- **Delay Line Mechanism Maturation**: This component follows a normal technology maturation pathway and does not present extraordinary development risks.

### 6.1.4 Path Forward

Building on the findings of our Phase I study, the next steps involve both internal research and development (IRAD) efforts and further refinement of *AeSI's* design to fully capitalize on Artemis infrastructure. Advancing key technologies will ensure that *AeSI* not only remains competitive with free-flying alternatives but also surpasses them in capability, operational longevity, and scientific yield.

Internal Research and Development Priorities
Several critical technology areas require near-term development to advance *AeSI* from concept to a fully optimized, deployable observatory:
- **Delay Line Mechanism** – Designing and developing a breadboard unit to validate the precision and stability required for the interferometric delay lines.
- **Thermal Management Systems** – Exploring pumped fluid loops and other heat transport methods to support large-scale lunar architectures, including maximizing the turn-down ratio on heat switches or other thermal control components to ensure effective operation during the lunar night.
- **Cryogenic Mechanisms** – Maturing cryogenic actuators and positioning systems to achieve nanometer-scale precision for key optical elements.
- **Detector Development** – Advancing UV/optical detectors with the necessary 2 µm pixel pitch at required wavelengths, or alternatively, updating the optical model to accommodate state-of-the-art detector technologies with slightly different specifications.

These IRAD efforts will address some of the primary technical challenges identified in our Baseline design and pave the way for a more robust, high-performance lunar interferometer.

Expanding *AeSI's* Scientific Potential
While *AeSI's* Baseline design is already competitive, further optimization—especially in sensitivity, stability, and operational efficiency—could unlock even greater scientific

---

[14] https://www.nasa.gov/space-technology-mission-directorate/tdm/fission-surface-power/

opportunities. Several key challenges, if addressed, could enable breakthrough discoveries:
- **Exoplanet Atmospheres & Lyman-α Imaging** – The current Baseline design does not achieve sufficient SNR in the wings of Lyman-α (see Section 3.4.5) to robustly detect ionized hydrogen escaping from exoplanet atmospheres. Improving sensitivity at 1216 Å would enable direct imaging of escaping hydrogen tails, allowing measurements of both their height and length. Additionally, by resolving blue-shifted Lyman-α absorption, *AeSI* could constrain the stellar wind ram pressure, providing insights into atmospheric escape processes driven by stellar winds.
- **Extrasolar Aurorae on Brown Dwarfs** – Brown dwarfs, with negligible thermal UV emission but powerful auroral UV output ($>10^{19}$ W), are promising targets for detecting extrasolar aurorae. A tentative detection was made using HST (Saur et al. 2021) in the 3065–3110 Å range, potentially linked to TiO emission excited by auroral electrons. By improving SNR in the 3000 Å region, *AeSI* could directly image auroral emissions on substellar objects, offering unprecedented insight into their magnetospheric processes.
- **Protoplanetary Disks & Star Formation** – UV/optical interferometry has the potential to revolutionize our understanding of planetary system formation by enabling high-resolution imaging of young stars, their disks, and associated outflows. Magnetic fields are crucial in these environments, governing angular momentum transport and influencing jet formation. *AeSI* could resolve accretion foot-points where disk material flows onto the star, shedding light on mass accretion processes and their role in shaping planetary systems. Additionally, imaging jet launch regions and magnetic field structures could clarify how disks evolve and interact with their host stars.

However, the feasibility of such observations depends on target accessibility. While eight nearby star-forming regions—including the Chamaeleon molecular cloud complex and Orion—fall within *AeSI's* field of regard, UV/optical observations of protoplanetary disks face significant limitations due to dust obscuration (see Ohashi et al. 2023). For instance, the proplyds in Orion, with fluxes on the order of $10^{-12}$ erg s$^{-1}$ cm$^{-2}$ (O'Dell & Wen 1994), are too faint for *AeSI's* current Baseline design. As a result, YSOs are not considered viable targets unless further advancements significantly enhance *AeSI's* sensitivity.

By continuing to refine *AeSI's* design—both through targeted technology maturation and deeper integration with Artemis infrastructure—we can ensure that a lunar-based UV/optical interferometer is not just feasible, but uniquely capable of addressing some of the most pressing questions in astrophysics.

## 6.2 Looking to the Future

*AeSI* stands at the intersection of cutting-edge technology and transformative science, poised to redefine our understanding of the universe. In April 2024, NASA published a document identifying 187 critical shortfalls—technology areas requiring further development to meet future exploration, science, and mission needs. The top two shortfalls, as ranked by the aerospace community, are surviving and operating through the lunar night and high-power energy generation on the Moon and Mars. These priorities, led by NASA's STMD, directly align with the innovations needed for *AeSI*'s success, ensuring that significant advancements will be made in the very areas we rely on. Interferometry has long held the promise of unlocking unprecedented detail in the cosmos, and *AeSI* will finally deliver on that promise, imaging in the ultraviolet and, for the first time, fully resolving the surfaces of distant stars. This leap forward will illuminate the physics of stellar activity and will also enable major advances in our understanding of AGN cores, supernovae, planetary nebulae, interacting binaries, stellar winds and pulsations, forming stars and disks, and evolved, dying stars. Beyond its scientific impact, *AeSI* will inspire the public by transforming the Moon into a new hub for astronomical discovery, demonstrating what humanity can achieve when exploration and science go hand in hand. As we take these steps toward the future, *AeSI* stands as a beacon of what is to come: a revolutionary new way to image the Universe.

# 7 Publications, Presentations, and Outreach Events

In Phase I, the *AeSI* team has generated a range of scientific contributions and community engagement efforts including publications, conference presentations, and media appearances.

We have worked with Paul Morris and the CI Lab at NASA Goddard Space Flight Center to create two outreach videos summarizing the mission concept at a high level:
https://vimeo.com/1035638528/63fdde18e2?share=copy
And another version that includes an interview with NIAC Fellow Dr. Kenneth Carpenter, which can be found on the *AeSI* website: https://hires.gsfc.nasa.gov/si/*AeSI*.html

## 7.1 Publications

Rau, G., Carpenter, K. G., Boyajian, T., Creech-Eakman, M., Foster, J., Karovska, M., Leisawitz, D., Morse, J. A., Mozurkewich, D., Peacock, S., Petro, N., Scowen, P., Sitarski, B., van Belle, G., Wilkinson, E. *Artemis-enabled Stellar Imager (AeSI): a Lunar long-baseline UV/optical imaging interferometer.* Proceedings of the SPIE, Volume 13095, id. 130951J, 6 pp. (2024).

## 7.2 Presentations

Naval Research Lab (NRL) | Colloquium | Feb. 2025
"The *Artemis-enabled Stellar Imager* (*AeSI*): UV/Optical Interferometry from the Lunar Surface", Kenneth Carpenter

AAS Conference | Talks | January 2025 | "*Artemis-enabled Stellar Imager* (*AeSI*): Observing the Universe in Ultra High Definition", Kenneth Carpenter
"Science Cases for the *Artemis-enabled Stellar Imager* (*AeSI*)", Gioia Rau

NOIRLab/Steward Observatory | Colloquium | December 2024
"The *Artemis-enabled Stellar Imager* (*AeSI*): UV/Optical Interferometry from the Lunar Surface", Kenneth Carpenter

NIAC Symposium | Talk & Poster | September 2024
"*Artemis-enabled Stellar Imager* (*AeSI*): Observing the Universe in High Definition", Kenneth Carpenter

SPIE Conference | Talk |  August 2024
"*Artemis-enabled Stellar Imager* (*AeSI*): a Lunar long-baseline UV/optical imaging interferometer", Gioia Rau

NIAC Orientation | Talk & Poster | March 2024
"Imaging the Surfaces of Distant Stars with Sub-Milli-Arcsec Resolution: *Artemis-enabled Stellar Imager* (*AeSI*)", Kenneth Carpenter

## 7.3 Outreach and Media Coverage

Universe Today Article | September 2024
"Artemis Missions Could Put the most Powerful imaging Telescope on the Moon"
https://www.universetoday.com/168511/artemis-missions-could-put-the-most-powerful-imaging-telescope-on-the-moon/

BBC Sky at Night Magazine Article | September 2024
"Proposed telescope on the Moon would give us close-up views of stars and black holes. But we'd better be quick…"
https://www.skyatnightmagazine.com/space-missions/artemis-enabled-stellar-imager

Phys.org Article | September 2024
"Artemis missions could put the most powerful imaging telescope on the moon"
https://phys.org/news/2024-09-artemis-missions-powerful-imaging-telescope.html

Planetary Radio Interview | September 2024
"2024 NASA Innovative Advanced Concepts Symposium: Part 2 - Stellar imaging and looking for life while mining water on Mars"
https://www.planetary.org/planetary-radio/2024-niac-part-2

Universe Today Article | January 2024
"NASA Wants to Put a Massive Telescope on the Moon"
https://www.universetoday.com/165397/nasa-wants-to-put-a-massive-telescope-on-the-moon/

Scott Manley YouTube Video | January 2024
"NASA Is Giving Money To Develop These Insane New Technologies"
https://www.youtube.com/watch?v=WFER_6uPwIk&ab_channel=ScottManley

Gold Pan Alumni Magazine, New Mexico Tech | Summer 2024
"Lunar Interferometry"
https://nmt.edu/advancement/gold_pan.php

# 8 Acknowledgements

The NIAC *AeSI* Concept Development Team wishes to express our thanks for the incredible amount and quality of work that the GSFC Integrated Design Center (IDC) put into our collaboration with them to ensure that we developed a feasible and credible mission architecture for our Phase I *AeSI* design. They deserve many thanks for the success of our Phase I Study, which showed that a lunar-based UV/optical Interferometer can indeed be built and for helping define our path forward to designing an even more capable observatory as we proceed into further, more detailed studies and development efforts.

**The IDC Team:**
**Discipline Engineers**
- Communications – George Bussey
- Detectors – Tilak Hewagama
- Power – Ryan Flora and Tim Petry
- Mechanical Systems/Design – Buddy Taylor and Walt Smith
- Mechanisms – Rajeev Sharma
- Optical – Bert Pasquale
- Thermal – Sharon Peabody
- Instrument Systems Engineering – Brian Ottens
- Deputy Team Lead – Beth Paquette
- Team Lead - Kan Yang
- MDL Systems – Craig Stevens

**IDC Staff**
- IDC Manager – Liz Matson
- Facilitators & IT Support – Will Smith and Henry Cao

**Consultants**
- Contamination - Therese Errigo
- Dust Mitigation – Mark Hasegawa

**Shadows**
- Caeden Burcham (Mechanical)
- Laiya Ackman (Systems)
- Arya Kazemnia (Communications)
- Cole Stephens
- Akesh Mallia
- Kruti Dharanipathi
- Samuel Obiorah
- Natalie Fullerman